\documentclass{iopart}
\usepackage{graphicx}

\usepackage{lineno}
\usepackage[title,titletoc]{appendix}

\usepackage{orcidlink}
\usepackage{hyperref}
\hypersetup
{
	colorlinks=true,		
	linkcolor=,				
	citecolor=red,			
	filecolor=green,		
	urlcolor=blue		
}
\usepackage{doi}
\usepackage{xcolor}
\usepackage{amsmath,amssymb}
\usepackage[utf8]{inputenc}
\usepackage[T1]{fontenc}
\usepackage{comment}
\usepackage{fancyhdr}
\usepackage[numbers]{natbib}
\def\newblock{\ }
\usepackage[range-phrase=\,--\,, range-units=single]{siunitx}
\DeclareSIUnit \count {counts}
\DeclareSIUnit \ppm {ppm}
\DeclareSIUnit \sqrthz {\ensuremath{\sqrt{\text{\hertz}}}}
\DeclareSIUnit \strain {\ensuremath{\text{strain}}}
\DeclareSIUnit \parsec {pc}
\DeclareSIUnit \radian {rad}
\DeclareSIUnit \angdeg {deg}

\usepackage{acronym}

\usepackage[normalem]{ulem} 

\begin{document}

\title{Advanced Virgo during the LIGO-Virgo-KAGRA fourth observing run}

\author{%
F Acernese$^{2,3}$ %
A Agapito\orcidlink{0009-0005-9004-3163}$^{4}$ %
D Agarwal\orcidlink{0000-0002-8735-5554}$^{5}$ %
I-L Ahrend$^{6}$ %
L Aiello\orcidlink{0000-0003-2771-8816}$^{7,8}$ %
A Ain\orcidlink{0000-0003-4534-4619}$^{9}$ %
W Ali$^{10,11}$ %
A Allocca\orcidlink{0000-0002-5288-1351}$^{12,3}$ %
W Amar\orcidlink{0009-0003-5623-8819}$^{13}$ %
A Amato\orcidlink{0000-0001-9557-651X}$^{14,15}$ %
F Amicucci\orcidlink{0009-0005-2139-4197}$^{16,17}$ %
C Amra$^{18}$ %
M Andia\orcidlink{0000-0003-3675-9126}$^{19}$ %
T Andri\'c\orcidlink{0000-0002-9277-9773}$^{20,21}$ %
S Antier\orcidlink{0000-0002-7686-3334}$^{19}$ %
F Arciprete\orcidlink{0000-0003-3602-3717}$^{7,8}$ %
F Armato\orcidlink{0000-0002-8856-8877}$^{10,11}$ %
N Arnaud\orcidlink{0000-0001-6589-8673}$^{22}$ %
L Asprea$^{23}$ %
M Assiduo$^{24,25}$ %
S Assis de Souza Melo\orcidlink{0000-0002-1550-1671}$^{26}$ %
P Astone\orcidlink{0000-0003-4981-4120}$^{16}$ %
F Attadio\orcidlink{0009-0008-8916-1658}$^{17,16}$ %
F Aubin\orcidlink{0000-0003-1613-3142}$^{27}$ %
G Avallone\orcidlink{0000-0001-5482-0299}$^{2}$ %
N Avdeev\orcidlink{0009-0005-0413-633X}$^{23}$ %
S Babak\orcidlink{0000-0001-7469-4250}$^{6}$ %
S Bagnasco\orcidlink{0000-0001-6062-6505}$^{23}$ %
S Baimukhametova\orcidlink{0009-0006-0971-8619}$^{28,29}$ %
T Baka\orcidlink{0000-0002-5629-3813}$^{30,15}$ %
G Balbi$^{31}$ %
G Baldi\orcidlink{0000-0001-8963-3362}$^{32,33}$ %
N Baldicchi\orcidlink{0009-0009-8888-291X}$^{34,35}$ %
G Ballardin$^{26}$ %
M Ballelli\orcidlink{0000-0003-1512-5423}$^{20,21}$ %
B Banerjee\orcidlink{0000-0002-8008-2485}$^{20}$ %
M Baratti\orcidlink{0009-0003-5744-8025}$^{36,37}$ %
F Barone\orcidlink{0000-0002-8069-8490}$^{38,3}$ %
M Barsuglia\orcidlink{0000-0002-1180-4050}$^{6}$ %
D Barta\orcidlink{0000-0001-6841-550X}$^{39}$ %
A Basti\orcidlink{0000-0003-2895-9638}$^{37,36}$ %
M Bawaj\orcidlink{0000-0003-3611-3042}$^{34,35}$ %
M Bazzan$^{40,41}$ %
F Beirnaert\orcidlink{0000-0002-4003-7233}$^{42}$ %
M Bejger\orcidlink{0000-0002-4991-8213}$^{43}$ %
C Bellani\orcidlink{0000-0003-3267-1450}$^{44}$ %
D Beltran-Martinez\orcidlink{0000-0003-4580-3264}$^{45}$ %
E Benedetti\orcidlink{0009-0008-5230-0597}$^{16}$ %
I Bentara\orcidlink{0009-0000-5074-839X}$^{22}$ %
S Bera\orcidlink{0000-0003-0907-6098}$^{46}$ %
D Bersanetti\orcidlink{0000-0002-7377-415X}$^{10}$ %
T Bertheas\orcidlink{0009-0005-4118-4170}$^{47}$ %
A Bertolini$^{15,14}$ %
J Bezerra-Sobrinho\orcidlink{0000-0003-2183-4488}$^{48}$ %
V Biancalana\orcidlink{0000-0002-1642-5391}$^{49}$ %
F Bianchi$^{35}$ %
M Bilicki\orcidlink{0000-0002-3910-5809}$^{50}$ %
A Binetti\orcidlink{0000-0001-6449-5493}$^{44}$ %
S Biot$^{51}$ %
M Bitossi\orcidlink{0000-0002-9862-4668}$^{26,36}$ %
M-A Bizouard\orcidlink{0000-0002-4618-1674}$^{52}$ %
M Bloch$^{53}$ %
G Boileau\orcidlink{0000-0002-3576-6968}$^{52}$ %
M Boldrini\orcidlink{0000-0001-9861-821X}$^{26}$ %
R Bonnand\orcidlink{0000-0001-5013-5913}$^{13,54}$ %
N Borghi\orcidlink{0000-0002-2889-8997}$^{55,31}$ %
V Boschi\orcidlink{0000-0001-8665-2293}$^{36}$ %
Y Bothra\orcidlink{0000-0002-9380-6390}$^{15,56}$ %
A Boudon$^{22}$ %
A Bozzi$^{26}$ %
C Bradaschia$^{36}$ %
M Branchesi\orcidlink{0000-0003-1643-0526}$^{20,21}$ %
T Briant\orcidlink{0000-0002-6013-1729}$^{57}$ %
A Brillet\footnote{Deceased, March 2026.}$^{52}$ %
M L Brozzetti\orcidlink{0000-0002-5260-4979}$^{34,35}$ %
G Bruno$^{5}$ %
F Bucci\orcidlink{0000-0003-1726-3838}$^{25}$ %
A Buchicchio$^{17}$ %
A Buggiani$^{26}$ %
O Bulashenko\orcidlink{0000-0003-1720-4061}$^{58,59}$ %
T Bulik$^{60}$ %
H J Bulten$^{15}$ %
R Buscicchio\orcidlink{0000-0002-7387-6754}$^{61,62}$ %
N Busdon$^{40}$ %
% ...
{(The Virgo Collaboration)}
}
% ...
\address{$^{2}$Dipartimento di Fisica ``E.R. Caianiello'', Universit\`a di Salerno, I-84084 Fisciano, Salerno, Italy}
\address{$^{3}$INFN, Sezione di Napoli, I-80126 Napoli, Italy}
\address{$^{4}$Centre de Physique Th\'eorique, Aix-Marseille Universit\'e, Campus de Luminy, 163 Av. de Luminy, 13009 Marseille, France}
\address{$^{5}$Universit\'e catholique de Louvain, B-1348 Louvain-la-Neuve, Belgium}
\address{$^{6}$Universit\'e Paris Cit\'e, CNRS, Astroparticule et Cosmologie, F-75013 Paris, France}
\address{$^{7}$Universit\`a di Roma Tor Vergata, I-00133 Roma, Italy}
\address{$^{8}$INFN, Sezione di Roma Tor Vergata, I-00133 Roma, Italy}
\address{$^{9}$Universiteit Antwerpen, 2000 Antwerpen, Belgium}
\address{$^{10}$INFN, Sezione di Genova, I-16146 Genova, Italy}
\address{$^{11}$Dipartimento di Fisica, Universit\`a degli Studi di Genova, I-16146 Genova, Italy}
\address{$^{12}$Universit\`a di Napoli ``Federico II'', I-80126 Napoli, Italy}
\address{$^{13}$Univ. Savoie Mont Blanc, CNRS, Laboratoire d'Annecy de Physique des Particules - IN2P3, F-74000 Annecy, France}
\address{$^{14}$Maastricht University, 6200 MD Maastricht, Netherlands}
\address{$^{15}$Nikhef, 1098 XG Amsterdam, Netherlands}
\address{$^{16}$INFN, Sezione di Roma, I-00185 Roma, Italy}
\address{$^{17}$Universit\`a di Roma ``La Sapienza'', I-00185 Roma, Italy}
\address{$^{18}$Aix Marseille Univ, CNRS, Centrale Med, Institut Fresnel, F-13013 Marseille, France}
\address{$^{19}$Universit\'e Paris-Saclay, CNRS/IN2P3, IJCLab, 91405 Orsay, France}
\address{$^{20}$Gran Sasso Science Institute (GSSI), I-67100 L'Aquila, Italy}
\address{$^{21}$INFN, Laboratori Nazionali del Gran Sasso, I-67100 Assergi, Italy}
\address{$^{22}$Universit\'e Claude Bernard Lyon 1, CNRS, IP2I Lyon / IN2P3, UMR 5822, F-69622 Villeurbanne, France}
\address{$^{23}$INFN Sezione di Torino, I-10125 Torino, Italy}
\address{$^{24}$Universit\`a degli Studi di Urbino ``Carlo Bo'', I-61029 Urbino, Italy}
\address{$^{25}$INFN, Sezione di Firenze, I-50019 Sesto Fiorentino, Firenze, Italy}
\address{$^{26}$European Gravitational Observatory (EGO), I-56021 Cascina, Pisa, Italy}
\address{$^{27}$Universit\'e de Strasbourg, CNRS, IPHC UMR 7178, F-67000 Strasbourg, France}
\address{$^{28}$D\'epartement de Physique Nucl\'eaire et Corpusculaire, Universit\'e de Gen\`eve, 24 quai E. Ansermet, CH-1211 Geneva, Switzerland}
\address{$^{29}$Gravitational Wave Science Center, UniGe, -, Switzerland}
\address{$^{30}$Institute for Gravitational and Subatomic Physics (GRASP), Utrecht University, 3584 CC Utrecht, Netherlands}
\address{$^{31}$Istituto Nazionale Di Fisica Nucleare - Sezione di Bologna, viale Carlo Berti Pichat 6/2 - 40127 Bologna, Italy}
\address{$^{32}$Universit\`a di Trento, Dipartimento di Fisica, I-38123 Povo, Trento, Italy}
\address{$^{33}$INFN, Trento Institute for Fundamental Physics and Applications, I-38123 Povo, Trento, Italy}
\address{$^{34}$Universit\`a di Perugia, I-06123 Perugia, Italy}
\address{$^{35}$INFN, Sezione di Perugia, I-06123 Perugia, Italy}
\address{$^{36}$INFN, Sezione di Pisa, I-56127 Pisa, Italy}
\address{$^{37}$Universit\`a di Pisa, I-56127 Pisa, Italy}
\address{$^{38}$Dipartimento di Medicina, Chirurgia e Odontoiatria ``Scuola Medica Salernitana'', Universit\`a di Salerno, I-84081 Baronissi, Salerno, Italy}
\address{$^{39}$HUN-REN Wigner Research Centre for Physics, H-1121 Budapest, Hungary}
\address{$^{40}$Universit\`a di Padova, Dipartimento di Fisica e Astronomia, I-35131 Padova, Italy}
\address{$^{41}$INFN, Sezione di Padova, I-35131 Padova, Italy}
\address{$^{42}$Universiteit Gent, B-9000 Gent, Belgium}
\address{$^{43}$Nicolaus Copernicus Astronomical Center, Polish Academy of Sciences, 00-716, Warsaw, Poland}
\address{$^{44}$Katholieke Universiteit Leuven, Oude Markt 13, 3000 Leuven, Belgium}
\address{$^{45}$Centro de Investigaciones Energ\'eticas Medioambientales y Tecnol\'ogicas, Avda. Complutense 40, 28040, Madrid, Spain}
\address{$^{46}$Aix-Marseille Universit\'e, Universit\'e de Toulon, CNRS, CPT, Marseille, France}
\address{$^{47}$Laboratoire des 2 infinis - Toulouse, Universit\'e de Toulouse, CNRS/IN2P3, Toulouse, France, Toulouse, France}
\address{$^{48}$Federal University of Rio Grande do Norte, Campus Universit\'ario - Lagoa Nova, Natal - RN, 59078-970, Brazil}
\address{$^{49}$Universit\`a di Siena, Dipartimento di Scienze Fisiche, della Terra e dell'Ambiente, I-53100 Siena, Italy}
\address{$^{50}$Center for Theoretical Physics, Polish Academy of Sciences, 02-668, Warsaw, Poland}
\address{$^{51}$Universit\'e libre de Bruxelles, 1050 Bruxelles, Belgium}
\address{$^{52}$Universit\'e C\^ote d'Azur, Observatoire de la C\^ote d'Azur, CNRS, Artemis, F-06304 Nice, France}
\address{$^{53}$Subatech, CNRS/IN2P3 - IMT Atlantique - Nantes Universit\'e, 4 rue Alfred Kastler BP 20722 44307 Nantes C\'EDEX 03, France}
\address{$^{54}$Centre national de la recherche scientifique, 75016 Paris, France}
\address{$^{55}$DIFA- Alma Mater Studiorum Universit\`a di Bologna, Via Zamboni, 33 - 40126 Bologna, Italy}
\address{$^{56}$Department of Physics and Astronomy, Vrije Universiteit Amsterdam, 1081 HV Amsterdam, Netherlands}
\address{$^{57}$Laboratoire Kastler Brossel, Sorbonne Universit\'e, CNRS, ENS-Universit\'e PSL, Coll\`ege de France, F-75005 Paris, France}
\address{$^{58}$Institut de Ci\`encies del Cosmos (ICCUB), Universitat de Barcelona (UB), c. Mart\'i i Franqu\`es, 1, 08028 Barcelona, Spain}
\address{$^{59}$Departament de F\'isica Qu\`antica i Astrof\'isica (FQA), Universitat de Barcelona (UB), c. Mart\'i i Franqu\'es, 1, 08028 Barcelona, Spain}
\address{$^{60}$Astronomical Observatory, University of Warsaw, 00-478 Warsaw, Poland}
\address{$^{61}$Universit\`a degli Studi di Milano-Bicocca, I-20126 Milano, Italy}
\address{$^{62}$INFN, Sezione di Milano-Bicocca, I-20126 Milano, Italy}
% ...

\begin{abstract}
  From April 10, 2024 to November 18, 2025
  Advanced Virgo participated
  in the fourth observing run of the network of gravitational-wave
  detectors, together with Advanced LIGO and KAGRA. For this observing
  run Advanced Virgo has completed its design optical configuration
  with the installation of a signal recycling mirror.
  In this paper we
  describe the challenges encountered in commissioning this optical
  configuration, alongside the other upgrades performed between the
  third and fourth observing run. The Virgo detector operated with a
  68.9\% duty cycle and with an angle-averaged median
  range to binary neutron star mergers of \qty{53}{Mpc}.
\end{abstract}

\tableofcontents

\section{Introduction}
\label{sec:introduction}

Advanced Virgo is a \ac{GW} detector in Cascina, Italy~\cite{aVirgo}. Following the
second and third observing runs joint with the two Advanced LIGO
detectors in the USA, the Virgo detector underwent the first phase of
the ``Advanced Virgo Plus'' upgrade project \cite{VIR-0596A-19, Flaminio:2020lqk}.
That upgrade affected most sub-systems of the detector,
most notably it changed the detector optical configuration from a power
recycled \ac{FP} Michelson interferometer to a dual-recycled
\ac{FP} Michelson as planned originally for Advanced Virgo, and
included the installation of a \qty{285}{m} filter cavity to produce frequency dependent
squeezing~\cite{virgocoll:FrequencyDependentSqueezedVacuum2023}. 

In this paper we describe the Virgo configuration during \ac{O4} in
which Virgo participated from April 10, 2024 to November 18, 2025,
alongside some of the upgrade and commissioning challenges that preceded
the observing run. We focus on the final state of the instrument used
for the observations, without chronologically retracing all the steps
that lead to it, in order to avoid an excessively long description.
This paper is complementary to the results on the
measurements of optical parameters of the Virgo interferometer during \ac{O4}
that have been published previously~\cite{VIRGO:2025sym}.

In section~\ref{sec:optical-configuration}
we
describe the upgraded optical configuration, then in
section~\ref{sec:nearly-unstable-recycling} we discuss how the
challenges of that optical configuration were overcome, and
in section~\ref{sec:injection-noise} how laser noises
coupling was reduced despite these challenges. We then
continue with a description of the dominant noise limiting the Virgo
sensitivity during \ac{O4} in section~\ref{sec:dominant-noise}.
 These four sections present the
difficulties in operating dual recycling with nearly unstable
recycling cavities, which among others required lowering the input laser power to
be overcome; and the consequences of that configuration on
interferometer noise levels, which resulted in Virgo being operated
with a misaligned signal recycling cavity during \ac{O4}.

This is followed by the description
of studies of scattered light and environmental noise coupling in
section~\ref{sec:scatter-env}.
There were also many effective upgrades to the
detector, which have achieved their goals and are described in
section~\ref{sec:instrument-upgrade}. However, most of them did not
result in overall detector improvements due to challenges of nearly
unstable dual-recycling masking their effect. Finally in
section~\ref{sec:performance} the overall detector performance during
\ac{O4} is summarized, and in section~\ref{sec:future} we conclude with plans
for detector improvements based on the \ac{O4} commissioning
experience.

\section{Nearly unstable optical configuration}
\label{sec:optical-configuration}

\begin{figure}
  \centering
  \includegraphics[width=\textwidth]{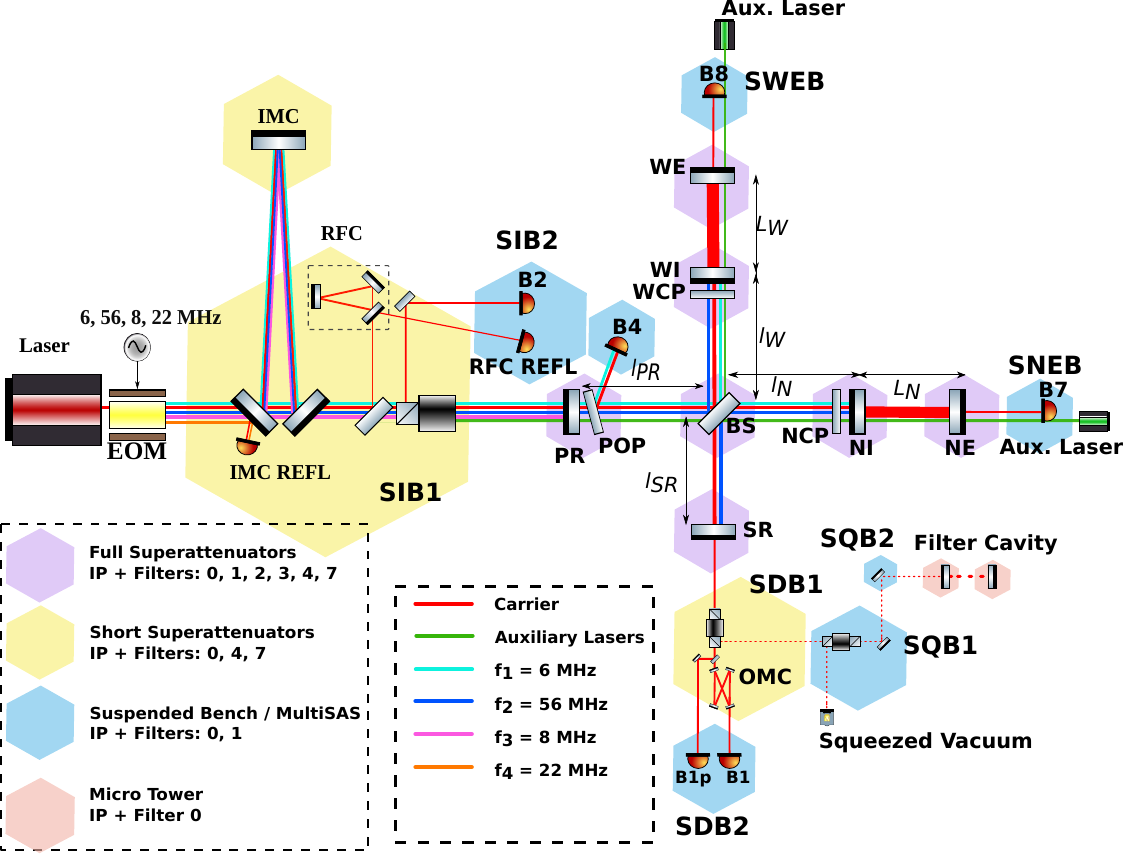}
  \caption{Optical configuration of the Advanced Virgo detector
    during \ac{O4}. The \qty{1064}{nm} laser is going
    through the \ac{IMC} towards the
    core optics: power recycling mirror, beam splitter, north input
    and end mirrors, west input and end mirrors, and signal recycling
    mirror. Different colored lines represent the carrier and
    sidebands, with the modulation frequencies listed in the caption
    at the bottom of the schematics.
  }
  \label{fig:Virgoscheme}
\end{figure}

During \ac{O4}, the Advanced Virgo detector~\cite{aVirgo}
operated in
a dual-recycled Michelson interferometer with \ac{FP}
arm cavities configuration as depicted in figure~\ref{fig:Virgoscheme}. The
interferometer has two three-kilometer-long perpendicular arms, the
north arm and the west arm. The laser light, from the main
laser source, is sent into the arms using a \ac{BS}. Each
arm is itself a \ac{FP} cavity, featuring one \ac{IM} and one
\ac{EM}, which increases the effective arm length. The input
mirrors are referred to as \ac{WI} and \ac{NI} for the west and north arms respectively,
while the end mirrors are referred to as \ac{WE} and \ac{NE}. A semi-reflective
mirror, called the \ac{PR} mirror, is present between the laser source and
the beam splitter, forming the \ac{PRC} with \ac{NI} and \ac{WI} to increase the
resonant circulating power inside the entire detector. Another
semi-reflective mirror, \ac{SR}, is placed between the beam splitter
and the output port of the interferometer. Together with the two input
mirrors, they form the \ac{SRC}, which is used to increase the
bandwidth of the interferometer. The \ac{SR} mirror was not present
during \ac{O3}, as Advanced Virgo was previously operated in the
power-recycled configuration.

In addition to the main optics mentioned above,
there are also two \acp{CP} installed in
front of the input mirrors of each arm cavity for wavefront
compensation, respectively the north (NCP) and west (WCP) \acl{CP}. More details on the \acp{CP} are given in Section \ref{sec:thermal-compensation}. A pick-off of the beam circulating inside the \ac{PRC},
extracted by
the pick-off-plate (POP), is sent to a set of sensors for control and
diagnostic purposes.

The laser source of the detector is a Nd:YAG non-planar ring oscillator, with a wavelength of
\qty{1064}{nm}. The beam emitted from the oscillator is fiber coupled into a
fiber amplifier. The beam is then coupled to free space in air,
amplified by a slave laser and a neoVAN free space optical amplifier,
and passes through a pre-mode cleaner.
Finally, three
custom \acp{EOM} add phase modulated
optical \ac{RF} sidebands, with each \ac{EOM} adding either one or two
distinct modulation frequencies. The beam is
then injected into the vacuum system, where it resonates in the \ac{IMC}
to reach its required power, shape, frequency and pointing
stability and is then transmitted using several mode matching optics
to finally inject \qty{17}{W} of laser power into the interferometer. The \ac{IMC}, a \qty{144}{m} long triangular cavity with a
finesse of 1200, is used to suppress the \acp{HOM}, to control the
beam jitter, and to reduce the laser frequency noise above
\qty{500}{Hz}. A
pick-off of the main beam is sent to the \ac{RFC}, to measure and
stabilize the frequency of the laser below a few hertz. For the
longitudinal control (length sensing) of optical cavities, the
\ac{PDH} technique~\cite{Black01,PDH_1983} is applied. Four modulation frequencies are injected,
\qty{6270778}{Hz}, \qty{8361037}{Hz}, \qty{22304000}{Hz} and \qty{56437004}{Hz}, as illustrated in figure~\ref{fig:Virgoscheme}.
They are referred to by only their integer frequency in MHz, rounded down.
The 22\,MHz is rejected by the \ac{IMC} and it is used to control the \ac{IMC}, while the
others are transmitted and used to control the rest of the interferometer as described in
Section~\ref{sec:lock-acquisition-procedure}.

In Virgo, both the \ac{PRC} and \ac{SRC} are nearly unstable
cavities. The stability of an optical cavity is characterized by the
cavity g-factor $g$. A cavity is said to be geometrically stable for
\begin{linenomath}
  \begin{equation}
    0 < g < 1 \,,
    \label{eq:cav_stab}
  \end{equation}
\end{linenomath}
where $g=g_1 g_2$ is the cavity g-factor with
$g_i = 1-L/R_i$, $R_i$ are the \acp{RoC} of the two mirrors forming the cavity, and
$L$ is the length of the cavity. In stable resonators whose
g-factor verifies this condition, the resonator transverse modes are
well-defined.  These modes, called the resonator eigenmodes, can be
described in a polynomial orthonormal basis, such as the Hermite-Gauss or Laguerre-Gauss bases
\cite{Kogelnik1966}.  Outside the limits of equation~\eqref{eq:cav_stab},
the resonator is unstable and there are no defined cavity
eigenmodes. This leads to unbound divergent beams in the cavity and
loss of light. In the case of the Virgo recycling cavities the nominal
g-factor is very close to 1, at about 0.999988, and the cavities are
nearly unstable.

For a \ac{FP} cavity, the round-trip Gouy phase can be written in
terms of the cavity g-factor \cite{siegman1986lasers} as
\begin{linenomath}
  \begin{equation}
    \phi_{\mathrm{Gouy}} = 2\arccos{(\pm \sqrt{g})}\,,
    \label{eq:gouy-roundtrip}
  \end{equation}
\end{linenomath}
where the $\pm$ sign is given by the sign of $g$.
For a \ac{HOM} with $m$ and $n$ being its transverse
mode indices, the round-trip Gouy phase~\cite{Kogelnik1966} is simply 
\begin{linenomath}
  \begin{equation}
    \phi_{\mathrm{Gouy}}^{mn} = (m+n+1)\,\phi_{\mathrm{Gouy}}\,.
  \end{equation}
\end{linenomath}

From equation~\eqref{eq:gouy-roundtrip}, it can be seen that at the edges of
stability the round-trip Gouy phase is equal to 0 for $g=1$.
Thus, for a nearly unstable cavity with $g\simeq1$, the mode
separation is 0 and all the \acp{HOM} resonate alongside the
fundamental mode. The Virgo recycling cavities are \qty{12}{m} long;
they are formed by either the \ac{PR} or the \ac{SR} mirror (with a design \ac{RoC} of \qty{1430}{m})
on one end of the cavity, and the compound \acp{IM} (with a design \ac{RoC} of
\qty{1420}{m}) on the other end.
The corresponding design round-trip Gouy
phase is \ang{0.39}.

The \ac{SRC} is tuned such that the carrier fundamental mode that is
resonant in the \ac{FP} arm cavities is anti-resonant in the
\ac{SRC}. This condition, combined with a g-factor very close to 1,
causes the carrier \acp{HOM}, which do not resonate in the arm
cavities, to resonate in the \ac{SRC}, as detailed below. As derived in
appendix~\ref{sec:hom-recycling}, the optical gain in the \ac{SRC} for any
\ac{HOM} of order $m+n$ can be described by the circulating Airy
function as follows: 
\begin{linenomath}
  \begin{equation}
    A_{\mathrm{SRC}}^{mn} = \frac{1-r_{\mathrm{SRC}}^2}{\left|1-r_{\mathrm{FP}}^{mn}r_{\mathrm{SRC}}e^{i\left[\pi/2 + 2(m+n+1)\phi_{\mathrm{Gouy}}\right]}\right|^2} \, ,
  \end{equation}
\end{linenomath}
where $r_{\mathrm{SRC}}$ is the amplitude field reflection coefficient
of \ac{SR}; while $r_{\mathrm{FP}}^{mn}$ is the
reflection coefficient of the \ac{FP} arm cavities for the
non-resonant \acp{HOM}.

The round-trip phase of a \ac{HOM} in the \ac{SRC} is given by
\begin{linenomath}
  \begin{equation}
    2\phi_{\mathrm{SRC}}^{mn} = \frac{\pi}{2} + 2(m+n+1)\phi_{\mathrm{Gouy}}\,,
  \end{equation}
\end{linenomath}
where $\pi/2$ is the working point of the \ac{SRC} such that the arm
cavity carrier, in the fundamental mode, is anti-resonant. The carrier
\acp{HOM}, however, do not resonate in the arm cavity and so
$r_{\mathrm{FP}}^{mn}$ has an opposite sign to that of
$r_{\mathrm{FP}}^{00}$ (see 
appendix~\ref{sec:hom-recycling}). This change of sign causes the \acp{HOM} to resonate in the
SRC while the carrier does not.

Additionally, due to the small Gouy
phase of the  reycling cavities, the \acp{HOM} are close to one another and with a similar
optical gain. This is especially the case for the \ac{SRC} due to its smaller finesse and thus larger linewidth. For reference, the first, second and third-order modes
have, respectively, an optical gain of 8.4, 8.3 and 8.2 in the \ac{SRC}, while the tenth-order modes still have an optical gain of
6.4. Resonating \acp{HOM} in the \ac{SRC} lead to excess light at the
detector dark port, thus decreasing the interferometer contrast.
Misaligning the \ac{SR} mirror adds losses in the \ac{SRC} and reduces
the optical gain of circulating \acp{HOM}. Thus, in order to reduce
the impact of the \ac{SRC} circulating \acp{HOM}, the detector was
operated with the \ac{SR} mirror misaligned during \ac{O4} as further
detailed in section~\ref{subsec:IntentionalMisalignment}.

For \ac{RF} sidebands, there is no change in sign when reflecting from
the arm cavities. This results  in sidebands \acp{HOM}
resonating in both recycling cavities alongside the fundamental mode due to the small round trip Gouy
phase. This impacts negatively the longitudinal and angular controls
of the interferometer cavities, for example by introducing offsets to
feedback control loops as will be discussed in section~\ref{sec:RF-offsets}.

With the recycling cavities g-factor so close to 1, a 2-meter change
in the \ac{RoC} of the arm cavities input mirrors is sufficient to bring
the recycling cavities to instability. The light reflected by the
\ac{IM} from inside the recycling cavity encounters twice the thermal
lens inside the substrate, in addition to the reflection from the
highly reflective surface of the mirror. The measured thermal lens
induced in the input mirrors by the coating absorption
of the light resonating in the \ac{FP} arm cavities is
\qty{5e5}{m} for \qty{1}{W} of input laser power, which corresponds
to an effective \ac{RoC} change of 4 meters.
Hence, an \qty{0.5}{W}
change of the input power can render the recycling cavities unstable. In order to cope
with thermal deformations an elaborate \ac{TCS} has been implemented in the
Virgo detector, which will be further described in
section~\ref{sec:thermal-compensation}.
Its goal is to find and maintain the optimal working point
of the interferometer in an independent way from the circulating power~\cite{paper_TCS}.

\section{Challenges}
\label{sec:nearly-unstable-recycling}

The nearly unstable dual-recycled optical configuration poses several
challenges in terms acquiring and maintaining the light resonance
inside the interferometer, which we will describe in the following subsections.

\subsection{Lock acquisition procedure}
\label{sec:lock-acquisition-procedure}

Setting the interferometer in its working point is a procedure with
multiple steps called \textit{lock acquisition}, which relies on error
signals generated, among others, with the \ac{PDH}
technique~\cite{PDH_1983}
for longitudinal sensing and with the Ward technique~\cite{Ward_1994} 
for alignment sensing. Both techniques use the \ac{RF} sidebands
described in section~\ref{sec:optical-configuration}, with
their resonance conditions shown in figure~\ref{fig:Virgoscheme}.

The lock acquisition procedure is automated using a software supervisor, called
\textsc{Metatron}. The same hierarchical finite state machine engine
was already used during \ac{O3}~\cite{Virgo-DetChar-O3-Results} and
is derived from the \textsc{Guardian} supervisor designed for
LIGO~\cite{GraefRollins:2016hki}. A more complete description of the
\textsc{Metatron} environment and its implementation for the \ac{O4}
run can be found in~\cite{Bersanetti_Metatron}.

The strategy adopted in \ac{O3}, called \textit{variable
  finesse}~\cite{Acernese_2006}, is no longer applicable due to the
introduction of the \ac{SRC}, which
brings the total number of longitudinal degrees of freedom in the
interferometer to five. Before continuing, let us recall their
definitions, referring to figure~\ref{fig:Virgoscheme}:
\begin{itemize}
	\item \ac{CARM}: $\frac{L_N + L_W}{2}$
	\item \ac{DARM}: $L_N - L_W$
	\item \ac{PRCL}: $l_{PR} + \frac{l_N + l_W}{2}$
	\item \ac{MICH}: $l_N - l_W$
	\item \ac{SRCL}: $l_{SR} + \frac{l_N + l_W}{2}$
\end{itemize}
From the point of view of alignment the introduction of the \ac{SR} mirror brings 
the total number of degrees of freedom (only considering main optics and input beam) to eighteen~\cite{VIR-0596A-19}:
\begin{itemize}
	\item pitch and yaw for \ac{PR}, \ac{SR} and \ac{BS};
	\item pitch and yaw for common and differential tilt and shift of the optical axes in the arms;
	\item pitch and yaw for tilt and shift of the input beam.
\end{itemize}
Due to the non-linearity of signals when far from the cavities' working point,
the control strategy in \ac{GW} detectors is to enable the control loops in stages which
progressively decouple the interconnected degrees of freedom through appropriate choices of error signals.
In particular, the lock acquisition strategy used in Advanced Virgo
Plus
is inspired by Advanced LIGO~\cite{Aasi_2015_ALIGO} and relies on an
\ac{ALS}~\cite{DeRossi2020} to decouple the degrees of freedom related
to the arms (\ac{CARM} and \ac{DARM}) from those related to the
\ac{DRMI} (also called \textit{central interferometer}):
\ac{PRCL}, \ac{MICH} and \ac{SRCL}.

\begin{figure}
	\centering
	\includegraphics[width=\textwidth]{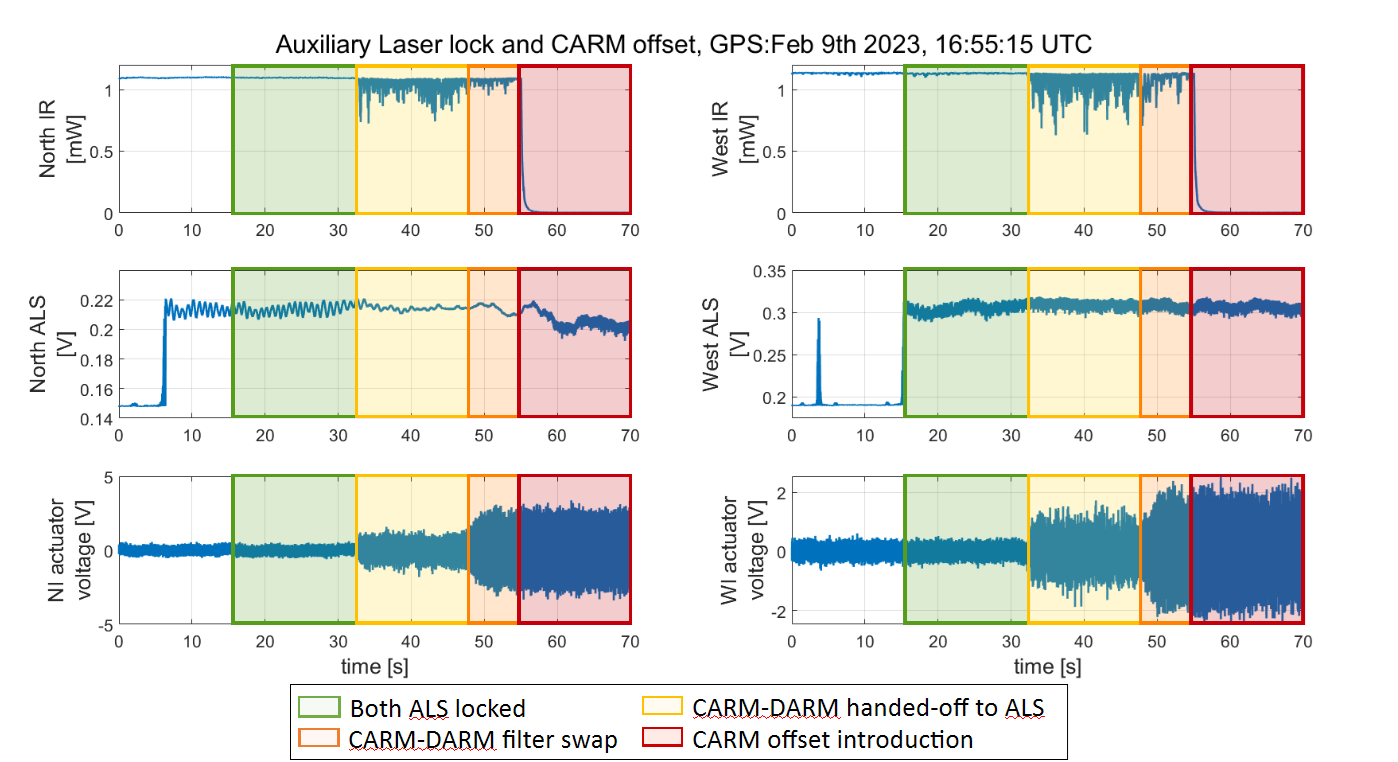}
	\caption{\ac{ALS} lock and introduction of the \ac{CARM} offset in the arm cavities of Virgo.  The top row shows the infrared
		power in transmission through the arms' end mirrors as a function of time,
		the middle row the \ac{ALS} beam power in transmission of the arms' input mirrors
		(the auxiliary beams are injected through the end mirrors, as per figure~\ref{fig:Virgoscheme}),
		and the bottom row the actuation voltage on the \ac{IM}.  
		The colored boxes show the different steps of the lock
		acquisition for the arms: the cavities start already locked independently on the infrared beam (top row, both stripcharts start at high and stable power
		in this plot), and the green laser frequency is then adjusted to resonate in them, maximizing its transmission
		(green box: the green beam is locked independently in each arm and the maximum of the resonant peaks is kept by the
		loop). Once the auxiliary laser frequency is locked, the arms control is handed off to signals based on this reference and in the
		\ac{CARM}/\ac{DARM} basis (yellow box: as these signals are noisier than the infrared-based \acp{PDH}, the actuation visibly increases in peak-to-peak amplitude  and
		the infrared transmission becomes noisier too). This excess noise is dampened by a more aggressive control filter, more
		capable of reducing higher frequency noise (orange box: the voltage in the actuators increases and the overall noise in
		the infrared transmission is reduced) before introducing the \ac{CARM} offset (red box: the arms move away from the infrared
		resonance and the corresponding powers in the top row drop to zero, while the transmitted power of the auxiliary beam remains constant).         
	}
	\label{fig:ALS_arms_control}
\end{figure}

The \ac{ALS} beam, which is generated by doubling the
frequency of a pick-off of the main laser, is injected in the arms
through the end mirrors and is used to control the arms length
without relying on the main infrared beam. By gradually
changing the frequency of the auxiliary beam, the input mirrors are moved
to make the arm cavities no longer resonant for neither the carrier nor the
sidebands, effectively introducing a common offset in both arms at the same 
time, called \textit{\ac{CARM} offset}.
In this condition the arm cavities remain under control and they
will be brought back to the carrier resonance at the right moment, leaving
the central degrees of freedom decoupled from the arms and able to be controlled independently.
These actions are demonstrated in figure~\ref{fig:ALS_arms_control}.

After the introduction of the \ac{CARM} offset the degrees of freedom in
the \ac{DRMI} can be locked but the correct working point must be 
selected with care. While resonant for the carrier, the arms 
behaved like an effective mirror that flips the phase of the reflected 
carrier by 180\textdegree. With the introduction of the \ac{CARM}
offset this effect
disappears and the carrier resonance condition within the \ac{DRMI}
changes
significantly, therefore it can no longer be used
to identify the working point of the central degrees of freedom. 
However the sidebands remain antiresonant in the arm cavities for the 
selected \ac{CARM} offset, therefore
their resonance conditions in the \ac{DRMI} 
remain unchanged and can be exploited both to identify the 
working point of the central degrees of freedom~\cite{Bersanetti_2022} 
and to build error signals based on the beat note between the first and 
second harmonics of each sideband~\cite{Arai_2002} on the symmetric B2
port of the interferometer.

This strategy renders the lock of the \ac{DRMI} insensitive 
to the \ac{CARM} offset which can then safely be removed to bring the 
interferometer in its final working point, where the resonance of the 
carrier in the arms and its resonance conditions in the \ac{DRMI} are 
restored. Notice that, as the carrier power circulating in the arms increases, the thermal lens effect inside the input mirrors becomes more relevant. Compensating for this effect becomes crucial to maintain the stability of the recycling cavities.

\subsection{Central interferometer thermal compensation}
\label{sec:thermal-compensation}

As introduced in section~\ref{sec:optical-configuration}, thermal 
effects become progressively more critical as the circulating power 
increases. 
During \ac{O4}, their impact was further enhanced by the transition 
to a nearly unstable dual-recycled optical configuration, which 
significantly reduces the tolerance to wavefront distortions and 
\ac{RoC} mismatches.  

The \ac{TCS} is a distributed system of sensors 
\cite{HWS_Lorenzo,PhaseCamera_Laura} and thermal actuators designed 
to mitigate thermally-induced optical aberrations and cold 
defects\footnote{Cold defects are due to imperfections in the 
production process of the mirrors, resulting in inhomogeneities of 
the refractive index, impurities or manufacturing errors such as 
mismatch of the \ac{RoC}.}, thereby preserving the optical configuration of the interferometer. In the \ac{FP} arm cavities, this includes tuning the optical parameters towards their design values -- in particular the Gouy phase -- and improving the symmetry between the two arms in order to minimize the contrast defect~\cite{VIRGO:2025sym}. In the nearly unstable recycling cavities, thermal compensation is also essential to keep the effective \ac{RoC} of the core optics within the margins required for stable operation.
The \ac{TCS} actuators include \acp{RH}, the \ac{CHRoCC} and CO$_2$ 
laser projectors, while the optical distortions in each \ac{FP} 
mirror are monitored using \ac{HWS}. 
The working principle of the \ac{RH} actuator consists in the 
contact-less tuning of the \ac{RoC} provided by heating elements 
radiatively
coupled to the mirror; a full description of the 
\ac{RH} actuators can be found in~\cite{VirgoRH}. 
The \ac{CHRoCC} actuator~\cite{Accadia2013_CHRoCC} projects a 
thermal radiation, from a 1 inch diameter high-temperature 
black-body emitter, onto the center of the high reflectivity side 
of the optic using an elliptical reflector.
The CO$_2$ laser projectors act on dedicated optics called \aclp{CP}
(CPs)
which are placed immediately upstream of the \ac{IM}s. These 
plates are heated with controlled patterns in order to generate 
compensating wavefront distortions before the beam enters the arm 
cavities. In particular, the \ac{DAS} actuator projects a heating 
pattern composed of two overlapping annuli designed to compensate 
the thermal lens induced in the \ac{IM} substrates by the main 
interferometer beam. 
A comprehensive description of the TCS architecture, actuation 
schemes, sensing strategy, and detailed modelling of 
thermally-induced wavefront distortions and substrate/surface 
deformations is presented in~\cite{paper_TCS}. The
present section focuses specifically on the operational 
configuration and tuning strategy adopted for the central
interferometer optics during \ac{O4}.

The \ac{TCS} architecture developed for \ac{O3} had sufficient actuation range to handle the planned increase in power; only minor extensions~-- described later~-- were installed or engaged for \ac{O4}. 
During the \ac{DRMI} lock acquisition, the arm cavities are 
not resonant and therefore no significant optical power is stored 
in the arms. 
At low input power, thermal effects in the central interferometer 
are negligible. In this regime, the \ac{TCS} is mainly used to 
correct the cold \ac{RoC} of the \ac{PR} and \ac{SR} mirrors.

\begin{figure}
    \centering
    \includegraphics[width=0.5\textwidth]{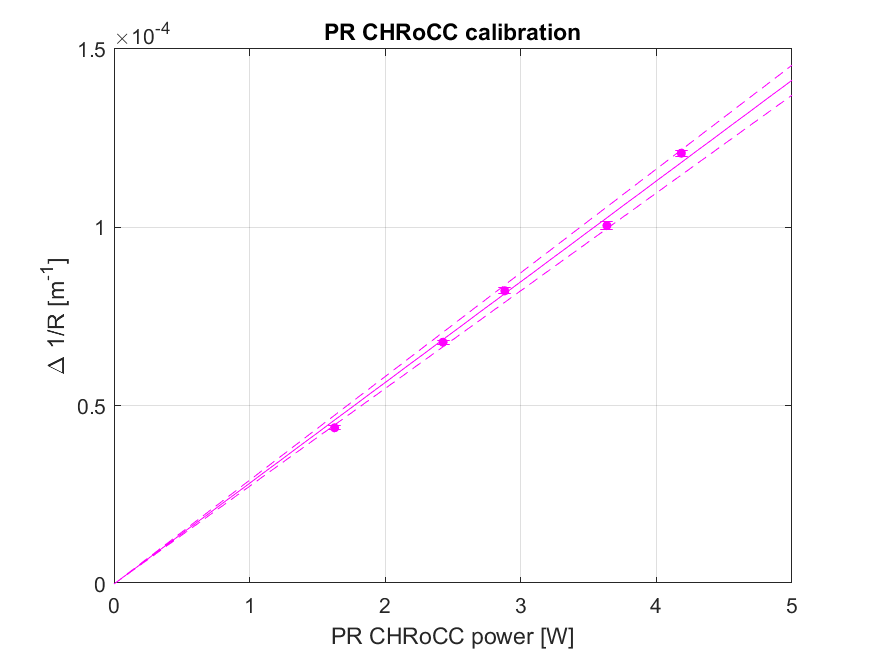} 
    \caption{Dynamics of \ac{PR} \ac{CHRoCC}. The points indicates the
      equivalent PR RoC correction, expressed as $\Delta\left(\frac{1}{R}\right)$, as a function of emitted thermal power, the solid line represents the linear fit and the dashed lines are the 95\% confidence bounds. 
    }
    \label{fig:dinamica CHRoCC}
\end{figure}

The optical layout was designed for a \ac{PR} mirror with a \ac{RoC} of
$R_{target}^{PR}=$\qty{1430}{m}; however, the \ac{PR} mirror was intentionally polished with a
 \ac{RoC} of $R_0^{PR}=$\qty{1477}{m}, in order to account for
the thermal lens expected in the \ac{POP} at high power (\qty{125}{W}
of input laser power). The \ac{PR} mirror is therefore
equipped with a \ac{RH}, which is intended to compensate for this
offset when the injected power is low.
However, in practice, the \ac{PR} \ac{RH} is not used as it overheats
the mirror actuation coils and strongly changes the input beam
mode-matching due to the induced thermal lens in the \ac{PR} substrate.
For this reason, a \ac{CHRoCC} was installed, which acts on the \ac{POP} by inducing a thermal
lens, which lowers the equivalent \ac{RoC} of the \ac{PRC}.
The relation between the equivalent PR RoC correction induced by the PR CHRoCC and the injected power
was characterized using the \ac{HWS}, and has the linear relation
\begin{linenomath}
\begin{equation}
    \Delta \left(\frac{1}{R}\right) = k P \, ,
\end{equation}  
\end{linenomath}
with $k=$\qty{2.859e-5}{m^{-1}.W^{-1}}, as shown in
figure~\ref{fig:dinamica CHRoCC}. During O4 the required curvature variation is 
\begin{linenomath}
\begin{equation}
    \Delta \left(\frac{1}{R}\right) = \frac{1}{R_{target}^{PR}} - \frac{1}{R_0^{PR}} \sim 2.23 \times 10^{-5} \mathrm{m}^{-1} \, ,
\end{equation}  
\end{linenomath}
corresponding to an injected power of $P\sim$\qty{0.78}{W}.\\

The \ac{SR} mirror is located at the output port of the 
interferometer; as a consequence, no significant thermal effects 
are expected on this optic even at high injected power. The optical
configuration was designed with $R_{target}^{SR}=$\qty{1430}{m} \ac{RoC} for the \ac{SR} 
mirror, but it was manufactured with a \ac{RoC} of $R_0^{SR}=$\qty{1440}{m} in order 
to ensure that the \ac{RoC} could always be adjusted to the nominal 
value using the \ac{RH}. 
Finite-element simulations provide the following relation between the \ac{RH} 
power and the resulting variation of the \ac{RoC} 
\begin{linenomath}
\begin{equation}
    R_{target}^{SR} = k' \times P + R_0^{SR}\,, 
\end{equation}  
\end{linenomath}
with $k'=$\qty{1.43}{m/W}. This corresponds to about \qty{7}{W} of \ac{RH} power needed to decrease the \ac{RoC} by \qty{10}{m}.\\

When the interferometer control progresses beyond the \ac{DRMI} stage, 
the arm cavities are brought to resonance and a large 
amount of optical power starts to circulate inside them.
In this condition, thermal effects become increasingly relevant, as 
absorption in the substrates of the \acp{IM} induces thermal lensing 
that modifies the optical path and the effective curvature seen by 
the interferometer beam. 
Switching the \ac{DAS} projectors on and off during each 
interferometer lock or unlock is not operationally feasible due to 
the different thermal time constants involved. The main 
interferometer beam has a thermal time constant of about 20 
minutes, while the \ac{DAS}-induced thermal lens exhibits an 
overshoot and reaches steady state only after approximately 2 
hours, as the heat from the annuli needs to propagate to the center of
the optic with a diffusion scale of \qty{5}{cm} in one hour.
This mismatch would introduce thermal transients, which would
temporarily render the recycling cavity unstable, and interrupt the
lock acquisition. For this reason, the 
\ac{DAS} projectors remain on throughout all the phases of the lock 
acquisition. Thus, to mimic the heating produced by the 
interferometer beam, the CO$_2$ laser projector system includes an 
additional actuator, 
the \ac{CH},
which produces a Gaussian-shaped
heating beam with a size comparable to that of the interferometer
beam. This doesn't induce a heat gradient as the heat is always
deposited on the same transverse position in the center of the optical
path, with a comparable distribution.

\begin{figure} 
    \centering
    \includegraphics[width=0.49\textwidth]{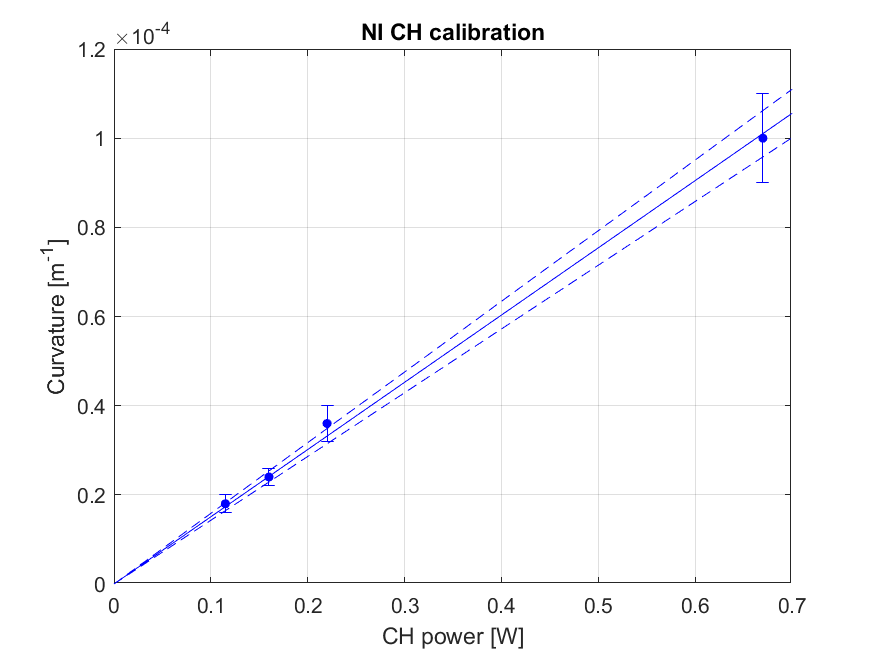} 
    \includegraphics[width=0.49\textwidth]{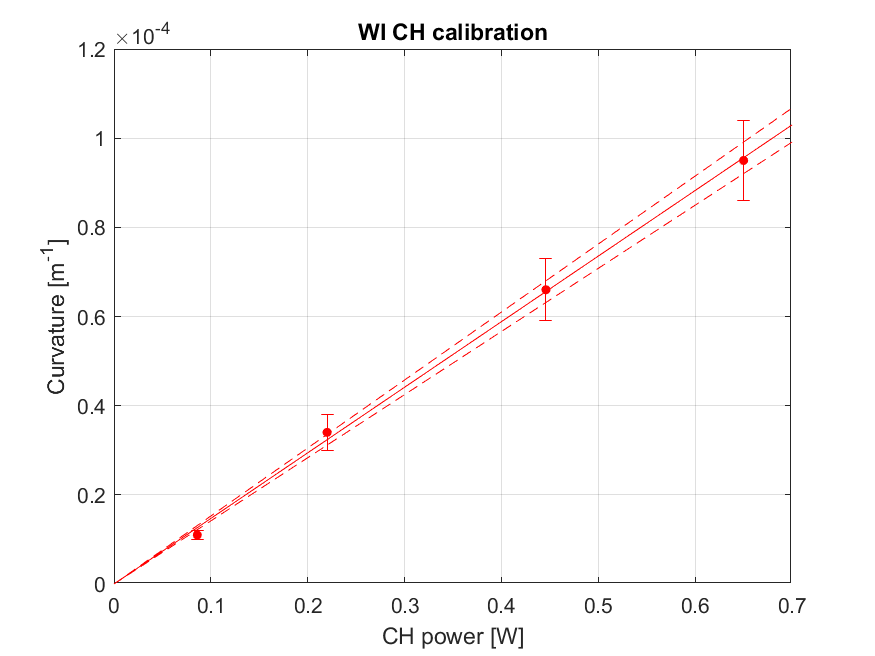} 
    \caption{Dynamics of \ac{NI} \ac{CH} on the left plot and of \ac{WI}
      \ac{CH} on the right one. The points are the values of \ac{HWS}-measured Gaussian weighted curvature 
	  related to different injected powers, the solid lines
      represent the linear fit and the dashed lines are the 95\%
      confidence bounds.  }
    \label{fig:CH dinamica}
\end{figure}

The \ac{CH}, projected on the same \ac{CP} optics used by the 
\ac{DAS}, is used to mimic the thermal load generated by the main 
beam when the arm cavities are not resonant.
For \ac{O4}, the \ac{CH} was operated with a dedicated CO$_2$ laser source, improving the flexibility and stability of the actuator.

\begin{figure} 
    \centering
    \includegraphics[width=\textwidth]{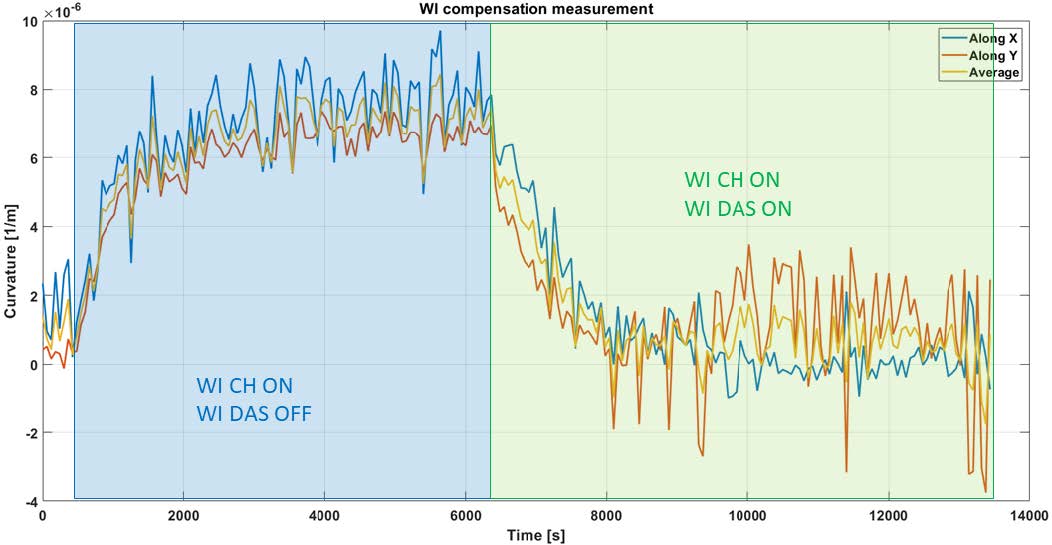} 
    \caption{The 1D Gaussian weighted curvatures induced by WI CH and compensated by \ac{WI} \ac{DAS}. The blue line represents the curvature extracted from the barycentre row and the red line from the column one. The yellow line is the average curvature.}
    \label{fig:compensazione}
\end{figure}

Within this framework, the \ac{CH} actuator plays a specific 
operational role, being the only thermal compensation actuator that 
is deliberately switched on and off during lock acquisition. Its 
function is to support a robust lock acquisition and to facilitate 
the recovery of the nominal working point after an unlock. 
The \ac{CH} actuators have been fully characterized 
by measuring with the \ac{HWS} the induced Gaussian weighted wavefront curvature as a function of the injected power -- see 
figure~\ref{fig:CH dinamica}. During operation, the \ac{CH} beams 
condition the thermal state of the \ac{IM}s by counterbalancing 
the action of the \ac{DAS} projectors and reproducing optical 
conditions close to those of the final interferometer working point.
This compensation mechanism is illustrated in  
figure~\ref{fig:compensazione}, where the induced curvatures produced by 
the two actuators are shown to counteract each other. 
By stabilizing the thermal state of the substrates, the system 
reduces thermal transients associated with power variations and 
shortens the time required to re-establish the optimal 
interferometer conditions, thereby improving the detector duty 
cycle.
Moreover, a dedicated tuning of the \ac{CH} power in the \ac{DRMI} 
increases the recycling gain of the 6 MHz and 56 MHz sidebands,
improving the robustness of the recycling-cavities control signals. 

Nevertheless, the resulting sideband build-up remained far from the
theoretical optimum: in the final O4 configuration, the hot-cavity
gains were measured to be $24 \pm 0.7$ and $17 \pm 0.9$ for the 6 MHz
and 56 MHz sidebands, respectively, compared to theoretical values of
72 and 53~\cite{VIRGO:2025sym}. Currently, this discrepancy is
explained by the following hypothesis: there are optical defects in
the optics which create losses of the fundamental mode to \acp{HOM}, and these losses are
amplified by the co-resonance of \acp{HOM} in the nearly unstable \ac{PRC}.
There are several sources of defects: the polishing of the optics
themselves, the thermal aberrations due laser absorption and thermal
compensation correction, and thermal deformation due to point
absorbers on the highly reflective surface of the input
mirrors~\cite{LIGOScientific:2021kro,cifaldiThesis,Ipatsia_studies}.

This strategy has been successful at lower input powers between
\qty{11}{W} and \qty{22}{W}. Initial attempts at successively
\qty{40}{W} and \qty{31}{W} were not successful.  
At \qty{40}{W}, attempts resulted in fields in the \ac{PRC} and \ac{SRC} that
did not remain stable during the thermal transient, while at
\qty{31}{W} the fields were in a bistable
state with rapid and repeated uncontrolled transitions between two
states.

\subsection{Automatic alignment}
\label{sec:automatic-alignment}

Once the final carrier resonance condition in the complete
interferometer is reached, the automatic alignment 
strategy is similar to the one used in Advanced Virgo
between 2017 and 2020~\cite{galaxies8040085}, with two main differences concerning the
alignment of the \ac{BS} and the \ac{SR} mirrors~\cite{Pinto_ASC}.

Due to the strong coupling of \ac{BS} and \ac{PR} alignment on all
standard wavefront sensors, an automatic alignment loop based on a
Ward signal~\cite{Ward_1994} was deemed unfeasible. Instead, following
indications from simulations, demodulating
the B1p asymmetric port (see figure~\ref{fig:Virgoscheme}) at the
beat note frequency between the \qty{6}{MHz} and the \qty{56}{MHz} sidebands provides a
better alignment error signal of the
\ac{BS}~\cite{martynov_phd_2015, boldrini:AutomaticAlignment2023}.
The beat notes between sidebands have low amplitude and thus
come at the cost of higher sensing noise, which becomes an issue
considering the typical alignment loops' unity gain frequency of a few \unit{Hz}.
To mitigate this drawback, this signal for the \ac{BS}
alignment is blended at a few hundreds mHz
with a signal generated by demodulating the
B4 pick-of from the \ac{PRC} at \qty{6}{MHz} for the pitch alignment,
and with a signal coming from the
mirror's optical levers for the yaw. These choices provide the best
overall outcome in terms of sensing noise at high frequency.

On the other hand, controlling the \ac{SR} alignment was problematic, as no signal,
neither based on beat between sidebands nor based on the power of carrier or sideband,
was sufficiently sensitive to its alignment to decouple it from other
degrees of freedom~\cite{boldrini:AutomaticAlignment2023}.
Instead, the adopted strategy relies on monitoring the change in shape
of the frequency response of the interferometer caused by the \ac{SRC} misalignment.
In particular, a change in the alignment of SR changes
the detector bandwidth given by the \ac{DARM} signal extraction from the arm
cavities by the \ac{SRC}.

This effect is exploited to build alignment signals for both the pitch
and the yaw of the \ac{SR}.  For pitch, the error signal is built
through \textit{double dithering}: the error signal for \ac{DARM}
is demodulated first at the frequency of a \ac{DARM} mechanical dither
at \qty{491.3}{Hz} and then at the frequency of an \ac{SR} pitch
mechanical dither at \qty{5.1}{Hz}. For the first demodulation both
the in-phase and the quadrature signal are used to calculate the phase
of the \ac{DARM} response, and then that phase signal is
demodulated a second time to maximize that phase as function of the
\ac{SR} alignment. In this way \ac{SR} is aligned to maximize the
detector bandwidth.

In principle, the same strategy could be used to control the mirror's
yaw as well, however during commissioning the amount of excess noise
in the interferometer could be reduced by intentionally misaligning
\ac{SR}, as detailed in
section~\ref{subsec:observational_properties_of_excess_noise}.
As the \ac{SR} misalignment also reduces the detector bandwidth,
the \ac{SR} yaw misalignment is maintained using  an ad-hoc error
signal: a pair of sinusoidal excitations is injected on \ac{DARM},
one at \qty{74.4}{Hz} and one at \qty{491.3}{Hz} in order to  measure the detector's bandwidth.
This signal is then directly fed to the \ac{SR}'s yaw to drive the interferometer's
bandwidth to a value that maximizes the detector's sensitivity: from $\sim$\qty{430}{Hz} to $\sim$\qty{190}{Hz}.

The \emph{double dithering} approach is also used for the \ac{OMC}
alignment~\cite{Smith-Lefebvre:2011joa}. In this case
the in-phase and quadrature signals are used to calculate the
magnitude of the \ac{DARM} response at
\qty{491.3}{Hz}, and that magnitude signal is 
demodulated a second time
at the frequency of a mechanical oscillation injected at
\qty{4.3}{Hz} into pitch and at \qty{2.1}{Hz} into yaw of the bench
hosting the \ac{OMC} in order to maximize the magnitude. In this way
the \ac{SR} and \ac{OMC} alignment are decoupled, as the \ac{SR}
alignment maximizes the phase of the \ac{DARM} response and the \ac{OMC}
maximizes the magnitude of the \ac{DARM} response.

These signals based on double mechanical dithering have a very poor
signal-to-noise ratio and effectively function as set-point trackers that only
provide very low frequency corrections, up to $\sim$\qty{100}{mHz},
while leaving the control at higher frequencies to optical levers,
which measure the orientation of the mirrors with regard to the
vacuum chambers.

\subsection{Error signal optical offsets}
\label{sec:RF-offsets}

Considering the near instability of the recycling cavities,
the main consequence from the control point of view is the co-resonance and
amplification of \acp{HOM} as described in section~\ref{sec:optical-configuration}. In
particular, the carrier \acp{HOM} are co-resonant in the \ac{SRC} and
sidebands \acp{HOM} are co-resonant in the \ac{PRC}.  

As the Gouy phase accumulated by each \ac{HOM} in the recycling cavities 
is low, their dephasing with respect to each other, in particular to the 
fundamental mode, is small enough that their resonant peak is still 
contained in each cavity's linewidth, allowing them to be at least 
partially recycled in either the \ac{PRC} and \ac{SRC}. The remaining 
\acp{HOM} pollute the sensing scheme in different ways, mainly:
\begin{itemize}
	\item \acp{HOM} alter the shape of the \ac{PDH} signals, introducing an offset 
          between their zero-crossing and the cavity resonance, which
          is the desired working point;
	\item modes above the first order pollute the quadrant signal used to build alignment 
	signals based on the Ward technique, making those signals sensitive to all the 
	defects lingering in the interferometer. They also increase the coupling between 
	angular degrees of freedom.
\end{itemize}

These effects are difficult to characterize and control. Furthermore,
these effects are time-dependent, due to the thermal transients in the
first hours of each lock, and to other drifts on longer time scales.
Therefore in order to subtract the resulting offsets from the error signals, it
is necessary to identify ways of measuring them in real time.

\paragraph{Central degrees of freedom:}
the cross-coupling among \ac{PRCL}, \ac{MICH} and \ac{SRCL} is
computed by dithering each degree of freedom at a specific frequency,
and measuring the magnitude of these lines in the error signals of the
other two degrees of
freedom. The coupling is then subtracted in real time 
by adjusting the respective weights in the longitudinal sensing matrix \cite{vanDael2024}.

\paragraph{Differential axis tilt:}
the differential tilt of the arms optical axis (called \emph{DIFFp}) 
disrupts the destructive interference in the Michelson interferometer. 
The loop that controls this degree of freedom is affected by a slowly moving offset 
in its error signal, that is measured by demodulating the \ac{DARM}
error signal at the frequency
of a pair of lines injected in pitch and yaw (\qty{3.3}{Hz} and \qty{5.2}{Hz} respectively) \cite{Pinto_ASC}.

\paragraph{\ac{BS} alignment:}
the loop that controls the yaw of the \ac{BS} has a setpoint 
adjusted to minimize the coupling between the laser frequency noise and 
the detector's output. This is achieved by dithering the laser
frequency at \qty{227.1}{Hz} 
and using the signal demodulated from \ac{DARM} error signal to adjust the \ac{BS} yaw position~\cite{Pinto_ASC}.

\paragraph{\ac{SRC} length tuning and alignment:}
the \ac{SRC} shapes the interferometer frequency response in two main 
ways: the frequency of the pole introduced by radiation pressure effects in the arms
is driven by the microscopic tuning of the \ac{SRC} length, while the bandwidth of the
detector is changed both by the \ac{SRC} length tuning and the \ac{SR} alignment~\cite{Boldrini_AdV+_SRCvsDARM}.

\begin{figure}
	\centering
	\includegraphics[width=\textwidth]{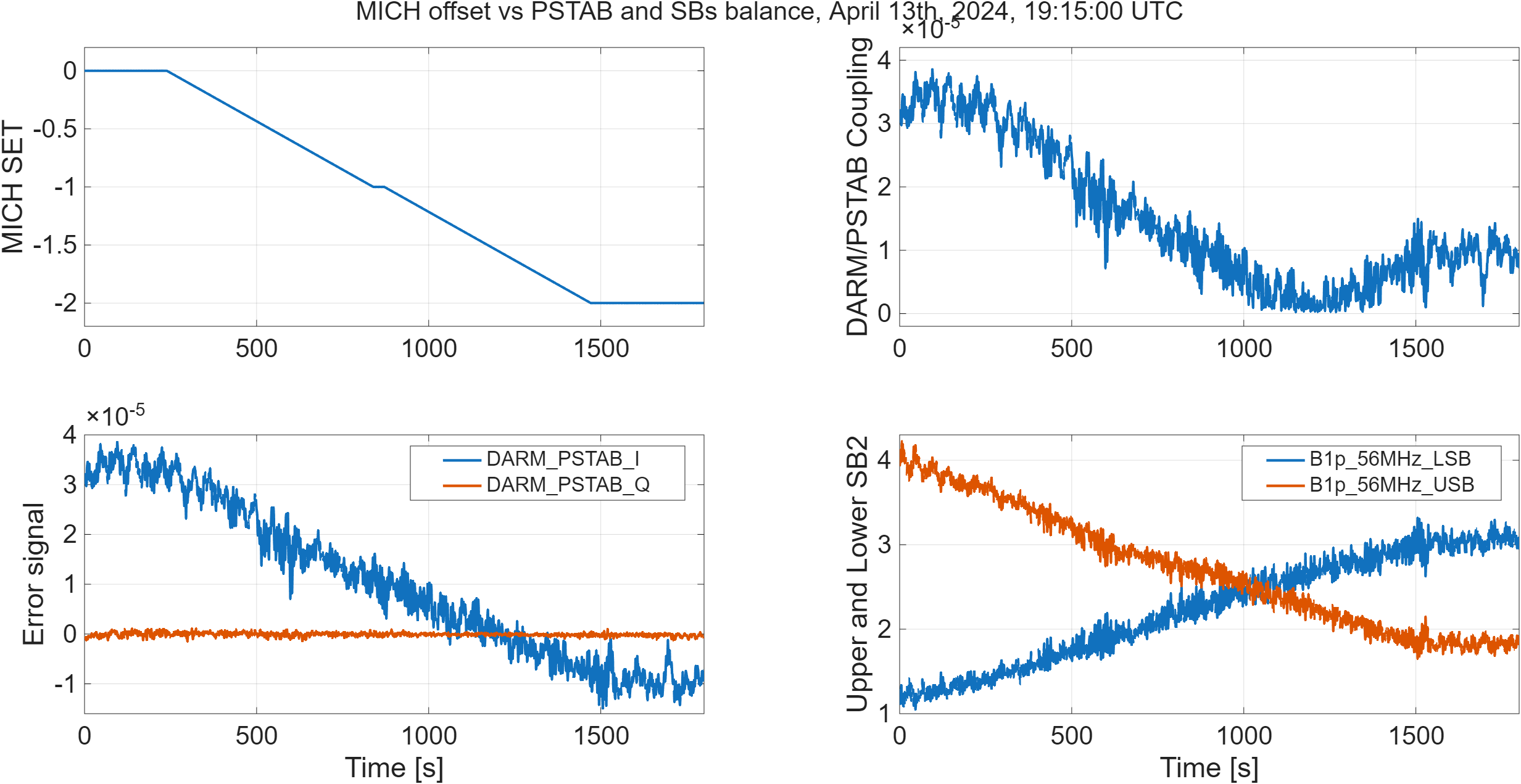}
	\caption{Effects of a scan of the setpoint of the \ac{MICH}
          control loop: \emph{top-left}: \ac{MICH} setpoint, \emph{top-right}: coupling between laser
          intensity noise and \ac{DARM},
          \emph{bottom-left}: error signal obtained by demodulating
          \ac{DARM} error signal at the frequency of the monitor line,
          \emph{bottom-right}: power of upper and lower \qty{56}{MHz}
          sidebands measured in transmission of the interferometer by
          phase-cameras. This scan shows an optimal working point,
          corresponding to the zero-crossing of the error signal,
          where the \ac{PSTAB} to \ac{DARM} coupling is minimum and the upper and lower \qty{56}{MHz} sidebands are roughly superposed.}
	\label{fig:MICH_SET}
\end{figure}

\paragraph{\ac{MICH} setpoint:} the frequency dependence of the \ac{MICH} to
\ac{DARM} coupling is consistent with the expectations from simulations of a
working point away from the destructive interference on \ac{BS}. This
coupling is measured directly at a single frequency by
introducing a sinusoidal
excitation at \qty{1501}{Hz} in the \ac{PSTAB} loop of the laser, 
and demodulating the \ac{DARM} error signal at that frequency. This  
yields an error signal for the
setpoint of the \ac{MICH} control, which minimizes the intensity noise
coupling to \ac{DARM} at all frequencies.

Figure \ref{fig:MICH_SET} shows a scan of this setpoint, to prove its 
effectiveness in reducing the aforementioned coupling between  the 
laser intensity noise and \ac{DARM}. As expected from simulations,
it also improves the balance between the upper and lower
\qty{56}{MHz} sidebands power in transmission of the interferometer.\\

All these loops based on dither signals are slow, with a response time
of tens of minutes, which extends the lock acquisition duration and
as consequence reduces the observation duty cycle 
as discussed in section~\ref{sec:performance}.

\section{Input laser noises}
\label{sec:injection-noise}
In section~\ref{sec:optical-configuration} we have shown how the nearly
unstable signal recycling cavity amplifies the power of \acp{HOM} at
the detector output. \acp{HOM} do not undergo noise filtering by the
arm cavities, hence they have a large laser noise content, and their
amplified power increases the coupling of the laser noises to the
strain noise curve. The couplings of these laser intensity noise and frequency
noise have been
reduced using the controls offsets described in
section~\ref{sec:RF-offsets}, and below we describe the projected
noise level following these adjustments.
  
\subsection{Intensity noise}
\label{PSTABnoise}

\begin{figure}
	\centering
        \includegraphics[width=\textwidth]{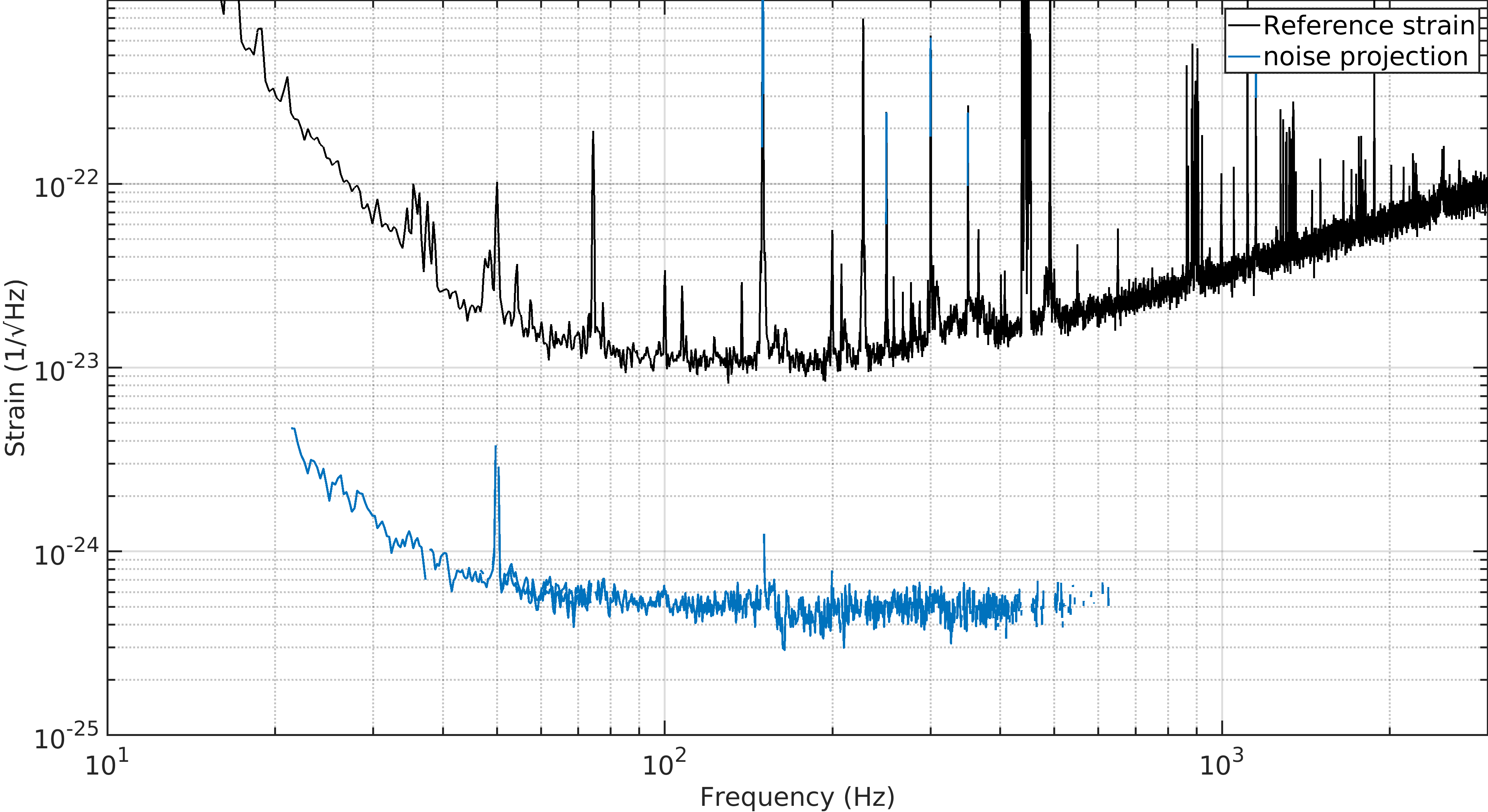}
	\caption{Projection of the laser intensity noise compared with the quiet
          strain noise curve. }
	\label{fig:pstab_proj}
\end{figure}

The power stabilization loop is designed to suppress laser relative intensity noise
below \num{2.5e-9}\,$\mathrm{/\sqrt{Hz}}$ (\qty{-172}{dB/\sqrt{Hz}})
  in the \qty{10}{Hz}\,--\,\qty{10}{kHz} bandwidth.  
The \ac{PSTAB} system received minor modifications after \ac{O3} by replacing single-ended in-air cabling with differential signals, which reduce the \qty{50}{Hz}
coupling. Its noise is around \qty{4e-9}{RIN/\sqrt{Hz}}, quite flat from 10 Hz to 5 kHz, described more in detail in~\cite{INJ_O4}.
The main modification with respect to \ac{O3} is the interferometer optical
configuration that results in a large offset in the \ac{MICH} error
signal, which increases the intensity noise coupling at the dark port.
As described in section~\ref{sec:RF-offsets}, 
that coupling is measured using an intentional laser intensity
excitation at \qty{1501}{Hz} to obtain an error signal, which acts on
the \ac{MICH} setpoint.
The residual coupling has been measured to be frequency
independent in the \qtyrange{50}{400}{Hz} range at level of \num{6.5e-17}
strain per relative intensity noise unit, and figure~\ref{fig:pstab_proj} shows the projected noise of the \ac{PSTAB}
loop, where power fluctuation
contributions remain at least a factor 10 below the strain noise curve across the full detection bandwidth.

\subsection{Frequency noise}

Laser frequency noise couples to the \ac{GW} signal through
asymmetries in the interferometer optical paths; therefore, a
dedicated frequency stabilization system is required to achieve
Advanced Virgo target sensitivity within the \qty{10}{Hz}\,--\,\qty{10}{kHz} detection
bandwidth. The stabilization architecture is based on a hierarchy of
nested control loops, each characterized by distinct
bandwidths and error signals, ensuring that both the \ac{IMC} and the interferometer remain resonant with the
fundamental mode of the carrier laser frequency~\cite{PhysRevA.79.053824}.

\begin{figure}
    \centering
    \includegraphics[width=\textwidth]{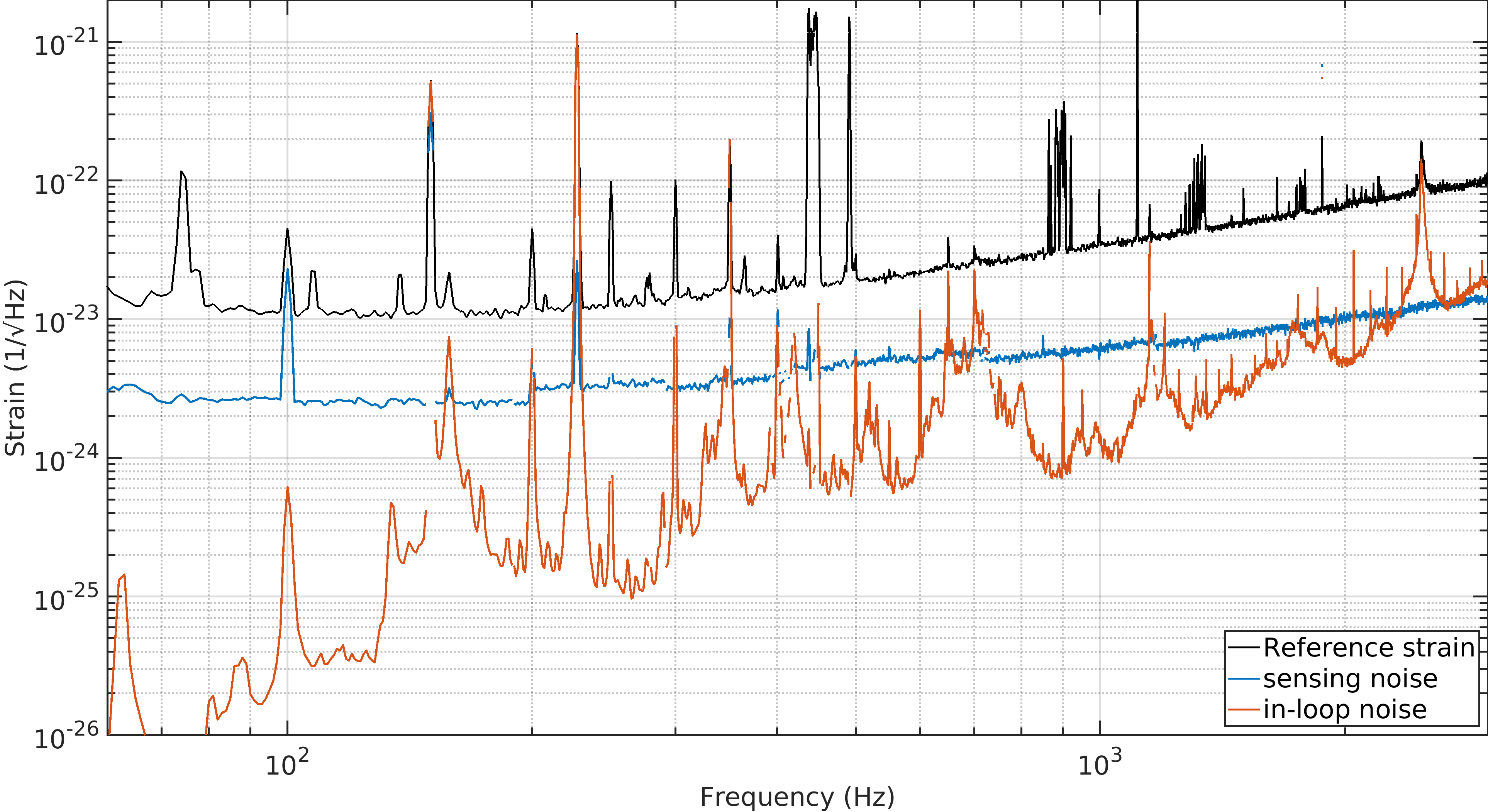} 
    \caption{Projection of the laser frequency stabilization in-loop signal and the loop sensing noise compared with the quiet
          detector noise curve. The total frequency noise is the
          quadratic sum of these two components.}
    \label{fig:SSFS}
\end{figure}

The in-loop error signal of the frequency noise control is obtained
from the B4 photodiode demodulated at \qty{6}{MHz}, which is a pick-off beam inside the \ac{PRC}
used for the \ac{SSFS} loop~\cite{SSFS}. A projection of the frequency
noise can be obtained by performing noise
injections into this loop, as shown in figure~\ref{fig:SSFS}. The
frequency noise is composed of two components, the in-loop error
signal which is the frequency noise at the input of the interferometer and that is
suppressed by the \ac{SSFS} loop, and the sensing noise of the
B4 photodiode which is derived from the quadrature of the in-loop
signal. The quadrature of the in-loop signal is a good estimate of
sensing noise as it has no signal, but it has
the same level of quantum shot noise and electronic noise as the in phase signal. In
addition, the laser phase noise at radio-frequency discussed in
section~\ref{sec:fiberedEOM} display the same level of noise for the two
quadratures of each radio demodulated photodiodes, hence the
quadrature of the in-loop signal includes this noise as part of
the sensing noise estimate.

The main source of peaks seen in the
spectrum are due to \ac{IMC} mechanical resonances, which are then
sensed and reduced by the \ac{SSFS} loop gain. The loop is able to
suppress most of the frequency noise to below the level of the B4
sensing noise. Moreover, 
an intentional frequency
sinusoidal excitation at \qty{227.1}{Hz} is used to obtain  an error signal used to control the \ac{BS}
yaw offset and reduce the coupling of the common noise in the detector
output. 

The coupling between the laser frequency noise and the detector strain
can be more naturally expressed as coupling between the common mode
and the differential mode of the two arms. The measured coupling is
around \num{3e-4} at \qty{100}{Hz} and around \num{1e-4} at
\qty{1}{kHz}, which is an order of magnitude higher than the expected
coupling from simulations of an interferometer with comparable
mode-mismatch between the arms of $\sim$1\% but no other defects~\cite{VIR-0062A-21}.

\section{Dominant broadband noise}
\label{sec:dominant-noise}
During \ac{O4}, a dominant broadband \emph{excess} noise was limiting
the Advanced Virgo sensitivity. In this section we present a simplified noise
budget of the detector, comprising of the sum of all the dominant noise
sources, which provides context for this \emph{excess} noise.
Furthermore, we summarize its main observational properties, and review
the experimental studies conducted to investigate the origin of the \emph{excess} noise. 
We also discuss the effect of the intentional \ac{SR} mirror misalignment introduced as a mitigation strategy.

\subsection{Simplified noise budget}
\label{subsec:simplified_noise_budget}

A complete noise budget of Virgo includes over a hundred different
noise source~\cite{Virgo-DetChar-O3-Results}. 
Here we present only a
simplified noise budget that includes the dominant fundamental and
technical noises. 
This model is not intended to provide a complete quantitative
description of all noise contributions, but rather to place the 
\emph{excess} noise in the context of the overall detector performance.

\begin{figure}
\centering
\includegraphics[width=\textwidth]{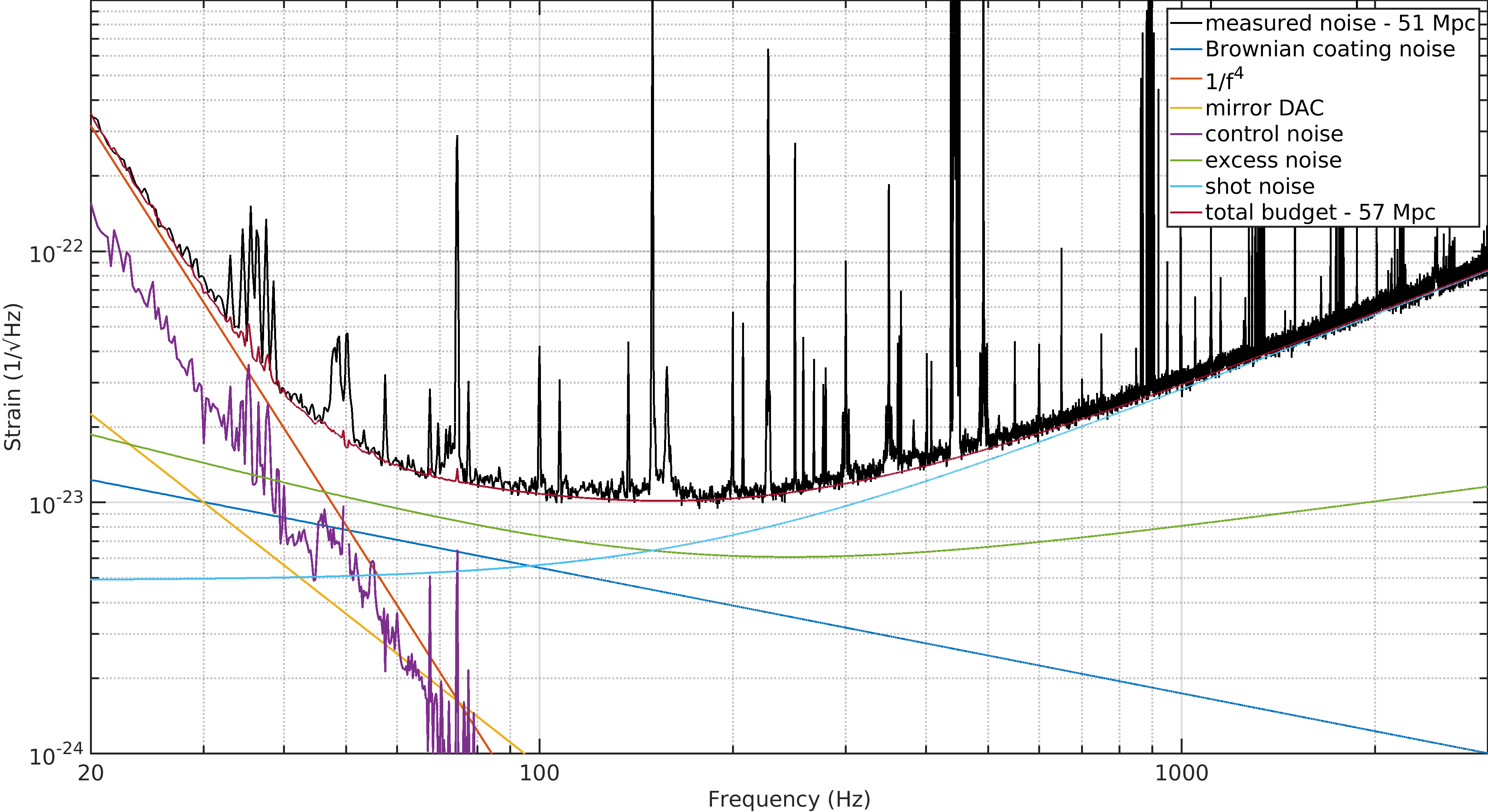}
\caption{Simplified noise budget of the Virgo strain noise curve
  on June~11,~2025, in the misaligned \ac{SR} configuration used during \ac{O4}.
The measured strain spectrum is shown together with the main modeled noise contributions, 
including coating thermal noise, quantum shot noise, control and \ac{DAC} noises, and other technical noise sources.
An additional broadband contribution with a $1/f^{0.67}$ frequency dependence,
multiplied by the detector optical response,
is included to account for the \emph{excess} noise observed in the \qtyrange{50}{200}{Hz} frequency range.
Narrow spectral features correspond to residual technical lines not
included in the broadband noise models. }
\label{fig:simplified_noise_budget}
\end{figure}

Figure~\ref{fig:simplified_noise_budget} summarizes the modeled
fundamental and technical noise contributions limiting the Virgo
sensitivity across the frequency band during \ac{O4}.
At high frequencies ($\gtrsim$ \qty{100}{Hz}), the sensitivity is primarily limited by quantum shot noise,
arising from the fundamental uncertainty in the arrival times of
photons at the output photodetector, governed by Poissonian statistics~\cite{aVirgo}. 
The shot noise has a frequency independent spectrum in
units of optical power, however its contribution to the strain noise curve increases with
frequency due to the interferometer's optical response. 

In the intermediate frequency range, Brownian coating thermal noise, caused by thermally driven displacement fluctuations in the mirror coatings~\cite{aVirgo}, is expected to dominate.
This noise, arising from mechanical dissipation in the coating materials, is described by the fluctuation--dissipation theorem and can be modeled using the Levin approach, 
which relates the dissipated power associated with an applied pressure profile to the resulting thermal noise spectrum~\cite{kubo:FluctuationdissipationTheorem1966, Levin98}.

At low frequencies, several control noise contributions are present~\cite{Pinto_ASC},
but the sensitivity is limited by a displacement noise of unknown
origin with an amplitude spectral density scaling as $1/f^4$.

Together, these modeled contributions reproduce the measured strain
noise curve over most of the spectrum.
However, in the \qtyrange{50}{200}{Hz} range, the measured noise curve exceeds the sum of all the modeled noises.
An additional broadband contribution, empirically described in
frequency domain by a power law with a $-0.6\pm0.1$ slope multiplied by the detector optical response, is
included as an \emph{excess} noise.
Figure~\ref{fig:simplified_noise_budget} shows that this \emph{excess} noise dominates in this band,
reaching a level comparable to quantum shot noise and thermal noise around \qty{100}{Hz}.
This simplified noise budget highlights the impact of the \emph{excess}
noise on the Virgo noise curve in the mid-frequency range
of the spectrum.

\subsection{Observational properties of the \emph{excess} noise} 
\label{subsec:observational_properties_of_excess_noise}
 
To investigate the origin of the \emph{excess} noise, we explored the
scaling of its amplitude spectral density as a function of several
parameters of the interferometer.  The most significant observations
were the scaling
with the differential arm offset and the optical gain.

The \emph{excess} noise in terms of power at the detector output scales linearly
with the differential arm offset used to create a local oscillator for DC readout of the arm differential length~\cite{Hild:2008pb}.
This excludes contributions from \ac{RF} sidebands,
whose noise would be independent of the differential arm offset and suggests that the noise originates from a fluctuating
optical field interfering with the DC readout local oscillator field at the output photodiode.

\begin{figure}
\centering
\includegraphics[width=0.8\columnwidth]{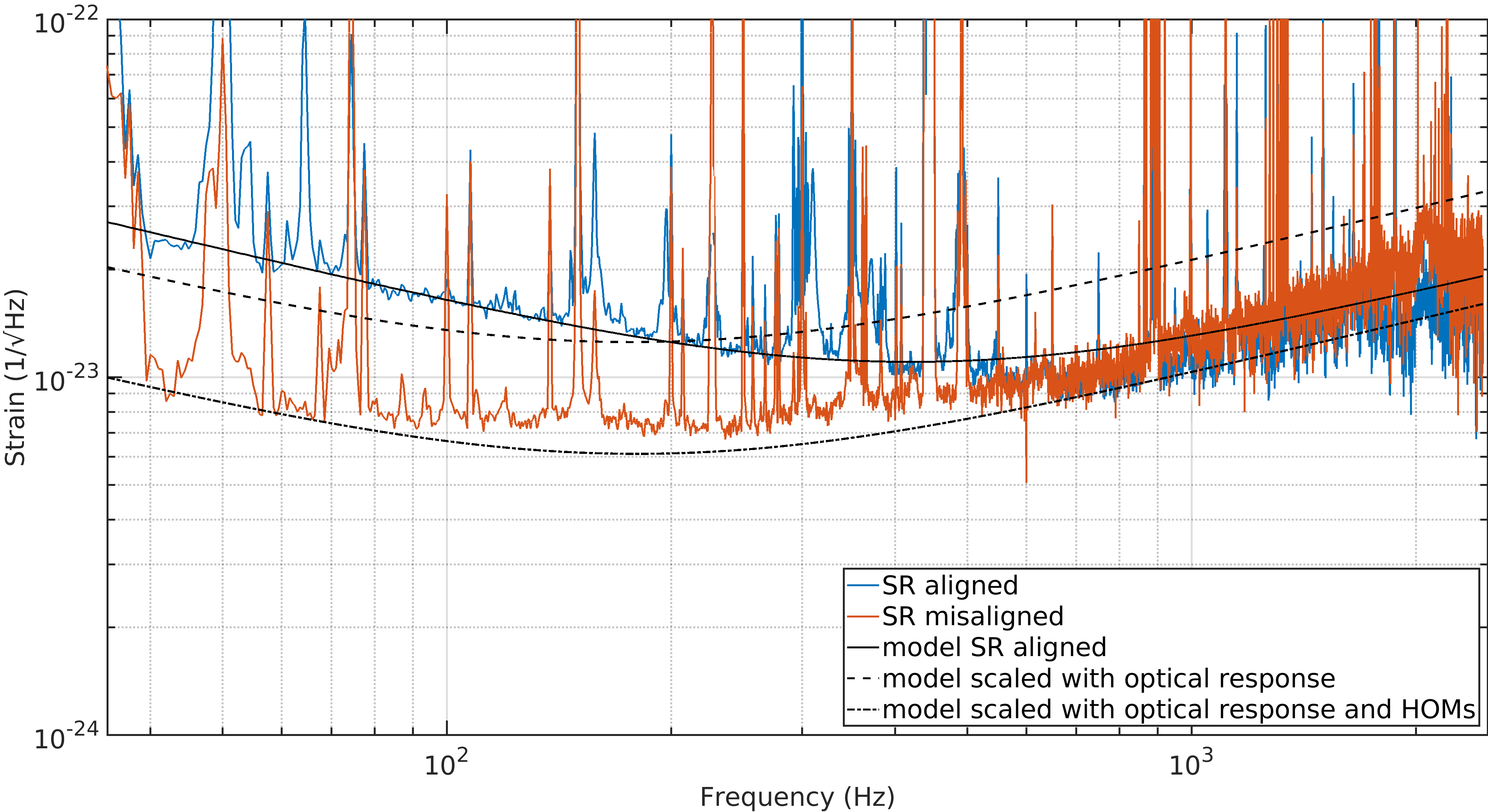}
\caption{Comparison of inferred \emph{excess} noise for two different conditions of the \ac{SR} mirror alignment.
The black solid line represents a simple model for the \emph{excess} noise in the aligned case,
which is a power law shaped by the optical response; the black dashed line represents the predicted \emph{excess} noise for the misaligned case,
by only scaling the model of the aligned case with the optical response; the black dashed-dotted line represents the predicted excess noise for the misaligned case,
by scaling the model of the aligned case with the optical response and the \ac{HOM}s amplification by \ac{SR}.}
\label{fig:excess1}
\end{figure}

The dependence of the \emph{excess} noise with the interferometer
optical response shape was investigated by misaligning the signal
recycling mirror, which
reduces the bandwidth of the detector optical response from
$\sim$\qty{430}{Hz} to $\sim$\qty{190}{Hz}.
This resulted in a change in the \emph{excess} noise level and shape in calibrated data as shown in figure~\ref{fig:excess1},
effectively excluding
arm mirror displacement noise as the origin of the \emph{excess} noise.
While the dependence of the \emph{excess} noise with the optical response of the interferometer excludes displacement noise of the test masses as a possible origin,
it is consistent with a noise carried by an optical
field that is resonant inside the \ac{SRC}. In particular, when the
\ac{SRC} is misaligned the resonant gain inside the \ac{SRC} is
reduced, while the interferometer optical response gain
increases. Including both changes allows to explain the change in
inferred \emph{excess} noise shown on figure~\ref{fig:excess1}.
As discussed in section~\ref{sec:optical-configuration},
the carrier light \ac{HOM}s are resonant inside the \ac{SRC} due to
the nearly unstable configuration of that cavity. This makes them the main
candidate for the noisy field carrying the \emph{excess} noise.

In addition, noise mechanisms directly affecting the carrier fundamental mode were also investigated. Laser frequency and intensity noise, polarisation fluctuations within the interferometer,
and length noise of the \ac{OMC} cavity, were experimentally probed and found incompatible with the observed noise behaviour.
The only noise vector consistent with all observations is the coupling
of the carrier \ac{HOM}s to the fundamental mode field at the interferometer output,
while the physical origin of the noise that couples through \ac{HOM}s remains unidentified.

\subsection{\emph{Excess} noise mitigation}
\label{subsec:IntentionalMisalignment}
To mitigate the \emph{excess} noise contribution to the total detector
noise the input laser power was reduced from \qty{22}{W} to eventually
\qty{17}{W}. This has reduced the \ac{HOM} power relative to the
circulating power in the arm cavities  by  a factor 2~\cite{VIRGO:2025sym}
and reduced modestly the \emph{excess} noise. 

Afterwards, an intentional misalignment of the \ac{SR} mirror
proved effective at reducing the \emph{excess} noise, and was
applied throughout the \ac{O4} data taking. The optimum between decreasing the \emph{excess} noise and
increasing the quantum shot noise, was achieved for a misalignment
of about \qty{2}{\micro rad} corresponding to a detector bandwidth
reduction from $\sim$\qty{430}{Hz} to $\sim$\qty{190}{Hz}. 
This
reduces the amplification of the noise by the \ac{SRC}, and
improves the
\ac{BNS} range\footnote{Defined by convention as the  sky- and
  orientation-averaged distance up to which a standard \ac{BNS} merger
  would be detected with a signal-to-noise ratio of 8}
from about \qty{40}{Mpc} to
\qty{55}{Mpc}.
The \acp{HOM} distribution does not have a cylindrical symmetry, and the
direction of the misalignment was chosen to minimize the \acp{HOM}
power, which also corresponds to minimizing the \emph{excess}
noise. After the replacement of the \ac{WE} mirror (which will be described in
section~\ref{subsec:end-mirror}), the \acp{HOM} distribution changed, and
the direction of the optimal misalignment was adjusted, consistently with
\acp{HOM} arising from localised mirror surface defects that break
cylindrical symmetry.  These observations further strengthen the
conclusion that \acp{HOM} are the noisy field carrying the \emph{excess}
noise.
However, this misalignment reduced the detector bandwidth from
$\sim$\qty{430}{Hz} to $\sim$\qty{190}{Hz}, and made the vacuum
squeezing injection ineffective, as it will be discussed in section~\ref{subsec:qnr}.

\section{Scattered light and environmental noise}
\label{sec:scatter-env}

In addition to the change of the optical configuration, several
improvements in the scattered light control were implemented in Advanced Virgo
between \ac{O3} and \ac{O4}. Scattered light is a major issue in gravitational wave detectors, as
it limits the optical gain of the interferometer and creates coupling
with the motion of the surface on which the scattered light shines on. In
the following subsection we will describe the improvement in baffling and
optics quality in order to reduce light scattering, and how it improved some of
the environmental noise couplings.

\subsection{Scattered light from optical benches}
\label{sec:scatter_bench}

The interferometer output beams are sensed by means of optical benches,
as shown on figure~\ref{fig:Virgoscheme}. In particular, two main
suspended optical benches are placed in series at the anti-symmetrical
port of the interferometer. The \ac{SDB1}
hosts a mode-matching telescope followed by a filtering cavity, called
\acl{OMC} (OMC), that rejects control sidebands and \acp{HOM},
while transmitting the carrier fundamental mode. This bench is also
equipped with a \ac{FI} that attenuates the light
back-reflected or back-scattered by the OMC or any component located
downstream.  The \ac{SDB2} hosts the
different sensors involved in the readout of the interferometer
signals (longitudinal and quadrant photodiodes) as well as the cameras
used to image the beams.  In addition, other suspended optical benches are
positioned at each inteferometer optical port. Thus, end benches
(\ac{SNEB} and \ac{SWEB}) are used to sense the beams in transmission of the
arm cavities, and the \ac{SPRB} is used to detect a small fraction
of the beam circulating inside the power recycling cavity.

Each optical bench can be a source of scattered light through
different mechanisms: light scattered by optics imperfections, light
partially hitting optics edges or mechanical mounts in case of
misalignment or insufficient clear aperture, and ghost beams induced
by anti-reflective surfaces or residual transmission through
reflective optics. A fraction of this light may be back-scattered in
the direction of the interferometer and interfere with the light
resonating inside it. As the phase of the back-scattered light is
imprinted by the optical bench motion, the interference with the main
beam produces a noise impacting the interferometer sensitivity. The main
optical benches are suspended to isolate them from ground motion above
a few hertz, and placed under vacuum to immune them against acoustic
noise, which helps in minimizing the impact of scattered light noise. 
Nevertheless, as the noise coupling mechanism is non-linear, an
elevated bench motion in the micro-seismic range (between \qty{0.01}{Hz} 
and \qty{1}{Hz}) induces upconverted noise that may limit the Advanced Virgo sensitivity at
frequencies up to \qty{30}{Hz}~\cite{poliniThesis, EB_scatteredlight}.

\subsubsection{Scattered light mitigation on the optical benches}\label{sl_mitigation}

Two main actions were undertaken before the O4 run in order to reduce
the impact of back-scattered light noise from the optical benches.
New diaphragms were installed in front of the telescope parabolic
mirrors on the \ac{SDB1} bench, with the goal of preventing beam clipping
on the mirror mounts and dumping light scattered around the main
beam. In addition, a detailed study of the ghost beams present on each optical bench
was performed and beam dumps were installed to properly block these
ghost beams.

\begin{figure}
\centering
\includegraphics[width=\textwidth]{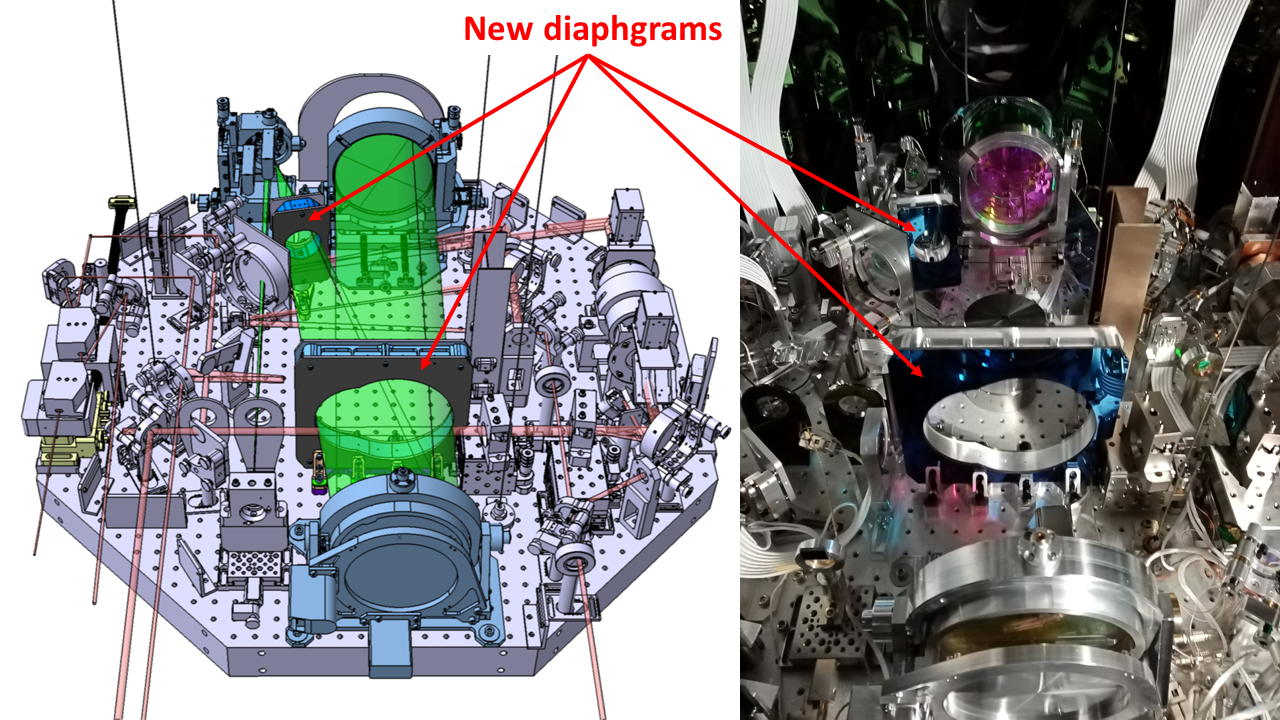}
\caption{Left: 3D drawing of the \ac{SDB1} bench where the newly
  installed diaphragms are indicated by the red arrows. The main
  infrared beams are shown as red tubes. The green beam represents the
  optical path of the Hartmann wavefront sensor light, which overlaps
  with the main infrared beam (not shown here). Right: Picture showing
  the central part of the \ac{SDB1} bench.}
\label{fig:Diaphragmes}
\end{figure}

The mode-matching telescope installed on the SDB1 bench is composed of
a meniscus lens and two parabolic mirrors that are upstream of the
\ac{FI}, as well as two other spherical lenses located
after the \ac{FI}~\cite{DF_telescope}. Light back-scattered at the level
of the first three components of the telescope is therefore not
attenuated by the \ac{FI}.
Moreover, the meniscus lens and the first parabolic mirror exhibit the smallest aperture-to-beam-radius ratio on the bench (aperture \qty{145}{mm}, 
beam radius w = \qty{23}{mm}), whereas all other optical components have a ratio at least twice as large.
This increases the likelihood of partial clipping of the beam on the mounts of the meniscus lens or the first parabolic mirror, particularly because more than \qty{50}{\%} of the beam power is distributed among \acp{HOM}, which exhibit a more extended spatial profile than the fundamental mode.

For these reasons a diaphragm was installed in front of each
parabolic mirror, which primarily aim at preventing light
from hitting the
mechanical mounts of the parabolic mirrors. This is in addition to the
diaphragm that was already integrated in front of the meniscus
lens. The second advantage of these diaphragms is that they absorb
part of the light back-scattered around the main beam by components located
downstream on the optical bench.

\begin{figure}
\centering
\includegraphics[width=\textwidth]{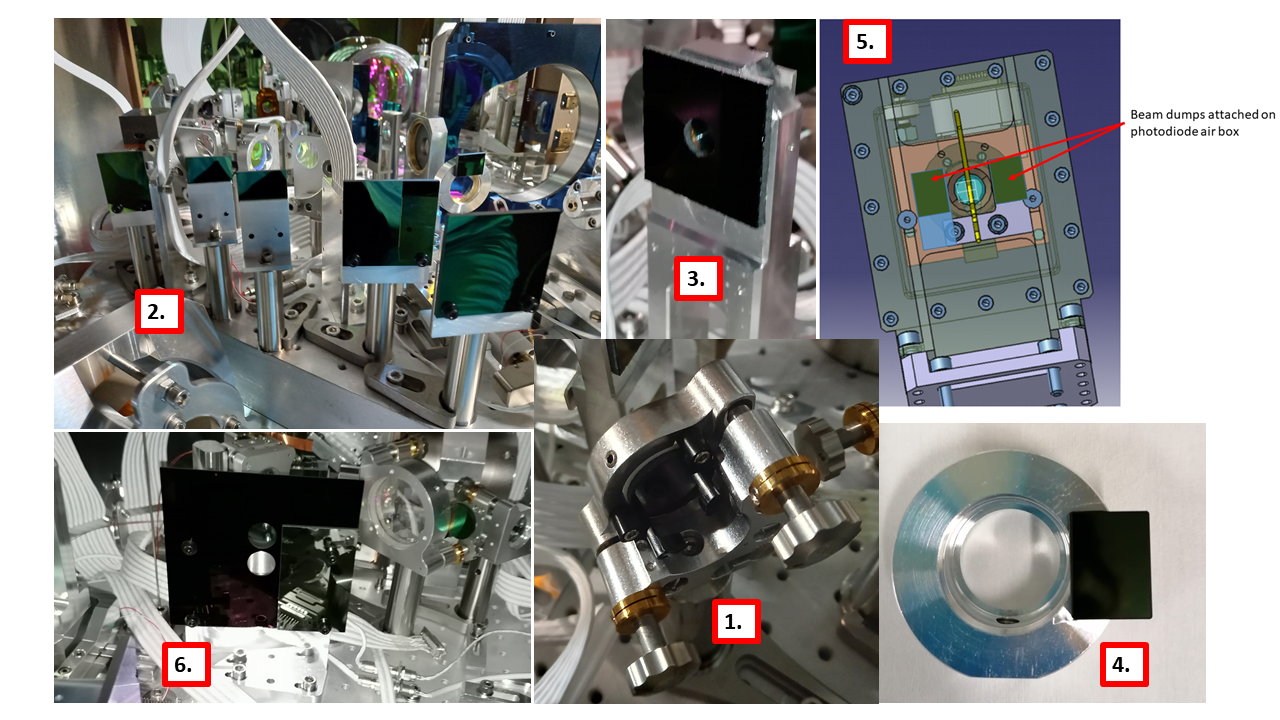}
\caption{Examples of beam dumps and diaphragms installed on the
  optical benches to block the ghost beams.
Circular absorbing glass beam dumps (1) were integrated on the highly
reflective mirror mounts to dump the ghost beams induced by the
spurious transmission through the highly reflective coatings.  Rectangular absorbing
glass beam dumps mounted on their own supports with absorbing black
screws (2) were added on the benches to dump the ghost beams induced
by spurious reflections on the anti-reflective coatings of
lenses, beam splitters, and the viewports installed at the vacuum
interface between the SDB1 and SDB2 benches. Absorbing glass diaphragm glued on the
mount of a quadrant photodiode (3), beam dump glued on a lens mount
(4), or on a photodiode enclosure (5). Silicon diaphragm blocking ghost beams exiting the interferometer (6).}
\label{fig:BeamDumps}
\end{figure}

Each diaphragm is composed of a central supporting structure made in
aluminum, with a \qty{3}{mm} thick plate of mirror-finished
anti-reflective coated stainless
steel attached on each side. 
A view of these diaphragms integrated on the SDB1 bench is shown on
figure~\ref{fig:Diaphragmes}. The external dimensions of the diaphragms
were constrained by the surrounding components on the bench, while the
inner diameter apertures were constrained by the radius of the beams
going through the diaphragms. In particular, the beam displayed in
green color on the left part of figure~\ref{fig:Diaphragmes} corresponds
to an auxiliary beam used for Hartmann wavefront sensing~\cite{HWS_Lorenzo},
that is copropagating with the main beam and has the most
limiting radius.

An effort was carried out to optimize the mechanical rigidity of the
diaphragms while not exceeding a total mass of a few kilograms. The
first resonance frequency has been measured to be around \qty{229}{Hz}
for the larger diaphragm (with a mass of \qty{2.4}{kg}) and around
\qty{325}{Hz} for the smaller one (with a mass of about \qty{1}{kg}).\\

A detailed study of ghost beams was carried out 
using a ray-tracing software to identify the ghost beams
on the most critical benches~\cite{poliniThesis} and implement solutions to properly dump
them. Figure~\ref{fig:BeamDumps} shows different examples of beam dumps
that were installed on the SDB1 and SDB2 benches. Most of these beams
dumps were made from \qty{3}{mm} thick plates of polished absorbing glass substrates 
with anti-reflective coating.
When the integration of
such beam dumps was too difficult due to space
constraints, the absorbing glass was directly glued on existing
mounts.

Most of the bench reflective optics have a wedge,
which separates
angularly the main reflection and the reflection from the second
surface. The orientation of the wedge angles were optimized given
the bench space constraints in order to ease the integration of the
beam dumps.

A silicon diaphragm (item (6) of figure~\ref{fig:BeamDumps}) was added on top of
an already existing diaphragm made of mirror-polished stainless steel
on the SDB1 bench. Silicon has a higher damage threshold as it
absorbs \qty{1064}{nm} light over several millimeters of depth, and
that part of the beam dump is used to absorb the reflections from
\ac{SR} when the mirror is misaligned during the first stages of the
lock acquisition.

Similar improvements were also implemented on the \ac{SNEB} and \ac{SWEB} benches. In
particular, absorbing glass beam dumps were added to absorb the light
reflected back by quadrant photodiodes as these components were found
to be critical in terms of scattered light noise during \ac{O3}~\cite{EB_scatteredlight}.

\subsubsection{Impact of the back-scattered light noise}\label{sl_noise}

The contribution of the
bench back-scattered light noise to the interferometer sensivity
$h_{r}(f)$ can be expressed as
\begin{linenomath}
\begin{equation}\label{equation_scatter}
h_{r}(f) =
\sqrt{f_{sc}} \,
\biggl\{
K_{n_{\phi}}(f)\,
\mathcal{F}\Bigl[
\sin\Bigl( \frac{4\pi}{\lambda}(x_{0}+ x(t)) \Bigr)
\Bigr]
+
K_{\frac{\delta P}{P}}(f)\,
\mathcal{F}\Bigl[
\cos\Bigl( \frac{4\pi}{\lambda}(x_{0}+ x(t)) \Bigr)
\Bigr]
\biggr\}\,,
\end{equation}  
\end{linenomath}
where $\mathcal{F}$ represents the Fourier Transform,
$K_{n_{\phi}}(f)$ and $K_{\frac{\delta P}{P}}(f)$ are the phase and
amplitude quadrature transfer functions (TFs) from a motion of the
optical bench where the light is scattered to the \ac{GW}
signal, $f_{sc}$ represents the fraction of back-scattered light that
recombines with the interferometer beam, $\lambda$ is the laser
wavelength, $x_{0}$ corresponds to the bench position at rest and
$x(t)$ is the bench displacement along the main interferometer beam
axis propagation.

The noise coupling transfer functions $K_{n_{\phi}}(f)$ and
$K_{\frac{\delta P}{P}}(f)$ for each optical bench were obtained from
simulations~\cite{finesse3_backscattering}.
The fraction of back-scattered light $f_{sc}$ was measured for each
optical bench during the commissioning of the interferometer and
during the observation run O4, by exciting the benches in order to
induce a large and controlled motion~\cite{EB_scatteredlight}.

Back-scattered light fractions measured at the \ac{SDB1}, \ac{SNEB}, and \ac{SWEB} benches during the O3 run, 
the pre-O4 commissioning, and the O4 run are shown in table~\ref{tab:fsc}.
The reduction of back-scattered light fractions from O3 to the pre-O4 commissioning is 
attributed to the mitigation actions described in section~\ref{sl_mitigation}.
However, an increase by about a factor 7 in back-scattered light fractions occurred from the pre-O4 commissioning
 to the O4 run estimations, particularly on SDB1.
Although this is not fully understood, a plausible explanation is that the \ac{SDB1} 
back-scattered light increased following an intervention on the bench
in July 2024, during which an 
optical waveplate was added upstream of the \ac{FI} in order to study
the polarization dependence of the \emph{excess} noise.

It has to be noted that the interferometer optical configuration was modified between the pre-O4 
commissioning and the \ac{O4} measurements of bench scattered light. The latter was performed 
with the \ac{SR} mirror misaligned by \qty{2}{\mu rad} (see section~\ref{sec:dominant-noise}), while the former 
was performed with the \ac{SR} mirror aligned. This change of optical configuration was accounted for in the simulation
used to obtain the back-scattered light noise coupling transfer functions.

\begin{table}
\centering
\caption{Measured back-scattered light fractions on Virgo optical benches}
\label{tab:fsc}

\begin{tabular}{lrrr}
\hline
\hline
\textbf{Optical bench} & \textbf{$f_{sc}$ 2020 (O3)} & \textbf{$f_{sc}$ 2023 (pre-O4)} & \textbf{$f_{sc}$ 2024 (O4)} \\

\hline
SDB1 & $2 \times 10^{-9}$ & $7 \times 10^{-10}$ & $5 \times 10^{-9}$ \\
SNEB & $2.5 \times 10^{-8}$ & $1.1 \times 10^{-8}$ & $1.2 \times 10^{-8}$ \\
SWEB & $4 \times 10^{-8}$ & $3 \times 10^{-9}$ & $6 \times 10^{-9}$ \\
\hline
\hline

\end{tabular}

\end{table}

\begin{figure}
\centering
\includegraphics[width=0.49\textwidth]{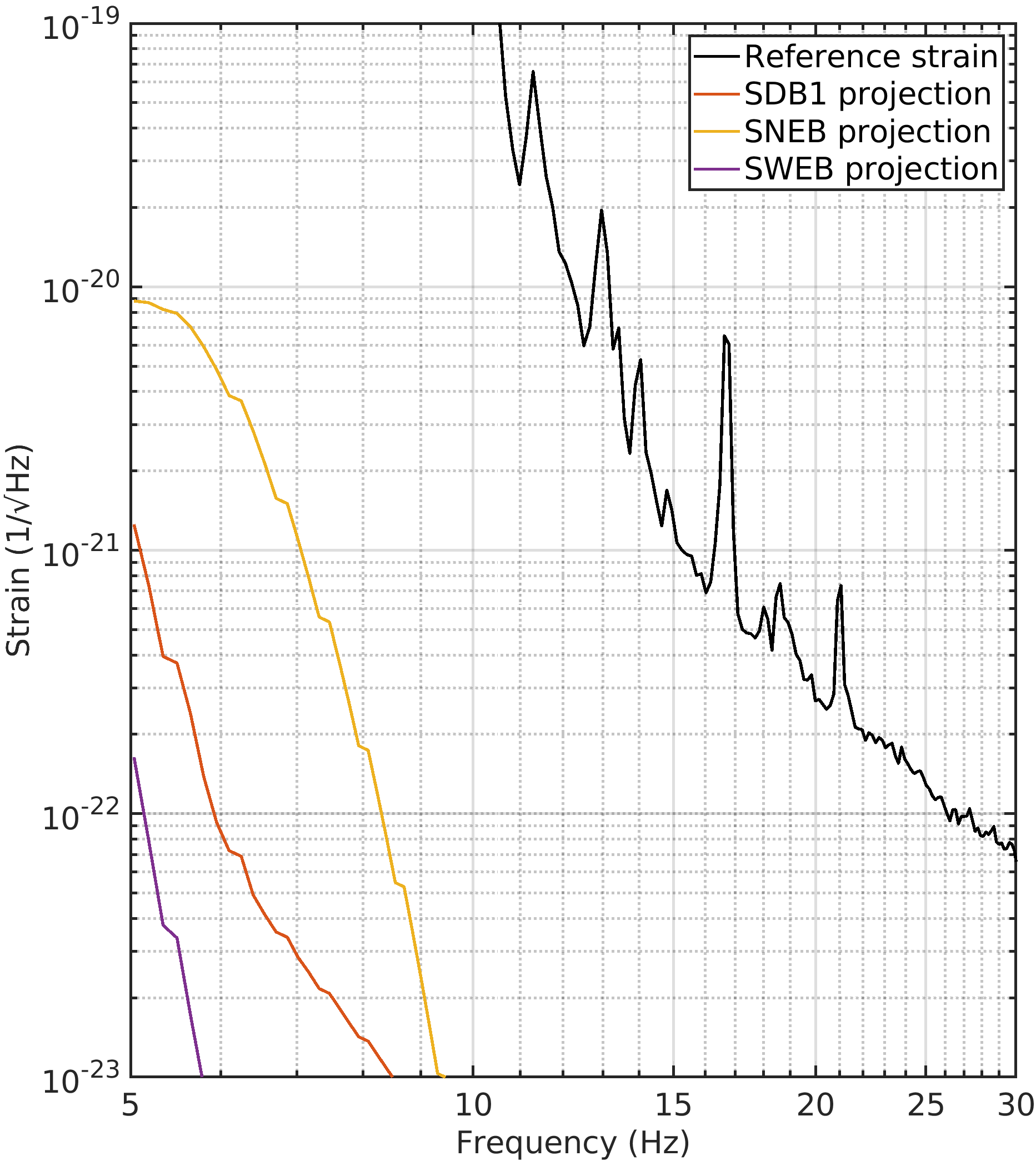}
\includegraphics[width=0.49\textwidth]{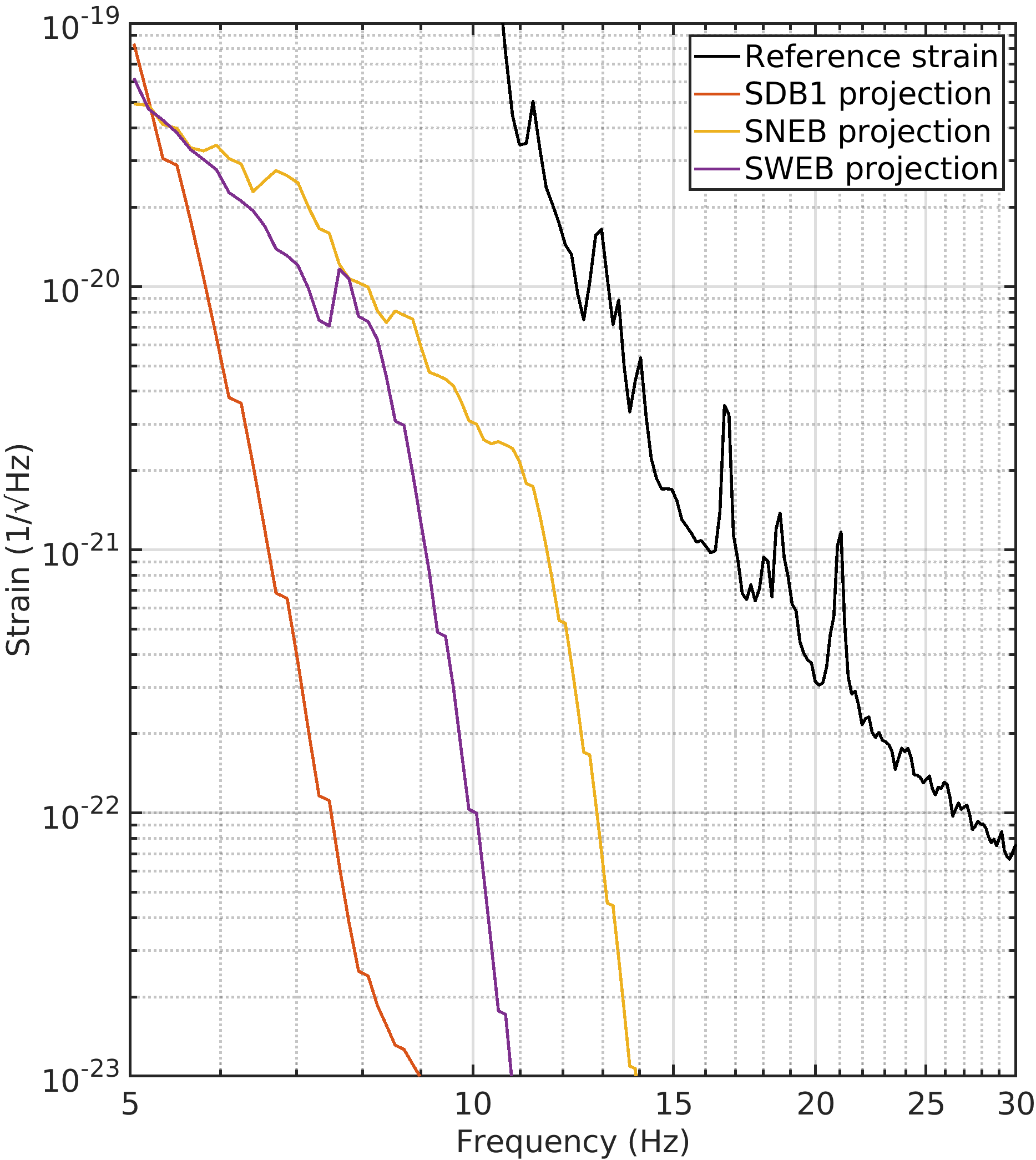}
\caption{Projection of back-scattered light noise from the optical
  benches \ac{SDB1}, \ac{SNEB}, \ac{SWEB} compared to the strain
  noise curve. Left: Data from February 12, 2025,
  during a period of low bench motion. The maximum bench speed measured
  was around the median
  of the annual speed distribution.
  Right: Data from March 2, 2025, during a period of high bench motion.
  For \ac{SNEB} and \ac{SWEB} the maximum speed exceeded the 99\textsuperscript{th} percentile,
  while \ac{SDB1} it is around the 87\textsuperscript{th} 
  percentile of the distribution.
}
\label{fig:scatter_weather}
\end{figure}

The contribution of back-scattered light to the detector noise curve
depends on bench displacement, which is typically related to micro-seismic
activity on site~\cite{paper_O3_environmentNoise}.
Projections of the back-scattered light noise from suspended benches
are compared with the strain noise curve for two different conditions of
residual bench motion in figure~\ref{fig:scatter_weather}.
During periods of low bench motion (figure~\ref{fig:scatter_weather},
left panel), 
back-scattered light noise from the suspended optical benches is entirely negligible. 
While during periods of elevated bench motion (figure~\ref{fig:scatter_weather}, right panel), 
the noise projection becomes significant, but with a cut-off  below
\qty{12}{Hz}, which is a factor two
improvement over the O3 performance for the end
benches~\cite{EB_scatteredlight}. This is the result of replacing an L-4C
geophone with a triaxial Nanometrics TC-120 seismometer, at the top of
each bench suspension~\cite{VIR-0596A-19}. The seismometer has orders of magnitude lower
sensing noise below \qty{0.5}{Hz} compared to the geophone,
considerably improving the inertial control during periods of high
ground motion.

Despite of the improvements made to the suspended benches between O3 and O4, 
back-scattered light noise remains a potential limitation to detector sensitivity 
during adverse weather conditions. This underscores the need for further improvements 
in preparation for future detector noise curve decreases, including additional modeling 
to better understand the influence of optical configuration and bench setups on this noise.

\subsection{End mirror replacement}
\label{subsec:end-mirror}

The losses of the mirrors of the arm cavities are the dominant source
of loss in the interferometer. To reduce dust contamination that can
be a major source of scattering losses, the mirrors are protected with
a layer of First Contact\texttrademark~polymer during installation, that is removed
just before pumping down the vacuum chamber hosting the mirror.
After installation, the measured round trip losses in the
arm cavities were $\sim$\qty{70}{ppm}~\cite{VIRGO:2025sym}. These
losses have been reduced between 2023
and 2025 and the 
recycling gain increased by 20\% with the following successive mirror changes.

The first replacement was in June 2023, when \ac{NE} was replaced by a
spare mirror.  It was contaminated by an uneven layer of pollution
that increased the scattering by a factor 10, and that layer was
tracing the imprint of First Contact\texttrademark.  After replacement
the arm cavity gain increased by 10\%. In addition, \ac{NE} had low
mechanical quality factor of $\sim$\num{1e5}, after replacement by a
spare mirror the quality factor increased and was measured to be
\num{9.2e6}.

The second replacement was for the \ac{WE} mirror that had a large point absorber located \qty{2}{cm} away from the
center of the mirror. This point absorber created a thermal
deformation of the mirror which scattered light into \acp{HOM},
in particular into modes order 8 and 9 which are close to
co-resonating with the fundamental mode of the arm cavities. This
near co-resonance amplified losses in the arms~\cite{LIGOScientific:2021kro}. That loss was
dominating the total arm losses, as displacing the position of the beam
away from the center of that mirror could change the arm power by
$\pm10$\%~\cite{VIR-1047A-19}.
To solve that issue, the \ac{WE} was replaced in May 2025
(during a LIGO downtime due to equipment failure~\cite{GWTC-5-intro})
by the cleaned
mirror that had been removed from the \ac{NE} tower back in 2023. After that
installation the arm cavity gain increased by another 10\%, and the
quality factor of the cleaned \ac{NE} mirror was measured to be
\num{2e6}, a factor 20 increase due to the cleaning. 

In addition, the missing anti-reflective outer baffle around the \ac{WE}
mirror payload was installed at the same time.
Prior to that change, the dominant source of scattered light
up-conversion during elevated micro-seismic times was correlated with
ground motion at the west end building, and tapping tests could
identify excess coupling from the vacuum vessel
several meters in
front of \ac{WE}, at approximately the location to which
the mirror suspension structure reflects light on the vacuum system walls.
The baffle installation reduced significantly the
acoustic noise coupling to the detector noise curve, as we will discuss in
section~\ref{sec:magnetic-acoustic}.

\subsection{Magnetic and acoustic noise}
\label{sec:magnetic-acoustic}

The effect of disturbances induced on the detector by magnetic and
acoustic noise can be estimated through noise injections performed in
selected locations. The aim is to produce a measurable effect on the
strain noise curve,
highly correlated with the injected noise that is measured by an
environmental witness sensor. In this way a coupling function can be
measured~\cite{Env_Couple}.

During \ac{O4}, recurrent  magnetic and acoustic noise injections were
performed to characterize the coupling of environmental disturbances
to the interferometer output and to derive the corresponding noise
projections.

Before the start of the run, a new magnetic noise injection system was designed, installed and characterized.
The aim was to have the possibility to perform more intense and repeatable noise injections, allowing a better
knowledge of the effect of the magnetic noise on the detector with
respect to \ac{O3}~\cite{Env_O3}. To this aim a large coil, with an
area of about 10~m$^2$, was mounted on the wall inside each of three
main detector buildings: \ac{CEB}, \ac{NEB} and \ac{WEB}.
The coil is driven by a high power amplifier
with a maximum output voltage of \qty{\pm 72}{V} and
a maximum output current of \qty{\pm 6}{A}. The signal is generated digitally and
drives the amplifier through a \ac{DAC}. 
The intensity of the induced magnetic field inside the experimental halls is of the order of a few tens of nT. 

The magnetic monitoring system was also expanded with the installation
of additional magnetic field sensors: three axis fluxgate sensors,
placed very close to the vacuum chambers
containing suspended mirrors (\ac{NI}, \ac{WI}, \ac{BS}, \ac{NE}, \ac{WE}) and suspended optical
benches (\ac{SIB1}, \ac{SDB1}) which are likely coupling locations due to the presence of magnetic sensitive devices, like magnetic actuators and Faraday isolators~\cite{Env_O3}.

Likewise, a recurrent acoustic noise injection setup was 
deployed.
The noise injection system was based on loudspeakers
installed in fixed
locations inside the \ac{CEB}, \ac{NEB} and \ac{WEB} experimental halls.  Toward the
end of \ac{O4}, the setup was further extended by adding 
loudspeakers in the Laser and Detection clean room laboratories (where optical benches are located),
improving the coverage of areas potentially relevant for acoustic
couplings.

During \ac{O4}, magnetic and acoustic injections were carried out as
part of coordinated campaigns automated using the same control
software supervisor used for the interferometer lock acquisition,
\textsc{Metatron}~\cite{Bersanetti_Metatron, Virgo-DetChar-O3-Results}.
The
injections were scheduled every two weeks and performed
sequentially in the three main buildings.  Magnetic injections were
performed by injecting a swept sine signal (\qtyrange{8}{1000}{Hz}), and
occasionally through a set of lines in the \qtyrange{2.5}{25}{Hz} range.  For
acoustic injections, the excitation consisted of band-limited colored
noise in the \qtyrange{20}{2000}{Hz} frequency range.

The injection amplitudes were tuned in order to maximize the response
of the witness sensors, while keeping the interferometer at its
working point. Incoherent coupling functions $CF(f)$ were computed
as the square root of the ratio of the increase in power spectral
density of the strain data and the increase in the power spectral
density of the witness sensor~\cite{Env_Couple}, 
whenever the injected disturbance produced a meaningful increase in
the strain data power spectral density.

Once the coupling function is known, the expected contribution
$h_{noise}(f)$ of the corresponding environmental noise under \emph{quiet} (no injections)
conditions can be estimated by projecting the witness sensor spectrum $X(f)$ through the
measured coupling function:
\begin{linenomath}
\begin{equation}
h_{noise}(f) = CF(f)\,X(f)\,.
\end{equation}  
\end{linenomath}

For magnetic injections, 
$X(f)$ corresponds to the magnetic field measured by one magnetometer
in each building, while for acoustic injections it corresponds 
to the sound pressure measured by the microphone centrally located in
each building. 

\begin{figure}
\centering
\includegraphics[width=\textwidth]{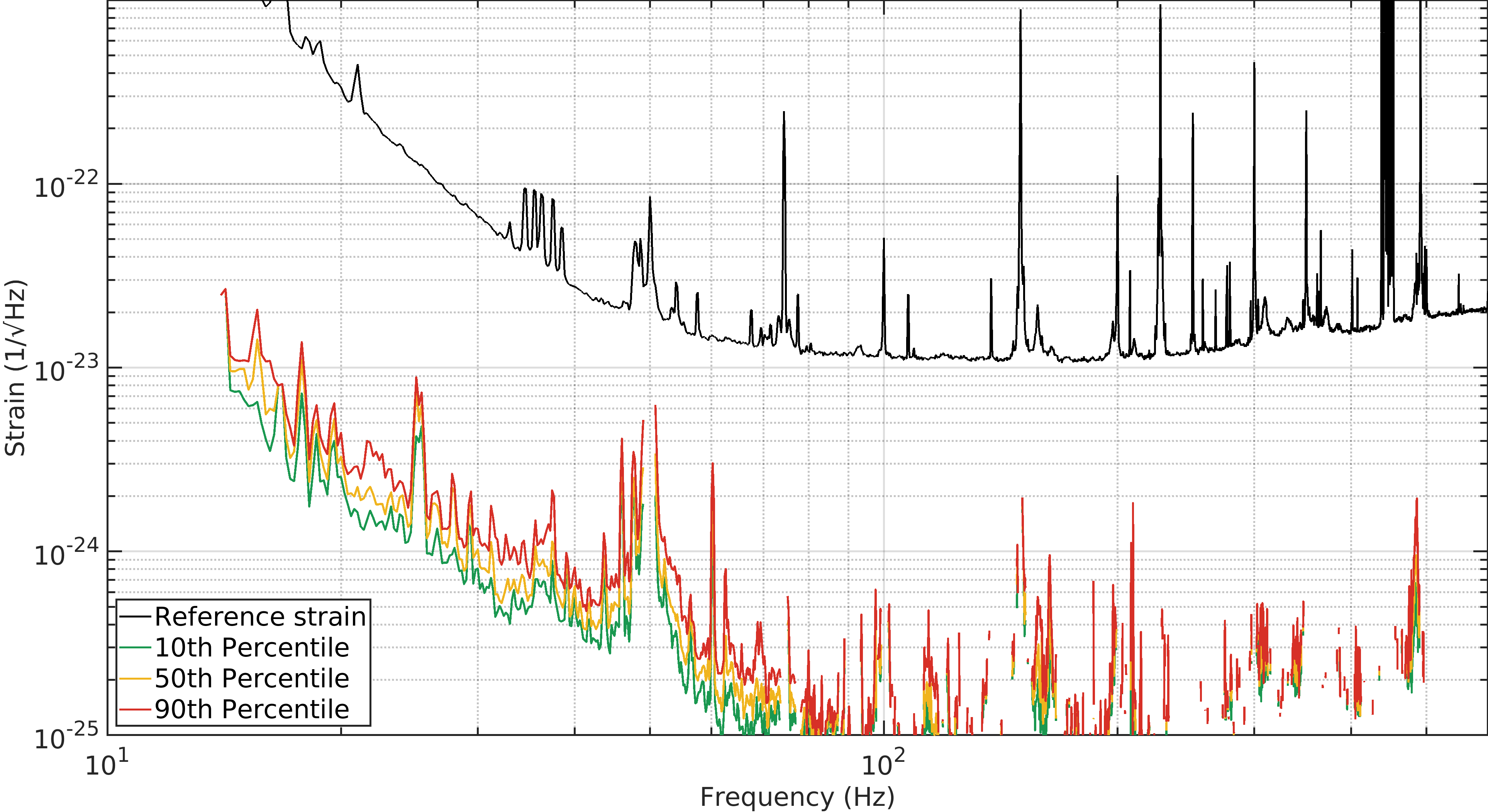}
\caption{Projections of magnetic noise in the central experimental halls, compared with quiet detector noise curve.}
\label{fig:MagProj}
\end{figure}

The magnetic noise projection for \ac{CEB} is shown in
figure~\ref{fig:MagProj}. The projections are not performed at \qty{50}{Hz} (mains frequency)
and its harmonics, as the injected field is lower than the magnetic field already 
present under quiet conditions, which prevents the measurement of the
coupling at these frequencies.
The magnetic noise does not appear as a limiting factor in CEB and similar results also
apply for NEB and WEB.

\begin{figure}
\centering
\includegraphics[width=\textwidth]{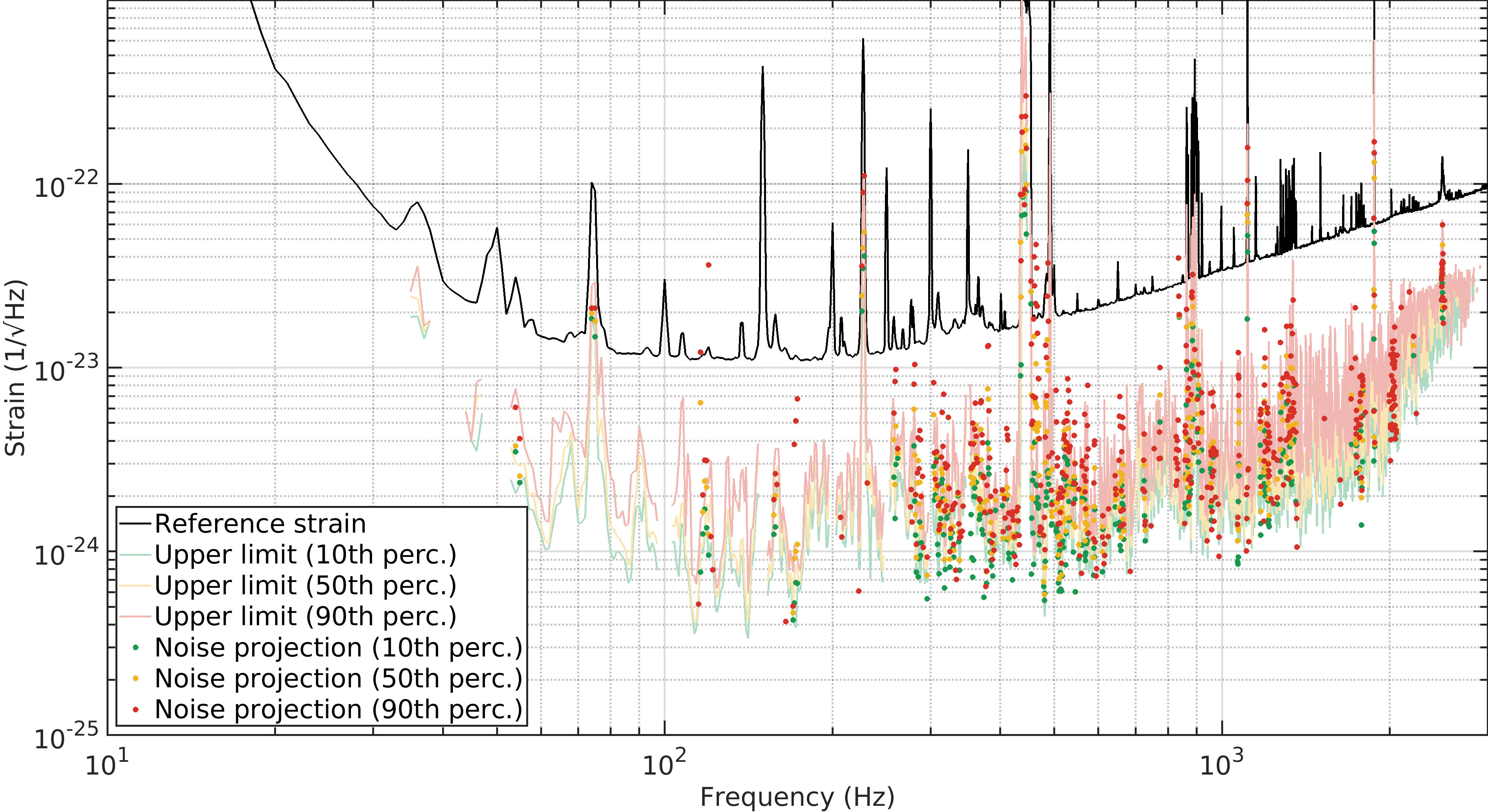}
\caption{Projections of acoustic noise for the central experimental hall compared with the quiet detector noise curve. 
The faint colored curves show upper-limit estimates, while the bold
dots are measurements.}
\label{fig:AcousticProj}
\end{figure}

The acoustic noise projection for \ac{CEB} is shown in
figure~\ref{fig:AcousticProj}.  The reference strain noise curve is
defined as the median strain spectrum computed over all the analyzed
quiet periods. The colored markers represent the 10th, 50th and 90th
percentiles of the acoustic noise projections, computed from the
measured coupling functions and the quiet microphone
spectra.
Overall, the projected acoustic contribution remains below the
reference strain noise curve over most of the frequency band,
although a limited number of points show higher values, with the upper
percentiles locally approaching the reference curve.

\begin{figure}
\centering
\includegraphics[width=\textwidth]{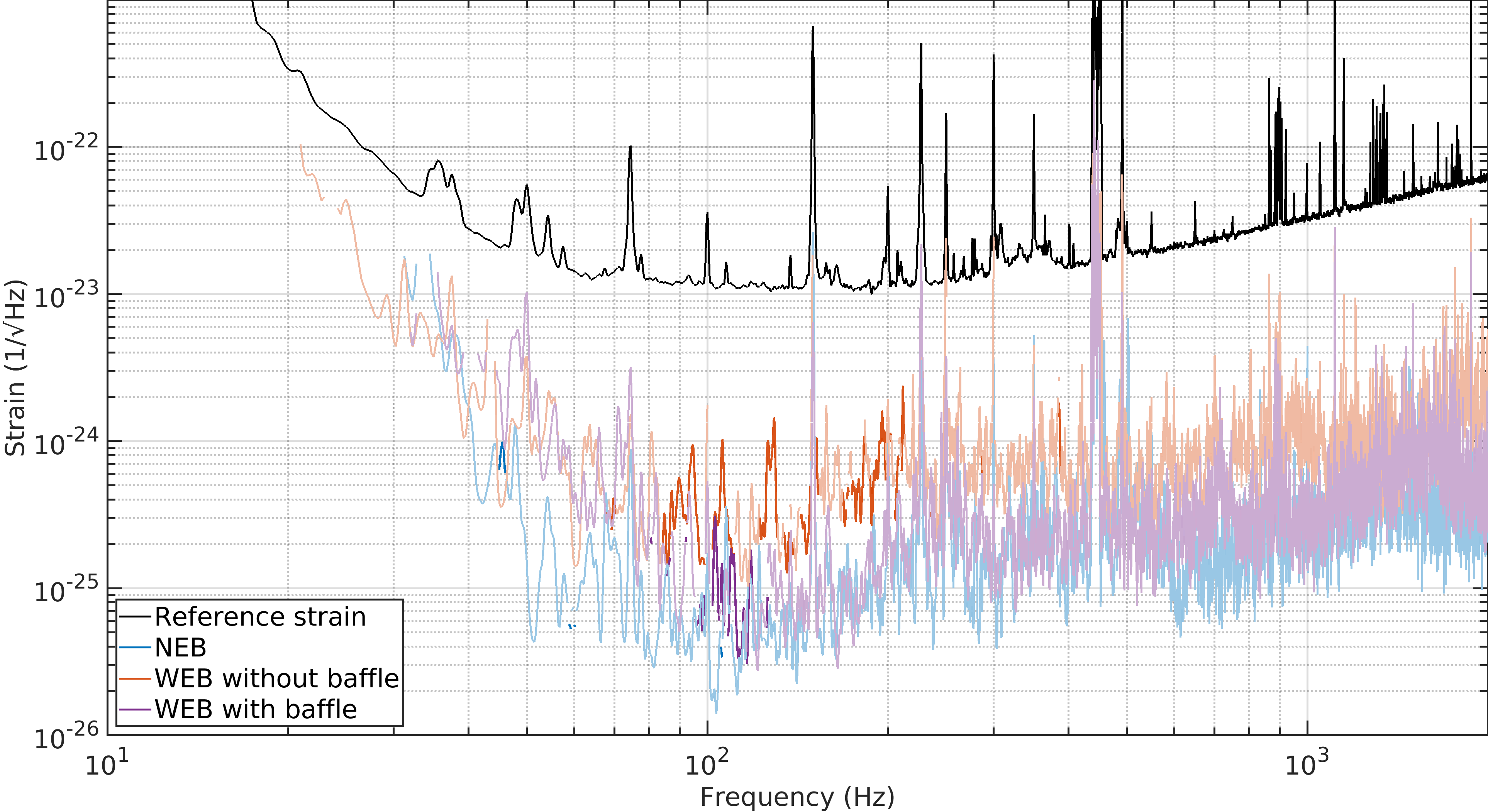}
\caption{Projections of acoustic noise based on injections performed
  at the \ac{NEB} and at the \ac{WEB}, respectively before
  and after  the installation of the \ac{WE} payload
  baffle. The black curve is a reference strain noise in quiet conditions. 
  The faint colored curves are upper-limit estimates, while the bold curves are measurements.}
\label{fig:AcousticProj_Baffle}
\end{figure}

The recurrent noise injection allows to monitor changes in coupling
over time. For example, after the installation of the baffle on the
\ac{WE} payload (see section \ref{subsec:end-mirror}) the acoustic noise projection in \ac{WEB} reduced by
about a factor 10, as shown on figure~\ref{fig:AcousticProj_Baffle},
and became comparable to the acoustic noise projection in \ac{NEB}.

\section{Effective instrumental upgrades}
\label{sec:instrument-upgrade}

In addition to the installation of the \ac{SR} mirror and the
improvement in scattered light control described previously, 
a large number of other upgrades took place. In the following
subsections, we describe some of the most notable instrumental changes
during the period between \ac{O3} and \ac{O4}. These upgrades were
effective in achieving their individual goals, however in many cases
they did not result in an overall detector improvement due to the
challenges of nearly unstable dual-recycling masking their effect.

\subsection{High finesse output mode cleaner}

The \acl{OMC} (OMC) is an optical cavity located at the
output port of the interferometer whose purpose is to filter the
carrier beam containing the \ac{GW} signal. The
incoming beam is filtered by the cavity so that only the fundamental
mode, which carries the GW information, is transmitted when the \ac{OMC} is
on resonance. The reflected beam instead contains higher-order modes
and \ac{RF} sidebands~\cite{ducrot:tel-01489175, poliniThesis}. The \ac{OMC} is
a bow-tie monolithic cavity made of fused silica (Suprasil~3001)
consisting of four polished surfaces, three flat and one
spherical.

One of the main goals of the first phase of Advanced Virgo Plus was to reduce optical losses
at the anti-symmetric port of the interferometer in order to improve
the performance of the squeezing system. The main reduction in losses was
achieved by replacing the \ac{OMC} system, previously composed of two
cavities in series (each with a finesse of $\sim 123$), by a single
\ac{OMC} cavity with a higher finesse ($\sim 1000$)~\cite{paper_OMC_O4}. This upgrade
eliminated the relative mismatch losses between the two cavities
($\sim 2.5\%$)~\cite{Bonnand:2017hdk}.\footnote{In particular, the
  losses associated with the mismatch between the two cavities in
  series were relative misalignment, mode mismatch, and polarization
  mismatch caused by substrate birefringence~\cite{VIR-0596A-19}.}

In addition, the half-wave plate located upstream of the \ac{OMC} was
mounted on a motorized rotator. This allows remote tuning of the
polarization matching between the incident beam and the cavity,
providing an additional reduction in losses of approximately
$\sim 1\%$~\cite{VIR-0596A-19}.

However, the internal losses of a cavity increase with its finesse,
since they scale with the number of round trips performed by the beam
inside the cavity. These losses are mainly due to scattering from the
four internal surfaces and are proportional to the square of the
surface micro-roughness. To compensate for the increase in internal
losses associated with the higher finesse, the root-mean-square micro-roughness of
the surfaces was improved by a factor of three. This improvement
compensates for the increase in finesse from 123 to 1000. The total
internal losses measured are $(2.05 \pm 0.10)\%$~\cite{poliniThesis}, which 
is roughly compatible with the expected losses of $(1.4 \pm 0.4)\%$.
\footnote{The internal loss budget includes absorption in the
  substrate ($7.5 \pm 5$ ppm per round trip), absorption in the
  coatings ($4 \pm 4$ ppm per round trip), scattering losses due to
  mirror defects ($3.5 \pm 0.5$ ppm per surface, corresponding to
  $14 \pm 2$ ppm for the four faces), Rayleigh scattering
  ($(17.0 \pm 0.5)$ ppm per round trip), and residual transmission of
  the high-reflectivity mirror ($2.5 \pm 0.4$ ppm). These
  contributions are summed and multiplied by the number of round
  trips, defined as $N = F/\pi \simeq 318$ for $F = 1000$. The
  expected total losses are therefore $(1.4 \pm 0.4)\%$.}
For comparison, the internal losses measured on the Virgo \ac{OMC}
 cavities of finesse $\sim 123$ were about $\sim 1\%$ per cavity, 
corresponding also to a total of $\sim 2\%$ for the two cavities in series. 
It should be noted that the \ac{OMC} initially installed for \ac{O4} was damaged prior to the start of \ac{O4}, and was replaced by a spare with higher losses (around $\sim5\%$).

The increase in \ac{OMC} finesse also fulfills another important
requirement, namely to provide a sufficient filtering of the \qty{6}{MHz}
sideband. With
the previous \ac{OMC} system, the relative intensity noise of the \qty{6}{MHz}
sidebands would have become a dominant noise source. The transmission
factor of the \qty{6}{MHz} sidebands through the former OMC system was 0.05,
whereas with the new single cavity of finesse $\sim1000$ it is
reduced to 0.0044, corresponding to an improvement of about one order
of magnitude~\cite{VIR-0596A-19}.

\subsection{Quantum noise reduction system}
\label{subsec:qnr}

The Virgo quantum noise reduction system was operational during
\ac{O4} and generated $\sim$\qty{8.5}{dB} of squeezing across \ac{O4}. 
The most relevant upgrade to the quantum noise reduction system
concerned the installation and commissioning of a \qty{285}{m}-long
filter cavity, and of in-vacuum suspended optical benches 
for the injection of squeezing.

The filter cavity has been demonstrated to yield a \qty{2}{dB} gain around the cavity 
resonance frequency, $\sim$\qty{50}{Hz}, and \qty{5.6}{dB} in the shot-noise-dominated band, above \qty{300}{Hz}. 
These gains have been measured in standalone configuration, using an internal balanced homodyne detector~\cite{virgocoll:FrequencyDependentSqueezedVacuum2023}. 
However, the radiation-pressure effects observed during \ac{O3}~\cite{Virgo:2020xlu} were still not visible in the sensitivity budget of the interferometer, 
and employing the filter cavity would result in an enhancement of optical losses at low frequencies, with negligible quantum noise suppression in this band. 
Therefore, for O4, we commissioned the squeezing system coupled to the detector by injecting frequency-independent squeezing~\cite{virgocoll:IncreasingAstrophysicalReach2019}. 

As stated in section~\ref{subsec:IntentionalMisalignment}, the
\emph{excess} noise is mitigated by intentionally misaligning the
\ac{SR} mirror in order to increase losses in the \ac{SRC} and reduce
the \acp{HOM} amplification. These losses are frequency-dependent and
highly detrimental to detected squeezing levels, particularly in the
high-frequency band ($f \gtrsim$ \qty{300}{Hz})~\cite{kwee:DecoherenceDegradationSqueezed2014,toyra17,mcculler:LIGOsQuantumResponse2021}.

\begin{figure}
    \centering
    \includegraphics[width=.7\textwidth]{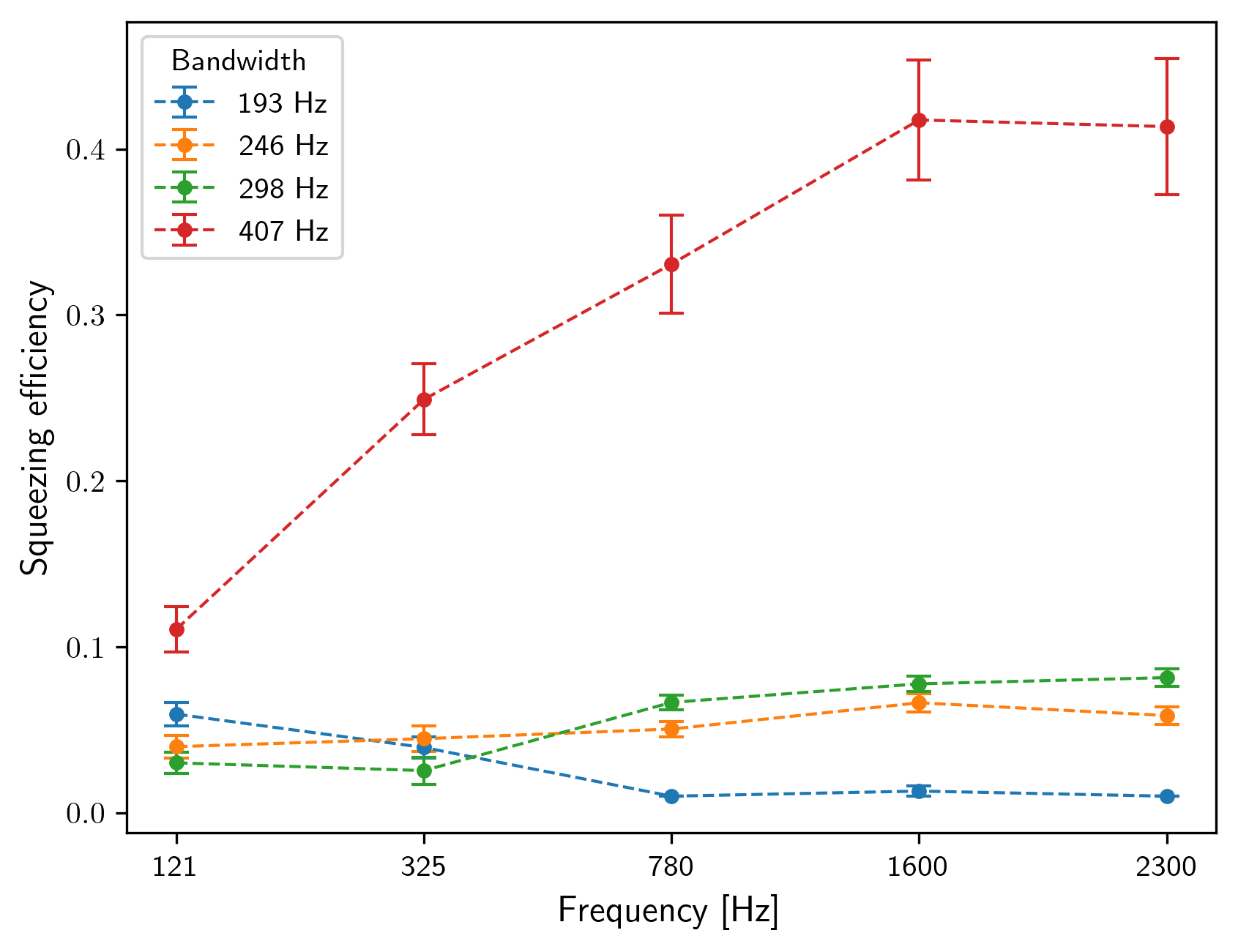} 
    \caption[\small{Effective squeezing efficiency in Virgo before and during the \ac{O4}b run. }]
    {Squeezing efficiency in Virgo before and during
        the \ac{O4}b run, for different detector bandwidth frequencies. The median values of the investigated 
        frequency bands are reported along the $x$-axis. The red line
        corresponds to the measurements taken with aligned
        configuration. Repeated scans with the same detector bandwidth
        are omitted for clarity, as they showed high consistency. 
        Adapted from \cite{demarco:EnhancingAstrophysicalReach2025}. }
    \label{fig:sqzeff_scans_all}
\end{figure}

The trade-off between \emph{excess} noise mitigation at mid-frequency
and quantum shot noise reduction at high-frequency has been studied,
by measuring the squeezing efficiency in selected frequency bands from 
\qty{110}{Hz} to \qty{2340}{Hz} as a function of \ac{SR} misalignment, 
quantified via the detector bandwidth~\cite{demarco:EnhancingAstrophysicalReach2025}. 
The efficiency quantifies the fraction of degree of squeezing effectively 
reducing the detector noise, with respect to the produced level of squeezing. 
These measurements have been repeated multiple times and showed high 
consistency over several months. 
The interval of detector bandwidth between \qty{300}{Hz} and 
\qty{400}{Hz} could not be steadily maintained due to the constraints 
of the error signal used to control the \ac{SR} misalignment. Hence, we 
investigated only misalignment values leading to a bandwidth lower than 
\qty{300}{Hz}, and the aligned configuration with a bandwidth of $\sim$~\qty{400}{Hz}. 
The overall squeezing efficiency varies drastically between the aligned and misaligned cases. 
Figure~\ref{fig:sqzeff_scans_all} shows that the squeezing efficiency attains a value 
of $(41 \pm 4)\, \%$ in the \qtyrange{2260}{2340}{Hz} band, only with the aligned configuration.

\begin{table*}
    \centering
    \renewcommand{\arraystretch}{1.2}
    \caption[\small{Squeezing loss budget in \ac{O4}b. }]
    {\small{Squeezing loss budget in \ac{O4}b, in the high frequency band $f \gtrsim$ \qty{1500}{Hz}, and with aligned \ac{SR} mirror. 
    Intervals are provided to account for the variation of the experimental conditions across several months. 
    The total loss is deduced from the product of the individual
    efficiencies, therefore it does not simply correspond to the
    direct sum of individual loss terms. }}
    \begin{tabular}{lc}
        \hline
        \hline
        Loss source & Loss value [\%] \\
        \hline
        Optical parametric amplifier escape inefficiency & \numrange{1.0}{1.5} \\
        \acp{FI} in injection & \numrange{4}{5} \\
        Injection losses & \numrange{6.0}{8.0} \\
        \ac{SRC} losses & \numrange{20}{25} \\
        \ac{FI} in detection & $\sim 1$ \\
        Mode-mismatch to the \ac{OMC} & \numrange{3}{5} \\
        \ac{OMC} intra-cavity losses & $\sim 5$ \\
        Control pick-off losses & $\sim 1$ \\
        Quantum inefficiency of the photodiodes & $\sim 1$ \\
        \hline
        \textbf{Total (only quantum)} & \textbf{\numrange{36}{43}} \\
        \hline
        Technical noise equivalent losses & \numrange{30}{40} \\
        \hline
        \textbf{Total (observed)} & \textbf{\numrange{55}{66}} \\
        \hline
        \hline
    \end{tabular}
    \label{tab:sqz_loss}
\end{table*}

A loss budget estimation was also carried out separately, to
disentangle the single loss mechanisms between each other, and inform
future upgrades to the squeezing system.  In particular, the
losses equivalent to non-quantum technical noise are obtained by
comparing the sum and the difference of the two DC readout photodiodes.
The latter
constitutes a real-time reference for quantum shot noise, allowing for
an assessment of both classical and quantum squeezing losses.

Table~\ref{tab:sqz_loss} lists the optical loss budget for the
squeezing beam path in the aligned case, from its optical parametric amplifier source,
through the interferometer and to the detection photodiodes. It refers
to the losses in the high-frequency band $f \gtrsim$ \qty{1500}{Hz}.
The \ac{SRC} is the most relevant source of optical
losses to the squeezing. Independent estimates of the \ac{SRC}
round-trip losses gave an upper bound of \qty{4}{\%}
\cite{SQZdegradation}, which is then amplified by a factor
8
by the resonance in the \ac{SRC}.
The amount of losses on cavity resonance has been estimated to be about \qtyrange{20}{25}{\%}, 
by injecting a bright laser probe through the same path of squeezed light, 
and comparing the amount of light reflected by the \ac{SRC} on and off
resonance. The total efficiency derived from the loss budget is $(40
\pm 5)\, \%$, in agreement with the measured efficiency of $(41 \pm
4)\, \%$. 

In the misaligned case, the squeezing efficiency drops below
\qty{10}{\%} independently of the frequency band and of the detector bandwidth. 
Misalignment losses are introduced, with an impact of at least \qty{30}{\%} on the total measured efficiency. 
We conclude that no trade-off is possible 
between \ac{SR} misalignment and squeezing providing a
reasonable sensitivity gain in the high frequency band.

\subsection{Input mode cleaner payload}

The replacement of the \ac{IMC} end mirror payload
was motivated by two reasons: 
contamination of the mirror during the installation (which resulted in increased
throughput losses of the cavity) and instability in the payload
control \cite{VIR-0435A-19}. In particular, the
original payload featured a gearbox mechanism to adjust the length
of the cavity by a few centimeters, which was seldom used and had a
spurious resonance at \qty{8}{Hz}. The payload had also a
small separation between the metallic suspension wires of the mirror
(\emph{cradle}) of about \qty{8}{mm}, which was too small 
and occasionally caused the mirror to slip, complicating stable
operation as laser power increased.

The new payload design was simulated using \textit{Octopus}
\cite{Octopus_sw} to validate and tune the design geometry \cite{Chiummo2019_sims}.
The \emph{cradle} separation was increased from \qty{8}{mm} to \qty{20}{mm}
in order to improve pitch stiffness; the height of the marionette suspension
point relative to its center-of-mass was modified to obtain a
slightly higher pitch and roll resonance frequencies, and the gearbox length
adjustment mechanism was removed. This resulted in a simplified
control strategy and improved controllability~\cite{logbook_50065}. 
The angular control loops exhibited higher
robustness, allowing for stable locking even during periods of
increased environmental noise or at the higher laser power initially
foreseen for the \ac{O4} run ($\approx$~\qty{50}{W}).

In addition, an instrumented baffle was installed around the \ac{IMC}
end
mirror~\cite{instr_baffle}. This system incorporates custom-made
photosensors to monitor small-angle scattering in real-time. It serves
as a tool for initial beam alignment, which is otherwise difficult due
to the high efficiency of standard baffles at trapping stray
light. Furthermore, stray light distribution as measured by the
instrumented baffle provided measurements of the mirror deformations~\cite{Ballester:2021bua}.

\subsection{Fibered EOM and broadband phase noise}
\label{sec:fiberedEOM}

Virgo had a long standing issue of saturation of the high frequency corrections
of the free space \ac{EOM} acting in the \ac{IMC} frequency loop,
causing the loss of control of the \ac{IMC} and the rest of the interferometer as a consequence.
To increase the dynamic of that actuator, a pigtailed fibered \ac{EOM}
was installed for \ac{O4}. This actuator provides a larger dynamic
range, \qty{60}{rad}, than the free-space phase modulator which had a dynamic
of \qty{3}{rad}, and the larger dynamic range prevents the high frequency correction
saturation. The origin of this phenomena, as well as the strategy
adopted to mitigate it, are extensively described in
reference~\cite{fastunlocks}.

However, the increased actuation dynamic also resulted in a higher
sensitivity to electronic noise at the actuator input. In particular,
the first version of the driving electronics introduced broadband
electronic noise at its output that resulted in optical phase noise in
the MHz region. This additional phase noise at
\qty{2}{MHz} was modulated by the \qty{6}{MHz} sideband resulting in a
pollution of the error signals based on the \qty{8}{MHz} modulation and
vice versa.  The
resulting disturbance was filtered by the \ac{IMC} cavity, and appeared as
broadband noise with a corner frequency corresponding to the \ac{IMC}
cavity pole (near \qty{520}{Hz}) on several interferometer control signals.

The issue was mitigated by modifying the driving electronics, notably
through a redesign of the gain chain and the addition of a low-pass
filter at the output stage. These changes reduced the electronic noise
at \qty{2}{MHz} by a factor of 14, down to \qty{36}{nV/\sqrt{Hz}}. The
upgrade significantly improved the noise floor of several
interferometer error signals, although it did not impact
the overall detector sensitivity since this noise source was not
limiting.

\subsection{Calibration}
\label{sec:calibration}

The production of the calibrated strain time series requires
devices able to produce a well known mirror displacement.  During O3,
this was done with the so-called photon calibrator using the
radiation pressure of an auxiliary laser~\cite{PcalO3}.  A new system,
the Newtonian calibrator, was deployed for O4.  This system,
made of spinning rotors, produces a modulated gravitation field which
moves one or two end mirrors.  During the commissioning phase, the
Newtonian calibrator has demonstrated an uncertainty of 0.17\%~\cite{NCal}, a factor
three smaller than for the photon calibrator~\cite{PcalO4}.  Therefore, just prior
the start of O4, the photon calibrator have been recalibrated on the
Newtonian calibrator, making
it the absolute reference for the Virgo calibration during O4.  The
overall stability of both calibrators during O4 was around
0.1\%~\cite{Calibration}.  While the Newtonian calibrator was used as an absolute reference
at a frequency of about \qty{36}{Hz}, the photon calibrator was the primary instrument to
scan a wide range of frequency and establish the calibration models of
the various mirror actuators.

From the calibration models, the detector strain time series is
reconstructed in low-latency.  Among the main novelties for O4, the
reconstruction software has been updated to tackle different optical
responses to the motions of the different interferometer mirrors, and
to take into account correlations of noise witness channels in the online noise
subtraction method. In addition, once the detector configuration was
stable in preparation of O4, the frequency-dependent bias of the
reconstructed strain was found to be sufficiently stable in time, thus
the bias has been
corrected online to provide an almost unbiased strain time series during
O4. Finally, the uncertainties in amplitude and phase on the
reconstructed detector strain have been provided, for the first time
for Virgo, as frequency-dependent uncertainties. They are of the order
of 2.5\% in amplitude, \qty{30}{mrad} in phase and \qty{8}{\mu s} in time stamp~\cite{Calibration}.

\subsection{Control electronics}
\label{subsec:SAT}

In order to control the newly installed systems -- such as squeezing,
auxiliary laser, Newtonian calibrator -- it was necessary to
extend the digital infrastructure. This included the
expansion of the timing system; the addition of about \num{1000}
analog-to-digital converters, \num{330} digital-to-analog converts and 92
digital \ac{RF} demodulation channels; also the computing power was increased with 11 additional real-time computers and
the associated real-time software configurations~\cite{VIR-0750C-19}. In total about
\num{250} new electronics board were installed for this purpose,
including fast digital-to-analog converters used in digital control
loops with unity-gain frequencies of a few kHz.

Moreover, the existing Superattenuator's electronics were expanded
with the  addition of a digital signal processing board to actuate on
the newly installed \ac{SR} mirror that replaced the large suspended
output lens. In total there are 10 Superattenuators operating in order to isolate the main
optical elements of Virgo from the ground motion~\cite{aVirgo}. 
These suspensions are controlled using 131 boards equipped 
by multiple core digital signal processing units, which monitor over 400 sensors and
drive 200 actuators.

In addition, in all Superattenuator towers, new remote-controlled double-pole, double-throw 
switches were introduced to physically disconnect both the phase and neutral 
wires of the motor power smart plug, mitigating the excitation of the
last isolation filter cross-bar mode at $\sim$\qty{48}{Hz} which is
near the
\qty{50}{Hz} mains alternate current frequency~\cite{Env_O3}.

An analysis of the Virgo detector downtime shows that Superattenuator
control issues 
caused down periods for at least 1\% of the total time of O4b and O4c
runs.
The downtime estimate includes unlocks and the time required to bring the 
interferometer back to operating conditions. Approximately half of this downtime 
was caused by
glitches in Superattanuator auxiliary channels 
that clearly triggered suspension trips, thus unlocking the interferometer. 
The remaining half resulted from operations needed to restore subsystem conditions, 
such as interventions to recover the Superattenuators working point, or to reboot 
or replace electronic boards. There has been no mechanical issues with
the Superattenuators themselves.

\section{Performance}
\label{sec:performance}

The Virgo detector took data during the \ac{LVK} O4b and O4c runs,
from April 10, 2024 to November 18, 2025. The O4b+O4c run was the
longest data-taking period for Virgo in the advanced detector era, and
the observations were performed jointly with the two LIGO detectors and
intermittently with the KAGRA detector.

In order to maximize the time during which Virgo and the two LIGO
detectors
observe simultaneously, which enables precise sky
localization of transient \ac{GW} events, most planned downtimes
(calibration and commissioning periods) were synchronous for all three
detectors. The exception is the regular weekly maintenance which, for
all detectors, is on Tuesday mornings local timezone. 

This section summarizes the performance of the Virgo detector, standalone
and within the \ac{LVK} network during the O4b and O4c runs. The two
main figures of merit are the instrument duty cycle and its \ac{BNS}
range.

\subsection{Duty cycle}

The transition between the O4b and O4c runs occurred on January 28,
2025 at 17:00 UTC. This was an administrative change to follow the
\ac{LVK} data public releases schedule with approximately 10-months
long chunks: no changes in the detectors occurred at that time.

A network-wide break occurred during O4c for 10 weeks from April 1,
2025 to June 11, 2025, in order to proceed with repairs and
upgrades. To account for that deliberate break in the following
section, we exclude that range from the duty cycle computations, and report
performance separately for O4b+O4c\_1 (until April 1, 2025) and O4c\_2 (from June
11), as the detector changes had relevant impact on detector performance.

\subsubsection{Duty cycle of the Virgo detector}

\begin{table*}
\centering
\caption{Breakdown of the activities on the Virgo detector during the O4b and O4c runs, excluding the 10-week break in spring 2025. The different categories listed in the first column are described in the text.}
\begin{tabular}{lrrr}
  \hline
  \hline
  & O4b + O4c & O4b + O4c\_1 & O4c\_2 \\
  \hline
\texttt{Science} & 68.9\% & 71.8\% & 62.3\% \\
\texttt{Locking} & 10.2\% & 9.0\% & 12.7\% \\
\texttt{Planned downtimes} & 10.2\% & 10.4\% & 9.6\% \\
\texttt{Detector problems} & 5.1\% & 2.8\% & 10.1\% \\
\texttt{Adverse environmental conditions} & 3.6\% & 3.6\% & 3.8\% \\
\texttt{Detector upgrade} & 2.1\% & 2.5\% & 1.4\% \\
  \hline
  \hline

\end{tabular}
\label{tab:Virgo_duty_cycle}
\end{table*}

As shown on table~\ref{tab:Virgo_duty_cycle}, the Virgo duty cycle
during the O4b and O4c runs 
was 69\% (\texttt{Science}),
somewhat lower than during the \ac{O3} run (${\sim} 75$\%)~\cite{Virgo-DetChar-O3-Results}. The
duty cycle was
higher during O4b+O4c\_1 (72\%) than during O4c\_2 (62\%), as the
final data taking period was impacted by more hardware problems than
the previous 12 months of running.

Besides taking data, 10\%
of the
time was spent bringing the detector to operating conditions (\texttt{Locking}). 
The O4 procedure for locking  the detector was very different from
the O3 procedure due to the addition of the \ac{SR} mirror. As
described in section~\ref{sec:nearly-unstable-recycling}, several
control loops were based on dither error signals which resulted in
very slow feedback control.
Overall, the median locking time was about 45~minutes,
(between two and three times longer than in O3), with most of the time
spent to adjust the longitudinal and angular configuration of the \ac{SR} cavity, and then
misalign the \ac{SR} mirror to optimize sensitivity.
Thus, Virgo spent
in average 10\% of the time locking during O4b and O4c, and one of the
reasons for unlocks were glitches in suspension electronics described
in section~\ref{subsec:SAT}.

However, the Virgo control procedures (both the locking phase and then
the control around the working point to allow data taking to continue)
remained robust, with the majority of the locking sequences requiring
a single attempt to be successful, and the segments during which Virgo
was fully controlled have a mean duration higher than 8~hours and
a median duration close to 4~hours.

An additional 10\% of the time 
was spent in scheduled downtimes (\texttt{Planned downtimes}): these
were planned at the network level and comprised of maintenance
that took 4.5 hours per week in average, for 4 hours scheduled;
calibration that took 2.7 hours per week, less than the budgeted 3
hours; and commissioning that took 10 hours per week, slightly more
than the 8 hours initially scheduled. The additional commissioning time
was mainly due to extended slots decided jointly at the
global \ac{LVK} level.

\begin{figure*}
  \center
  \includegraphics[width=\textwidth]{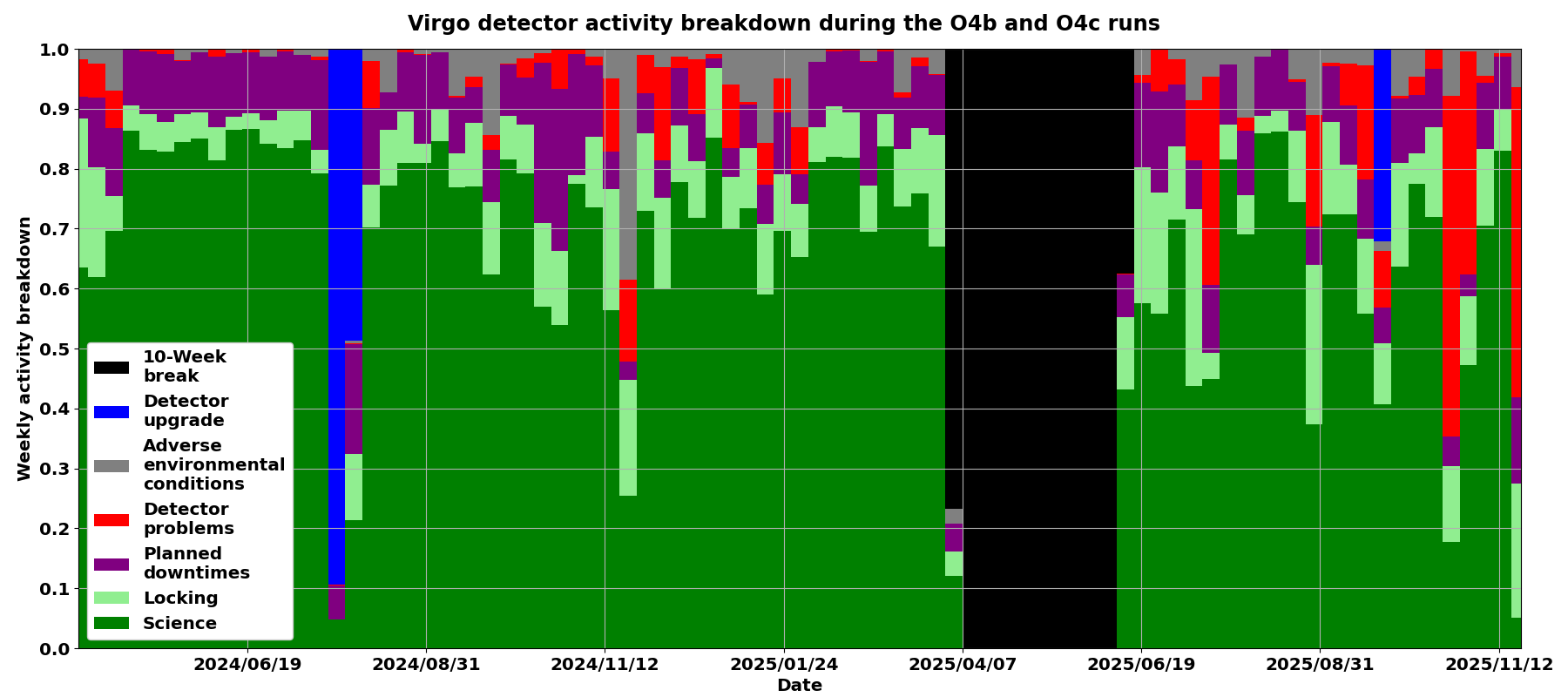}
  \caption{Weekly breakdown of the Virgo detector activities during the O4b and O4c runs.}
  \label{fig:duty_cycle}  
\end{figure*}

There were also down periods induced by problems that accounted for
9\% of the time, these are split between 5\% of
\texttt{Detector problems} and
4\% of environmentally-driven problems (\texttt{Adverse
environmental conditions}, such as local bad weather periods and global
strong earthquakes). The last
2\% of the time were associated with short Virgo downtimes required to
make more invasive actions on the Virgo detector (\texttt{Detector
upgrade}).
The two main periods of this kind were a 10-day break at the end of July
2024 (while the LIGO Hanford detector was down for problems with some
optics) and a couple days in
September 2025. Figure~\ref{fig:duty_cycle}
displays the same information with a weekly granularity.

\subsubsection{LVK network duty cycle}

The LIGO Hanford, LIGO Livingston and Virgo duty cycles during the
O4b+O4c runs were respectively 54\%, 69\% and 69\%~-- to be compared
with ${\sim} 75$\% during the O3 run.
Therefore, the LIGO-Virgo
3-detector duty cycle was only 33\% (compared to 47\% in average
during the 11 months of O3) despite the coordination  among
the detectors to prioritize synchronous data taking.
Two out of three detectors were taking data 36\% of the time (with
the LIGO Livingston - Virgo pair accounting for half of that time,
while the
LIGO Hanford - LIGO Livingston and LIGO Hanford - Virgo pairs counting
for a quarter of that time each); a single detector was taking data
19\% of the time (9\% for Virgo alone).
Finally, the \ac{LVK} network was blind 11\% of the time, with
most of this global downtime being a direct consequence of the choice
to synchronize downtimes in order to maximize the synchronous
observations of all three detectors in the network.

Thanks to the policy of aligning most downtimes of the LIGO and Virgo detectors, Virgo was online during O4b and O4c 77\% of the time the two LIGO detectors were taking data (above 80\% of the time during O4b and O4c\_1, only 70\% of the time during O4c\_2 due to a lower duty cycle). During O4b+O4c, Virgo data was analyzed in 80\% (138/173) of the public alerts.

\subsection{Astrophysical range}

\begin{figure*}
  \center
  \includegraphics[width=\textwidth]{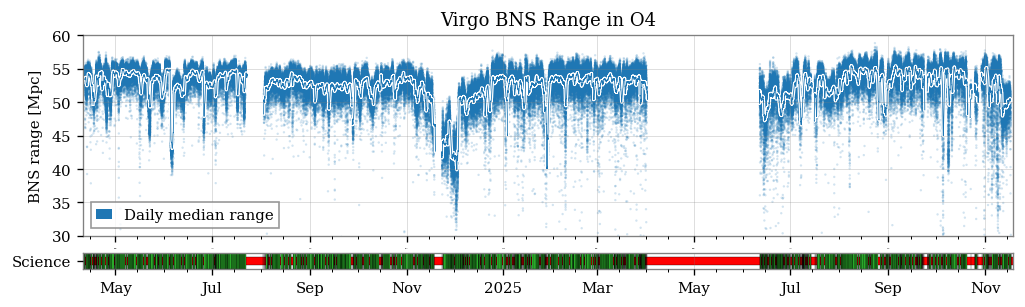}
  \caption{Evolution of the Virgo \ac{BNS} range in O4, from the start of O4b at 15:00 UTC on April 10, 2024, to the end of O4c at 16:00 UTC on November 18, 2025. The data has been taken from the online range computation
, which uses $84$-second sliding estimates with FFTs of 4 seconds and $50\%$ overlap, and is represented by dot markers. The continuous line is the running median range in a day.}
  \label{fig:bns_range_evolution}  
\end{figure*}

\begin{figure}
  \center
  \includegraphics[width=\textwidth]{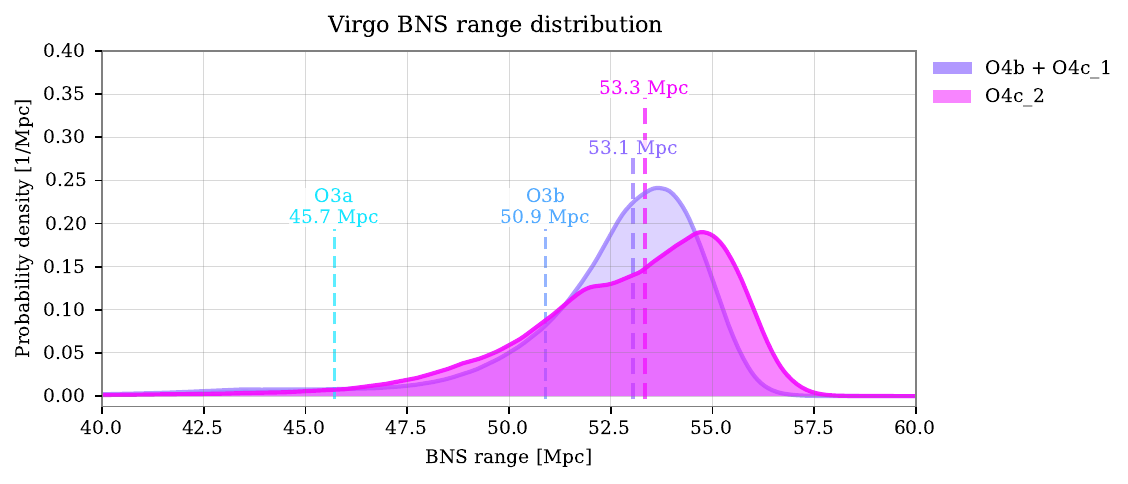}
  \caption{Distribution of the Virgo \ac{BNS} range in O4, divided into O4b plus the first part of O4c until spring break, and the rest of O4c. 
  Vertical dashed lines represent the median ranges for the O3 and O4 subruns.
  These curves have been produced with the same data shown in figure~\ref{fig:bns_range_evolution}.
  }
  \label{fig:bns_range_distribution}  
\end{figure}

The  Virgo \ac{BNS} range  was
consistently above \qty{50}{Mpc} during O4b+O4c, with a median value of
\qty{53}{Mpc}~-- as shown on figures~\ref{fig:bns_range_evolution} and \ref{fig:bns_range_distribution}. 
This is $\sim 10-15$\% higher than during the O3 run, except
for the final weeks of O3b during which the median Virgo \ac{BNS}
range was about \qty{56}{Mpc}. The presence of the {\it excess} noise described in
section~\ref{sec:dominant-noise} prevented a significant improvement
in sensitivity compared to O3. Following the replacement of the \ac{WE} mirror during the Spring 2025 O4c
break, the optimal direction of the \ac{SR} misalignment to reduce the
excess noise changed. That direction was modified in August,
which lead to a gain in sensitivity of \qtyrange{1}{2}{Mpc} on average.

\section{Future prospects}
\label{sec:future}
As a result of the experience of commissioning the nearly unstable
dual-recycled configuration of Virgo, the collaboration has proposed
an upgrade plan to replace these two recycling cavities with two
stable recycling cavities~\cite{Virgo:2026fpe}.
In that configuration the \ac{PR} and \ac{SR}
mirrors would be replaced by three mirrors each, with two mirrors
forming a telescope to reduce the beam radius, as a smaller beam
accumulates Gouy phase on a shorter distance, and the third mirror
closing the cavity~\cite{StableCavityPrinciple}.

The use of stable recycling cavities is expected to strongly reduce
the issues of high sensitivity to small thermal deformation of
mirrors, the lack of a \ac{RF} error signal for \ac{SR} alignment and
the appearance of \ac{RF} offsets described in
section~\ref{sec:nearly-unstable-recycling}. In addition, it should
also significantly reduce the dominant noise which is associated with
\acp{HOM} as described in section~\ref{sec:dominant-noise}.

This mature proposal of a major modification of the Virgo detector is
currently evaluated by funding agencies, and if approved and financed
could be implemented over the next few years. The sensitivity
expectation in the first phase of that upgrade would correspond to a
\ac{BNS} range of \qtyrange{90}{130}{Mpc}, and includes many other
improvements in addition to the change of topology of the recycling
cavities.

\section*{Acknowledgements}
The authors gratefully acknowledge the Italian Istituto Nazionale di Fisica Nucleare (INFN),  
the French Centre National de la Recherche Scientifique (CNRS),
the Netherlands Organization for Scientific Research (NWO), 
the Belgian Fonds de la Recherche Scientifique (FRS-FNRS), 
Actions de Recherche Concertées (ARC) and
Fonds Wetenschappelijk Onderzoek – Vlaanderen (FWO), Belgium,
for the construction and operation of the Virgo detector
and the creation and support of the EGO consortium.
The authors also gratefully acknowledge research support from these agencies as well as by 
the Spanish  Agencia Estatal de Investigaci\'on, 
the Consellera d'Innovaci\'o, Universitats, Ci\`encia i Societat Digital de la Generalitat Valenciana and
the CERCA Programme Generalitat de Catalunya, Spain,
the National Science Centre of Poland and the European Union – European Regional Development Fund; Foundation for Polish Science (FNP),
the Polish Minister of Science,
the Hungarian Scientific Research Fund (OTKA),
the French Lyon Institute of Origins (LIO),
the Aristotle University of Thessaloniki (AUTH),
the European Commission.
The authors gratefully acknowledge the support of the NSF, STFC, INFN, CNRS and Nikhef for provision of computational resources.
	
 We would like to thank all of the essential workers who put their health at risk during the COVID-19 pandemic, without whom we would not have been able to complete this work.

\begin{appendices}

\section{Higher order mode recycling}

\label{sec:hom-recycling}

The Airy function of the circulating power inside a \ac{FP} cavity with respect to the incident power is given by \cite{ismail2016fabry}
\begin{linenomath}
  \begin{equation}
    A_{\mathrm{circ}} = \frac{1-r_1^2}{\left|1-r_1r_2e^{-2i\phi}\right|^2}\,,
    \label{eq:airy_circ}
  \end{equation}
\end{linenomath}
where $r_1$ and $r_2$ are the field amplitude reflection coefficients of the input and end mirrors of the cavity and $\phi$ the single-pass phase shift between the two cavity mirrors. Equation~\eqref{eq:airy_circ} describes the so-called cavity optical gain as a function of the single-pass phase $\phi$.

Considering that the two \ac{FP} arm cavities are equal, the \ac{SRC} can be modelled as a coupled cavity between the signal recycling mirror and a \ac{FP} cavity. The \ac{FP} cavity behaves as a phase or frequency-dependent mirror. In this case, the circulating Airy function for the \ac{SRC} can be written as
\begin{linenomath}
  \begin{equation}
    A_{\mathrm{SRC}}= \frac{1-r_{\mathrm{SRC^2}}}{\left|1-r_{\mathrm{FP}}(\theta)r_{\mathrm{SRC}}e^{-2i\phi_{\mathrm{SRC}}}\right|^2}\,,
    \label{eq:airy_circ_src}
  \end{equation}
\end{linenomath}
where $r_{\mathrm{SRC}}$ and $r_{\mathrm{FP}}$ are the amplitude field reflection coefficients for the signal recycling mirror and \ac{FP} arm cavities. In the absence of losses, the latter can be written as \cite{VIR-0030B-08}
\begin{linenomath}
  \begin{equation}
    r_{\mathrm{FP}}(\theta) = i\frac{r_{\mathrm{IM}}+r_{\mathrm{EM}}e^{2i\theta}}{1+r_{\mathrm{IM}}r_{\mathrm{EM}}e^{2i\theta}}\,,
    \label{eq:fp_r}
  \end{equation}
\end{linenomath}
where $r_{\mathrm{IM}}$ and $r_{\mathrm{EM}}$  are the field amplitude reflection coefficients of the input and end mirrors of the cavity and $\theta$ the single-pass phase shift. Thus, we can retrieve the \ac{FP} reflectivity seen by the resonating carrier by introducing the resonance condition $e^{2i\theta}=-1$ in equation~\eqref{eq:fp_r}, resulting in
\begin{linenomath}
  \begin{equation}
    r_{\mathrm{FP}}^{00} = i \frac{r_{\mathrm{IM}}-r_{\mathrm{EM}}}{1-r_{\mathrm{IM}}r_{\mathrm{EM}}}\,.
    \label{eq:fp_r_00_app}
  \end{equation}
\end{linenomath}

Assuming that any \ac{HOM} with order $m+n$ does not resonate in the \ac{FP} cavity, we can write the cavity reflectivity for an \ac{HOM} using the anti-resonance condition $e^{2i\theta}=1$, resulting in
\begin{linenomath}
  \begin{equation}
    r_{\mathrm{FP}}^{mn} = i \frac{r_{\mathrm{IM}}+r_{\mathrm{EM}}}{1+r_{\mathrm{IM}}r_{\mathrm{EM}}}\,.
    \label{eq:fp_r_mn_app}
  \end{equation}
\end{linenomath}

\begin{figure}
  \centering
  \includegraphics[width=0.7\textwidth]{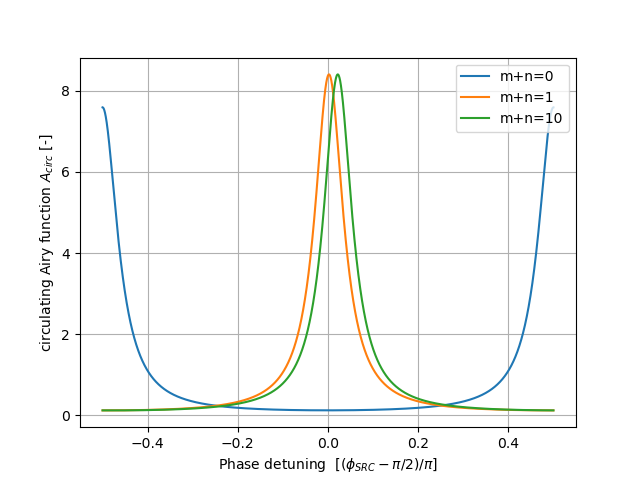}
  \caption{Circulating Airy distributions of equation~\eqref{eq:airy_circ_src_app} inside the \ac{SRC} for different detunings of the \ac{SRC} round-trip phase $\phi_{\mathrm{SRC}}$ for the fundamental, 1st and 10th order modes.}
  \label{fig:SRC_Airy_app}
\end{figure}

Replacing equation~\eqref{eq:fp_r_mn_app} in equation~\eqref{eq:airy_circ_src}, the optical gain of any \ac{HOM} in the \ac{SRC} can be retrieved. To do this, the phase of the \ac{SRC} is written as
\begin{linenomath}
  \begin{equation}
    2\phi_{\mathrm{SRC}}^{mn} = \pi/2 + 2(m+n+1)\phi_{\mathrm{Gouy}}\,,
  \end{equation}
\end{linenomath}
where $\pi/2$ is the working point of the \ac{SRC}  and $\phi_{\mathrm{Gouy}}$ is the Gouy phase of the \ac{SRC}, which in this case takes the value of 0.0068 or \ang{0.39}. In this way, the optical gain of any \ac{HOM} with order $m+n$ in the \ac{SRC} can be written as
\begin{linenomath}
  \begin{equation}
    A_{\mathrm{SRC}}^{mn} = \frac{1-r_{\mathrm{SRC^2}}}{\left|1-r_{\mathrm{FP}}^{mn}r_{\mathrm{SRC}}e^{i\left[\pi/2 + 2(m+n+1)\phi_{\mathrm{Gouy}}\right]}\right|^2}\,.
    \label{eq:airy_circ_src_app}
  \end{equation}
\end{linenomath}

The working point of the \ac{SRC} of $\pi/2$  is chosen such that the carrier fundamental mode, which resonates in the arm cavities, is anti-resonant in the \ac{SRC}. However, since the carrier \acp{HOM} do not resonate in the arm cavities, and considering overcoupled cavities with $r_{\mathrm{EM}}>r_{\mathrm{IM}}$ as is the case in Virgo, $r_{\mathrm{FP}}^{mn}$ has an opposing sign to that of $r_{\mathrm{FP}}^{00}$, as shown in equations \eqref{eq:fp_r_mn_app} and \eqref{eq:fp_r_00_app}. This change of sign causes the carrier \acp{HOM} to resonate in the \ac{SRC}, as shown in figure~\ref{fig:SRC_Airy_app}, which plots equation~\eqref{eq:airy_circ_src_app} for the fundamental, 1st and 10th order modes.

\section{Acronyms}

\begin{acronym}[ ITF]
  \acro{ALS}{auxiliary laser system}
  \acro{BNS}{binary neutron star}
  \acro{BS}{beam splitter}
  \acro{CARM}{common arm motion}
  \acro{CEB}{central building}
  \acro{CH}{central heating}
  \acro{CHRoCC}{central heating radius of curvature correction}
  \acro{CP}{compensation plate}
  \acro{DAC}{digital analog converter}
  \acro{DARM}{differential arm motion}
  \acro{DAS}{double axicon system}
  \acro{DRMI}{dual-recycled Michelson interferometer}
  \acro{EM}{end mirror}
  \acro{EOM}{electro-optic modulator}
  \acro{FI}{Faraday isolator}
  \acro{FP}{Fabry-Perot}
  \acro{GW}{gravitational wave}
  \acro{HOM}{higher-order mode}
  \acro{HWS}{Hartmann wavefront sensor}
  \acro{IM}{input mirror}
  \acro{IMC}{input mode cleaner}
  \acro{LVK}{LIGO-Virgo-KAGRA}
  \acro{MICH}{short Michelson interferometer}
  \acro{NE}{north end}
  \acro{NEB}{north end building}
  \acro{NI}{north input}
  \acro{OMC}{output mode-cleaner}
  \acro{O3}{the third observing run}
  \acro{O4}{the fourth observing run}
  \acro{PDH}{Pound Drever Hall}
  \acro{POP}{pick off plate}
  \acro{PR}{power recycling}
  \acro{PRC}{power recycling cavity}
  \acro{PRCL}{power recycling cavity length}
  \acro{PSTAB}{power stabilization}
  \acro{RF}{radio frequency}
  \acro{RFC}{reference cavity}
  \acro{RH}{ring heater}
  \acrodefplural{RoC}{radii of curvature}
  \acro{RoC}{radius of curvature}
  \acro{SIB1}{first suspended injection bench}
  \acro{SDB1}{first suspended detection bench}
  \acro{SDB2}{second suspended detection bench}
  \acro{SPRB}{suspended power recycling bench}
  \acro{SNEB}{suspended north end bench}
  \acro{SWEB}{suspended west end bench}
  \acro{SR}{signal recycling}
  \acro{SRC}{signal recycling cavity}
  \acro{SRCL}{signal recycling cavity length}
  \acro{SSFS}{second stage frequency stabilization}
  \acro{TCS}{thermal compensation system}
  \acro{WE}{west end}
  \acro{WEB}{west end building}
  \acro{WI}{west input}
\end{acronym}

\end{appendices}

\bibliographystyle{unsrturl}

\begin{thebibliography}{10}

\bibitem{aVirgo}
F.~Acernese et~al.
\newblock {Advanced Virgo: a second-generation interferometric gravitational
  wave detector}.
\newblock {\em Classical and Quantum Gravity}, 32(2):024001, December 2015.
\newblock URL:
  \url{https://iopscience.iop.org/article/10.1088/0264-9381/32/2/024001}, \href
  {https://doi.org/10.1088/0264-9381/32/2/024001}
  {\path{doi:10.1088/0264-9381/32/2/024001}}.

\bibitem{VIR-0596A-19}
{The Virgo Collaboration}.
\newblock {Advanced Virgo Plus Phase I: Design Report}.
\newblock Technical Report VIR-0596A-19, June 2019.
\newblock URL: \url{https://tds.virgo-gw.eu/ql/?c=14430}.

\bibitem{Flaminio:2020lqk}
Raffaele Flaminio.
\newblock {Status and plans of the Virgo gravitational wave detector}.
\newblock {\em Proc. SPIE Int. Soc. Opt. Eng.}, 11445:1144511, 2020.
\newblock \href {https://doi.org/10.1117/12.2565418}
  {\path{doi:10.1117/12.2565418}}.

\bibitem{virgocoll:FrequencyDependentSqueezedVacuum2023}
The~Virgo Collaboration, H.~Vahlbruch, M.~Mehmet, H.~Lück, and K.~Danzmann.
\newblock {Frequency-dependent squeezed vacuum source for the Advanced Virgo
  gravitational-wave detector}.
\newblock {\em Physical Review Letters}, 131(4):041403, July 2023.
\newblock URL: \url{https://link.aps.org/doi/10.1103/PhysRevLett.131.041403},
  \href {https://doi.org/10.1103/PhysRevLett.131.041403}
  {\path{doi:10.1103/PhysRevLett.131.041403}}.

\bibitem{VIRGO:2025sym}
F.~Acernese et~al.
\newblock {Optical characterization of the Advanced Virgo gravitational wave
  detector for the O4 observing run}.
\newblock {\em Applied Optics}, 64(17):4710--4726, June 2025.
\newblock URL: \url{https://opg.optica.org/ao/abstract.cfm?URI=ao-64-17-4710},
  \href {https://doi.org/10.1364/AO.555312} {\path{doi:10.1364/AO.555312}}.

\bibitem{Black01}
E.~D. {Black}.
\newblock {An introduction to Pound-Drever-Hall laser frequency stabilization}.
\newblock {\em American Journal of Physics}, 69(1):79--87, January 2001.
\newblock \href
  {http://arxiv.org/abs/https://pubs.aip.org/aapt/ajp/article-pdf/69/1/79/10115998/79_1_online.pdf}
  {\path{arXiv:https://pubs.aip.org/aapt/ajp/article-pdf/69/1/79/10115998/79_1_online.pdf}},
  \href {https://doi.org/10.1119/1.1286663} {\path{doi:10.1119/1.1286663}}.

\bibitem{PDH_1983}
R.~W.~P. Drever, J.~L. Hall, F.~V. Kowalski, J.~Hough, G.~M. Ford, A.~J.
  Munley, and H.~Ward.
\newblock {Laser phase and frequency stabilization using an optical resonator}.
\newblock 31(2):97--105, June 1983.
\newblock \href {https://doi.org/10.1007/BF00702605}
  {\path{doi:10.1007/BF00702605}}.

\bibitem{Kogelnik1966}
H.~Kogelnik and T.~Li.
\newblock {Laser beams and resonators}.
\newblock {\em Applied Optics}, 5(10):1550--1567, October 1966.
\newblock URL: \url{https://opg.optica.org/ao/abstract.cfm?URI=ao-5-10-1550},
  \href {https://doi.org/10.1364/AO.5.001550} {\path{doi:10.1364/AO.5.001550}}.

\bibitem{siegman1986lasers}
A.~E. Siegman.
\newblock {\em {Lasers}}.
\newblock {University Science Books}, 1986.

\bibitem{paper_TCS}
V.~Fafone et~al.
\newblock {Wavefront sensing and aberration mitigation in Advanced Virgo}.
\newblock {\em in preparation}, 2026.

\bibitem{Ward_1994}
E.~Morrison, B.~J. Meers, D.~I. Robertson, and H.~Ward.
\newblock {Automatic alignment of optical interferometers}.
\newblock 33(22):5041--5049, August 1994.
\newblock URL: \url{https://opg.optica.org/ao/abstract.cfm?URI=ao-33-22-5041},
  \href {https://doi.org/10.1364/AO.33.005041}
  {\path{doi:10.1364/AO.33.005041}}.

\bibitem{Virgo-DetChar-O3-Results}
F.~Acernese et~al.
\newblock {Virgo detector characterization and data quality: results from the
  O3 run}.
\newblock {\em Classical and Quantum Gravity}, 40(18):185006, August 2023.
\newblock \href {https://doi.org/10.1088/1361-6382/acd92d}
  {\path{doi:10.1088/1361-6382/acd92d}}.

\bibitem{GraefRollins:2016hki}
J.~Graef~Rollins.
\newblock {Distributed state machine supervision for long-baseline
  gravitational-wave detectors}.
\newblock {\em Review of Scientific Instruments}, 87(9):094502, September 2016.
\newblock \href
  {http://arxiv.org/abs/https://pubs.aip.org/aip/rsi/article-pdf/doi/10.1063/1.4961665/15760069/094502_1_online.pdf}
  {\path{arXiv:https://pubs.aip.org/aip/rsi/article-pdf/doi/10.1063/1.4961665/15760069/094502_1_online.pdf}},
  \href {https://doi.org/10.1063/1.4961665} {\path{doi:10.1063/1.4961665}}.

\bibitem{Bersanetti_Metatron}
D.~Bersanetti et~al.
\newblock {Metatron: the Virgo implementation of the LIGO Guardian finite state
  machine environment}.
\newblock Technical Report VIR-0199A-26, May 2026.
\newblock URL: \url{https://tds.virgo-gw.eu/ql/?c=22605}.

\bibitem{Acernese_2006}
F.~Acernese et~al.
\newblock {The status of VIRGO}.
\newblock 23(8):S63, March 2006.
\newblock \href {https://doi.org/10.1088/0264-9381/23/8/S09}
  {\path{doi:10.1088/0264-9381/23/8/S09}}.

\bibitem{Aasi_2015_ALIGO}
J.~Aasi et~al.
\newblock {Advanced LIGO}.
\newblock 32(7):074001, March 2015.
\newblock \href {https://doi.org/10.1088/0264-9381/32/7/074001}
  {\path{doi:10.1088/0264-9381/32/7/074001}}.

\bibitem{DeRossi2020}
C.~De~Rossi, J.~Brooks, J.~Casanueva~Diaz, A.~Chiummo, E.~Genin, M.~Gosselin,
  N.~Leroy, M.~Mantovani, B.~Montanari, F.~Nocera, and G.~Pillant.
\newblock {Development of a frequency tunable green laser source for Advanced
  Virgo+ gravitational waves detector}.
\newblock 8(4):87, December 2020.
\newblock URL: \url{https://www.mdpi.com/2075-4434/8/4/87}, \href
  {https://doi.org/10.3390/galaxies8040087}
  {\path{doi:10.3390/galaxies8040087}}.

\bibitem{Bersanetti_2022}
D.~Bersanetti, M.~Boldrini, J.~Casanueva~Diaz, A.~Freise, R.~Maggiore,
  M.~Mantovani, and M.~Valentini.
\newblock {Simulations for the locking and alignment strategy of the DRMI
  configuration of the Advanced Virgo Plus Detector}.
\newblock 10(6):115, December 2022.
\newblock URL: \url{https://www.mdpi.com/2075-4434/10/6/115}, \href
  {https://doi.org/10.3390/galaxies10060115}
  {\path{doi:10.3390/galaxies10060115}}.

\bibitem{Arai_2002}
K.~Arai et~al.
\newblock {Sensing and controls for Power-Recycling of TAMA300}.
\newblock 19(7):1843--1849, March 2002.
\newblock \href {https://doi.org/10.1088/0264-9381/19/7/383}
  {\path{doi:10.1088/0264-9381/19/7/383}}.

\bibitem{HWS_Lorenzo}
L.~Aiello, P.~P. Palma, M.~Lorenzini, E.~Cesarini, M.~Cifaldi, C.~Di~Fronzo,
  D.~Lumaca, Y.~Minenkov, I.~Nardecchia, A.~Rocchi, C.~Taranto, and V.~Fafone.
\newblock {Thermal defocus-free Hartmann Wavefront Sensors for monitoring
  aberrations in Advanced Virgo}.
\newblock {\em Classical and Quantum Gravity}, 41(12):125001, May 2024.
\newblock URL:
  \url{https://iopscience.iop.org/article/10.1088/1361-6382/ad4508}, \href
  {https://doi.org/10.1088/1361-6382/ad4508}
  {\path{doi:10.1088/1361-6382/ad4508}}.

\bibitem{PhaseCamera_Laura}
L.~van~der Schaaf, K.~Agatsuma, M.~van Beuzekom, M.~Gebyehu, and J.~van~den
  Brand.
\newblock {Advanced Virgo phase cameras}.
\newblock {\em Journal of Physics: Conference Series}, 718(7):072008, May 2016.
\newblock URL:
  \url{https://iopscience.iop.org/article/10.1088/1742-6596/718/7/072008},
  \href {https://doi.org/10.1088/1742-6596/718/7/072008}
  {\path{doi:10.1088/1742-6596/718/7/072008}}.

\bibitem{VirgoRH}
I.~Nardecchia, Y.~Minenkov, M.~Lorenzini, L.~Aiello, E.~Cesarini, D.~Lumaca,
  V.~Malvezzi, F.~Paoletti, A.~Rocchi, and V.~Fafone.
\newblock {Optimized radius of curvature tuning for the virgo core optics}.
\newblock {\em Classical and Quantum Gravity}, 40(5):055004, February 2023.
\newblock URL:
  \url{https://iopscience.iop.org/article/10.1088/1361-6382/acb632}, \href
  {https://doi.org/10.1088/1361-6382/acb632}
  {\path{doi:10.1088/1361-6382/acb632}}.

\bibitem{Accadia2013_CHRoCC}
T.~Accadia et~al.
\newblock {Central heating radius of curvature correction (CHRoCC) for use in
  large scale gravitational-wave interferometers}.
\newblock {\em Classical and Quantum Gravity}, 30(5):055017, February 2013.
\newblock \href {https://doi.org/10.1088/0264-9381/30/5/055017}
  {\path{doi:10.1088/0264-9381/30/5/055017}}.

\bibitem{LIGOScientific:2021kro}
A.~F. Brooks et~al.
\newblock {Point absorbers in Advanced LIGO}.
\newblock {\em Appied Optics}, 60(13):4047--4063, April 2021.
\newblock \href {https://doi.org/10.1364/AO.419689}
  {\path{doi:10.1364/AO.419689}}.

\bibitem{cifaldiThesis}
M.~Cifaldi.
\newblock {\em {Mitigation of anomalous absorptions in the Virgo core optics}}.
\newblock PhD thesis, {Tor Vergata University of Rome}, July 2023.
\newblock URL: \url{https://tds.virgo-gw.eu/ql/?c=19759}.

\bibitem{Ipatsia_studies}
I.~Nardecchia, M.~Cifaldi, S.~Melo, M.~Lorenzini, and P.~Spinicelli.
\newblock {IPATSiA studies}.
\newblock Technical Report VIR-0138A-26, February 2026.
\newblock URL: \url{https://tds.virgo-gw.eu/ql/?c=22544}.

\bibitem{galaxies8040085}
A.~Allocca, D.~Bersanetti, J.~Casanueva~Diaz, C.~De~Rossi, M.~Mantovani,
  A.~Masserot, L.~Rolland, P.~Ruggi, B.~Swinkels, E.~N. Tapia San~Martin,
  M.~Vardaro, and M.~Was.
\newblock {Interferometer sensing and control for the Advanced Virgo experiment
  in the O3 scientific run}.
\newblock {\em Galaxies}, 8(4):85, December 2020.
\newblock URL: \url{https://www.mdpi.com/2075-4434/8/4/85}, \href
  {https://doi.org/10.3390/galaxies8040085}
  {\path{doi:10.3390/galaxies8040085}}.

\bibitem{Pinto_ASC}
M.~Pinto, D.~Bersanetti, M.~Boldrini, J.~Casanueva~Diaz, M.~Mantovani, and
  P.~Ruggi.
\newblock {Automatic alignment in Advanced Virgo + during O4}.
\newblock {\em in preparation}, 2026.

\bibitem{martynov_phd_2015}
D.~Martynov.
\newblock {\em {Lock Acquisition and Sensitivity Analysis of Advanced LIGO
  Interferometers}}.
\newblock PhD thesis, California Institute of Technology, 2015.

\bibitem{boldrini:AutomaticAlignment2023}
M.~Boldrini.
\newblock {\em {Automatic alignment in Advanced Virgo + Phase I and effects of
  radiation pressure}}.
\newblock PhD thesis, Sapienza Università di Roma, 2023.

\bibitem{Smith-Lefebvre:2011joa}
Nicolas Smith-Lefebvre, Stefan Ballmer, Matt Evans, Sam Waldman, Keita Kawabe,
  Valery Frolov, and Nergis Mavalvala.
\newblock {Optimal Alignment Sensing of a Readout Mode Cleaner Cavity}.
\newblock {\em Opt. Lett.}, 36:4365, 2011.
\newblock \href {http://arxiv.org/abs/1110.4122} {\path{arXiv:1110.4122}},
  \href {https://doi.org/10.1364/OL.36.004365}
  {\path{doi:10.1364/OL.36.004365}}.

\bibitem{vanDael2024}
M.~van Dael, G.~Witvoet, B.~Swinkels, M.~Pinto, D.~Bersanetti, J.~Casanueva,
  P.~Ruggi, M.~Mantovani, P.~Spinicelli, C.~De~Rossi, M.~Boldrini, and
  T.~Oomen.
\newblock {Online decoupling of the time-varying longitudinal feedback loops
  for improved performance in Advanced Virgo Plus}.
\newblock {\em Classical and Quantum Gravity}, 41(21):215008, October 2024.
\newblock \href {https://doi.org/10.1088/1361-6382/ad7cb9}
  {\path{doi:10.1088/1361-6382/ad7cb9}}.

\bibitem{Boldrini_AdV+_SRCvsDARM}
M.~Boldrini, D.~Bersanetti, J.~Casanueva~Diaz, M.~Mantovani, M.~Pinto, and
  P.~Ruggi.
\newblock {Interaction of Signal Recycling Cavity and DARM in Advanced Virgo+
  during O4}.
\newblock {\em in preparation}, 2026.

\bibitem{INJ_O4}
M.~Gosselin, C.~De~Rossi, S.~Melo, and P.~Spinicelli.
\newblock {Injection System for the observing run O4}.
\newblock {\em in preparation}, 2026.

\bibitem{PhysRevA.79.053824}
F.~Acernese et~al.
\newblock {Laser with an in-loop relative frequency stability of
  $1.0\ifmmode\times\else\texttimes\fi{}{10}^{\ensuremath{-}21}$ on a 100-ms
  time scale for gravitational-wave detection}.
\newblock {\em Physical Review A}, 79:053824, May 2009.
\newblock URL: \url{https://link.aps.org/doi/10.1103/PhysRevA.79.053824}, \href
  {https://doi.org/10.1103/PhysRevA.79.053824}
  {\path{doi:10.1103/PhysRevA.79.053824}}.

\bibitem{SSFS}
M.~van Dael, J.~Casanueva, G.~Witvoet, B.~Swinkels, D.~Bersanetti, M.~Pinto,
  P.~Ruggi, M.~Mantovani, C.~De~Rossi, P.~Spinicelli, M.~Boldrini, and
  T.~Oomen.
\newblock {Control of the laser frequency in the Virgo interferometer: dynamic
  noise budgeting for controller optimization}.
\newblock {\em Astroparticle Physics}, 164:103028, January 2025.
\newblock URL:
  \url{https://www.sciencedirect.com/science/article/pii/S0927650524001051},
  \href {https://doi.org/10.1016/j.astropartphys.2024.103028}
  {\path{doi:10.1016/j.astropartphys.2024.103028}}.

\bibitem{VIR-0062A-21}
A.~Freise and M.~Was.
\newblock {ITM etalon modelling for Advanced Virgo Plus}.
\newblock Technical Report VIR-0062A-21, Jan 2021.
\newblock URL: \url{https://tds.virgo-gw.eu/?r=18192}.

\bibitem{kubo:FluctuationdissipationTheorem1966}
R.~Kubo.
\newblock {The fluctuation-dissipation theorem}.
\newblock {\em Reports on Progress in Physics}, 29(1):255, January 1966.
\newblock \href {https://doi.org/10.1088/0034-4885/29/1/306}
  {\path{doi:10.1088/0034-4885/29/1/306}}.

\bibitem{Levin98}
Y.~Levin.
\newblock {Internal thermal noise in the LIGO test masses: a direct approach}.
\newblock {\em Physical Review D}, 57(2):659--663, January 1998.
\newblock URL: \url{https://link.aps.org/doi/10.1103/PhysRevD.57.659}, \href
  {https://doi.org/10.1103/PhysRevD.57.659}
  {\path{doi:10.1103/PhysRevD.57.659}}.

\bibitem{Hild:2008pb}
S.~Hild et~al.
\newblock {DC-readout of a signal-recycled gravitational wave detector}.
\newblock {\em Classical and Quantum Gravity}, 26(5):055012, February 2009.
\newblock \href {https://doi.org/10.1088/0264-9381/26/5/055012}
  {\path{doi:10.1088/0264-9381/26/5/055012}}.

\bibitem{poliniThesis}
E.~Polini.
\newblock {\em {Broadband quantum noise reduction in AdV+ : from
  frequency-dependent squeezing implementation to detection losses and
  scattered light mitigation}}.
\newblock PhD thesis, {Université Savoie Mont Blanc}, December 2022.
\newblock URL: \url{https://theses.hal.science/tel-04124206}.

\bibitem{EB_scatteredlight}
M.~Was, R.~Gouaty, and R.~Bonnand.
\newblock {End benches scattered light modeling and subtraction in Advanced
  Virgo}.
\newblock {\em Classical and Quantum Gravity}, 38(7):075020, March 2021.
\newblock \href {https://doi.org/10.1088/1361-6382/abe759}
  {\path{doi:10.1088/1361-6382/abe759}}.

\bibitem{DF_telescope}
C.~Buy, E.~Genin, M.~Barsuglia, R.~Gouaty, and M.~Tacca.
\newblock {Design of a high-magnification and low-aberration compact
  catadioptric telescope for the Advanced Virgo gravitational-wave
  interferometric detector}.
\newblock {\em Classical and Quantum Gravity}, 34(9):095011, April 2017.
\newblock \href {https://doi.org/10.1088/1361-6382/aa65e3}
  {\path{doi:10.1088/1361-6382/aa65e3}}.

\bibitem{finesse3_backscattering}
A.~Demagny.
\newblock Backscattering simulation.
\newblock GIT:
  \url{https://git.ligo.org/augustin.demagny/backscattering_simulation}, 2026.
\newblock Commit 0520205e, accessed 2026-03-04.

\bibitem{paper_O3_environmentNoise}
F.~Acernese et~al.
\newblock {The Virgo O3 run and the impact of the environment}.
\newblock {\em Classical and Quantum Gravity}, 39(23):235009, November 2022.
\newblock \href {https://doi.org/10.1088/1361-6382/ac776a}
  {\path{doi:10.1088/1361-6382/ac776a}}.

\bibitem{VIR-1047A-19}
A.~Allocca, A.~Chiummo, P.~Ruggi, and H.~Yamamoto.
\newblock {Transient power drop in dark fringe lock acquisition during the
  commissioning before O3}.
\newblock Technical Report VIR-1047A-19, October 2019.
\newblock URL: \url{https://tds.virgo-gw.eu/ql/?c=14881}.

\bibitem{GWTC-5-intro}
The LIGO~Scientific Collaboration, the Virgo~Collaboration, and the
  KAGRA~Collaboration.
\newblock {GWTC-5.0: an introduction to version 5.0 of the gravitational-wave
  transient catalog}.
\newblock {\em in preparation}, 2026.
\newblock \href {http://arxiv.org/abs/2605.27223} {\path{arXiv:2605.27223}}.

\bibitem{Env_Couple}
P.~Nguyen et~al.
\newblock {Environmental noise in advanced LIGO detectors}.
\newblock {\em Classical and Quantum Gravity}, 38(14):145001, June 2021.
\newblock \href {https://doi.org/10.1088/1361-6382/ac011a}
  {\path{doi:10.1088/1361-6382/ac011a}}.

\bibitem{Env_O3}
I.~Fiori et~al.
\newblock {The hunt for environmental noise in Virgo during the third observing
  run}.
\newblock {\em Galaxies}, 8(4):82, December 2020.
\newblock URL: \url{https://www.mdpi.com/2075-4434/8/4/82}, \href
  {https://doi.org/10.3390/galaxies8040082}
  {\path{doi:10.3390/galaxies8040082}}.

\bibitem{ducrot:tel-01489175}
M.~Ducrot.
\newblock {\em {Etude des cavités optiques de filtrage de sortie du détecteur
  d'ondes gravitationnelles Advanced Virgo}}.
\newblock PhD thesis, {Université Grenoble Alpes}, September 2016.
\newblock URL: \url{https://theses.hal.science/tel-01489175}.

\bibitem{paper_OMC_O4}
W.~Amar et~al.
\newblock {A high finesse output mode cleaner cavity for Advanced Virgo +}.
\newblock {\em in preparation}, 2026.

\bibitem{Bonnand:2017hdk}
R.~Bonnand, M.~Ducrot, R.~Gouaty, F.~Marion, A.~Masserot, B.~Mours, E.~Pacaud,
  L.~Rolland, and M.~Wąs.
\newblock {Upper-limit on the Advanced Virgo output mode cleaner cavity length
  noise}.
\newblock {\em Classical and Quantum Gravity}, 34(17):175002, July 2017.
\newblock \href {https://doi.org/10.1088/1361-6382/aa7f64}
  {\path{doi:10.1088/1361-6382/aa7f64}}.

\bibitem{Virgo:2020xlu}
F.~Acernese et~al.
\newblock {Quantum backaction on kg-scale mirrors: observation of radiation
  pressure noise in the Advanced Virgo detector}.
\newblock {\em Physical Review Letters}, 125(13):131101, September 2020.
\newblock URL: \url{https://link.aps.org/doi/10.1103/PhysRevLett.125.131101},
  \href {https://doi.org/10.1103/PhysRevLett.125.131101}
  {\path{doi:10.1103/PhysRevLett.125.131101}}.

\bibitem{virgocoll:IncreasingAstrophysicalReach2019}
{The Virgo Collaboration}, H.~Vahlbruch, M.~Mehmet, H.~Lück, and K.~Danzmann.
\newblock {Increasing the astrophysical reach of the Advanced Virgo detector
  via the application of squeezed vacuum states of light}.
\newblock {\em Physical Review Letters}, 123(23):231108, December 2019.
\newblock URL: \url{https://link.aps.org/doi/10.1103/PhysRevLett.123.231108},
  \href {https://doi.org/10.1103/PhysRevLett.123.231108}
  {\path{doi:10.1103/PhysRevLett.123.231108}}.

\bibitem{kwee:DecoherenceDegradationSqueezed2014}
P.~Kwee, J.~Miller, T.~Isogai, L.~Barsotti, and M.~Evans.
\newblock {Decoherence and degradation of squeezed states in quantum filter
  cavities}.
\newblock {\em Physical Review D}, 90(6):062006, September 2014.
\newblock URL: \url{https://link.aps.org/doi/10.1103/PhysRevD.90.062006}, \href
  {https://doi.org/10.1103/PhysRevD.90.062006}
  {\path{doi:10.1103/PhysRevD.90.062006}}.

\bibitem{toyra17}
D.~Töyrä, D.~D. Brown, M.K. Davis, S.~Song, A.~Wormald, J.~Harms, H.~Miao,
  and A.~Freise.
\newblock {Multi-spatial-mode effects in squeezed-light-enhanced
  interferometric gravitational wave detectors}.
\newblock {\em Physical Review D}, 96(2):022006, July 2017.
\newblock URL: \url{https://link.aps.org/doi/10.1103/PhysRevD.96.022006}, \href
  {https://doi.org/10.1103/PhysRevD.96.022006}
  {\path{doi:10.1103/PhysRevD.96.022006}}.

\bibitem{mcculler:LIGOsQuantumResponse2021}
L.~McCuller et~al.
\newblock {LIGO's quantum response to squeezed states}.
\newblock {\em Physical Review D}, 104(6):062006, September 2021.
\newblock URL: \url{https://link.aps.org/doi/10.1103/PhysRevD.104.062006},
  \href {https://doi.org/10.1103/PhysRevD.104.062006}
  {\path{doi:10.1103/PhysRevD.104.062006}}.

\bibitem{demarco:EnhancingAstrophysicalReach2025}
F.~De~Marco.
\newblock {\em {Enhancing the astrophysical reach of present and future
  gravitational-wave detectors via quantum squeezing}}.
\newblock PhD thesis, Sapienza Università di Roma, 2025.

\bibitem{SQZdegradation}
R.~Flaminio.
\newblock {Squeezing degradation in a degenerate signal recycling cavity?}
\newblock Technical Report VIR-0239A-23, March 2023.
\newblock URL: \url{https://tds.virgo-gw.eu/ql/?c=19109}.

\bibitem{VIR-0435A-19}
A.~Chiummo, EGO~Optics Group, INFN Pisa, and IFAE.
\newblock {AdV+: IMC payload and instrumented baffles @VW}.
\newblock Technical Report VIR-0435A-19, April 2019.
\newblock URL: \url{https://tds.virgo-gw.eu/ql/?c=14263}.

\bibitem{Octopus_sw}
P.~Ruggi, M.~Pinto, L.~Trozzo, G.~Cella, E.~Majorana, G.~Losurdo, P.~Chessa,
  A.~Longo, and A.~Viceré.
\newblock {Mechanical simulation tool based on impedance matrices}.
\newblock {\em Physical Review D}, 112:022002, July 2025.
\newblock URL: \url{https://link.aps.org/doi/10.1103/PhysRevD.112.022002},
  \href {https://doi.org/10.1103/PhysRevD.112.022002}
  {\path{doi:10.1103/PhysRevD.112.022002}}.

\bibitem{Chiummo2019_sims}
P.~Ruggi, A.~Basti, A.~Chiummo, and F.~Frasconi.
\newblock {AdV+ IMC payload replacement - Octopus Transfer Functions}.
\newblock Technical Report VIR-0111A-20, January 2020.
\newblock URL: \url{https://tds.virgo-gw.eu/ql/?c=15225}.

\bibitem{logbook_50065}
P.~Ruggi.
\newblock {MC local controls: TX}.
\newblock Logbook entry: \url{https://logbook.virgo-gw.eu/virgo/?r=50065},
  2020.
\newblock Accessed 2026-05-08.

\bibitem{instr_baffle}
M.~Andrés-Carcasona, O.~Ballester, O.~Blanch, J.~Campos, G.~Caneva,
  L.~Cardiel, M.~Cavalli-Sforza, P.~Chiggiato, A.~Chiummo, V.~Dattilo, et~al.
\newblock {Instrumented baffle for the Advanced Virgo input mode cleaner end
  mirror}.
\newblock {\em Physical Review D}, 107:062001, March 2023.
\newblock URL: \url{https://link.aps.org/doi/10.1103/PhysRevD.107.062001},
  \href {https://doi.org/10.1103/PhysRevD.107.062001}
  {\path{doi:10.1103/PhysRevD.107.062001}}.

\bibitem{Ballester:2021bua}
O.~Ballester et~al.
\newblock {Measurement of the stray light in the Advanced Virgo input mode
  cleaner cavity using an instrumented baffle}.
\newblock {\em Class. Quant. Grav.}, 39(11):115011, 2022.
\newblock \href {http://arxiv.org/abs/2111.09312} {\path{arXiv:2111.09312}},
  \href {https://doi.org/10.1088/1361-6382/ac6a9d}
  {\path{doi:10.1088/1361-6382/ac6a9d}}.

\bibitem{fastunlocks}
M.~Turconi et~al.
\newblock {Fast unlocks saga in Virgo: an experimental investigation and
  mitigation strategy}.
\newblock {\em in preparation}, 2026.

\bibitem{PcalO3}
D.~Estevez, P.~Lagabbe, A.~Masserot, L.~Rolland, M.~Seglar-Arroyo, and
  D.~Verkindt.
\newblock {The Advanced Virgo photon calibrators}.
\newblock {\em Classical and Quantum Gravity}, 38(7):075007, February 2021.
\newblock \href {https://doi.org/10.1088/1361-6382/abe2db}
  {\path{doi:10.1088/1361-6382/abe2db}}.

\bibitem{NCal}
F.~Aubin, E.~Dangelser, D.~Estevez, A.~Masserot, B.~Mours, T.~Pradier, A.~Syx,
  and P.~Van~Hove.
\newblock {The Virgo newtonian calibration system for the O4 observing run}.
\newblock {\em Classical and Quantum Gravity}, 41(23):235003, October 2024.
\newblock \href {https://doi.org/10.1088/1361-6382/ad869c}
  {\path{doi:10.1088/1361-6382/ad869c}}.

\bibitem{PcalO4}
C.~Grimaud et~al.
\newblock {Calibration of the Advanced Virgo Photon Calibrator for the
  observing run O4}.
\newblock {\em in preparation}, 2026.

\bibitem{Calibration}
{The Virgo Collaboration}.
\newblock Calibration of the {Advanced Virgo Plus} gravitational wave detector
  and reconstruction of the detector strain h(t) during the observing run {O4}.
\newblock {\em in preparation}, 2026.

\bibitem{VIR-0750C-19}
A.~Dalmaz, N.~Letendre, A.~Masserot, B.~Mours, E.~Pacaud, S.~Petit, and
  L.~Rolland.
\newblock {DaqBox and Mezzanines user manual }.
\newblock Technical Report VIR-0750C-19, Jan 2023.
\newblock URL: \url{https://tds.virgo-gw.eu/?r=21502}.

\bibitem{Virgo:2026fpe}
F.~Acernese et~al.
\newblock {Advanced Virgo Plus for O5 -- Design Report Overview}.
\newblock March 2026.
\newblock \href {http://arxiv.org/abs/2603.20342} {\path{arXiv:2603.20342}},
  \href {https://doi.org/10.48550/arXiv.2603.20342}
  {\path{doi:10.48550/arXiv.2603.20342}}.

\bibitem{StableCavityPrinciple}
W.~Amar, R.~Bonnand, R.~Flaminio, and E.~Tournefier.
\newblock Optical design of stable recycling cavities for the virgo
  gravitational wave detector.
\newblock {\em in preparation}, 2026.

\bibitem{ismail2016fabry}
N.~Ismail, C.~C. Kores, D.~Geskus, and M.~Pollnau.
\newblock {Fabry-Perot resonator: spectral line shapes, generic and related
  Airy distributions, linewidths, finesses, and performance at low or
  frequency-dependent reflectivity}.
\newblock {\em Optics Express}, 24(15):16366--16389, July 2016.
\newblock \href {https://doi.org/10.1364/OE.24.016366}
  {\path{doi:10.1364/OE.24.016366}}.

\bibitem{VIR-0030B-08}
G.~Vajente.
\newblock {Signal recycling I: Field equations}.
\newblock Technical Report VIR-0030B-08, June 2008.
\newblock URL: \url{https://tds.virgo-gw.eu/ql/?c=2002}.

\end{thebibliography}

\end{document}